\documentclass{aa}  

\usepackage{graphicx}
\usepackage{txfonts}

\usepackage{color} 

\usepackage{lscape}
\usepackage{rotating}

\begin{document}

   \title{HADES RV Programme with HARPS-N at TNG\thanks{Based on observations collected at the European Southern Observatory, Chile
             under programme ID 0102.D-0119(A) and 0102.D-0119(B)}}

  \subtitle{XII. The abundance signature of M dwarf stars with planets}

   \author{J. Maldonado\inst{1}
          \and G. Micela    \inst{1}
          \and M. Baratella \inst{2,3}
	  \and V. D'Orazi   \inst{3}
	  \and L. Affer     \inst{1}
	  \and K. Biazzo    \inst{4}
          \and A. F. Lanza  \inst{4}
	  \and A. Maggio    \inst{1}
	  \and J. I. Gonz\'alez Hern\'andez \inst{5,6}
	  \and M. Perger    \inst{7,8}
	  \and M. Pinamonti \inst{9}
	  \and G. Scandariato \inst{4}
	  \and A. Sozzetti  \inst{9}
	  \and D. Locci     \inst{1}
	  \and C. Di Maio   \inst{10,1}
	  \and A. Bignamini \inst{11}
	  \and R. Claudi    \inst{3}
	  \and E. Molinari  \inst{12}
	  \and R. Rebolo    \inst{5,6}
	  \and I. Ribas     \inst{7,8}
	  \and B. Toledo-Padr\'on \inst{5,6}
	  \and E. Covino    \inst{13}
	  \and S. Desidera  \inst{3}
	  \and E. Herrero   \inst{7,8}
	  \and J. C. Morales \inst{7,8}
	  \and A. Su\'arez-Mascare\~{n}o \inst{5,6}
          \and I. Pagano     \inst{4}
	  \and A. Petralia   \inst{1}
	  \and G. Piotto     \inst{2}
	  \and E. Poretti    \inst{14,15}
}

   \institute{INAF - Osservatorio Astronomico di Palermo,
              Piazza del Parlamento 1, 90134 Palermo, Italy\\
              \email{jesus.maldonado@inaf.it}
         \and 
	     Dipartimento di Fisica e Astronomia Galileo Galilei, Vicolo Osservatorio 3, 35122 Padova, Italy
	 \and 
             INAF - Osservatorio Astronomico di Padova, vicolo dell'Osservatorio 5, 35122 Padova, Italy  
	 \and 
	     INAF - Osservatorio Astrofisico di Catania, Via S. Sofia 78, 95123, Catania, Italy
	 \and 
	      Instituto de Astrof\'isica de Canarias, 38205 La Laguna, Tenerife, Spain
         \and 
              Universidad de La Laguna, Departamento Astrof\'isica, 38206 La Laguna, Tenerife, Spain
	 \and 
	     Institut de Ci\'encies de l'Espai (ICE, CSIC), Campus UAB, Carrer de Can Magrans s/n, 08193 Bellaterra, Spain
	 \and 
	     Institut d'Estudis Espacials de Catalunya (IEEC), 08034 Barcelona, Spain
	 \and 
	     INAF - Osservatorio Astrofisico di Torino, Via Osservatorio 20, 10025 Pino Torinese, Italy
	 \and 
             Universit\`a degli Studi di Palermo, Dipartimento di Fisica e Chimica, via Archirafi 36, Palermo, Italy 
	 \and 
	     INAF - Osservatorio Astronomico di Trieste, Via Tiepolo 11, 34143 Trieste, Italy
	 \and 
	     INAF - Osservatorio Astronomico di Cagliari \& REM, Via della Scienza, 5, 09047 Selargius CA, Italy
	 \and 
	     INAF - Osservatorio Astronomico di Capodimonte, Salita Moiariello 16, 80131 Napoli, Italy
	 \and 
	    Fundaci\'on Galileo Galilei-INAF, Rambla Jos\'e Ana Fern\'andez P\'erez 7, 38712 Bre\~{n}a Baja, TF, Spain 
	 \and 
	    INAF - Osservatorio Astronomico di Brera, Via E. Bianchi 46, 23807 Merate, Italy
             }
%

   \date{Received September 15, 1996; accepted March 16, 1997}

 
  \abstract
    {Most of our current knowledge on planet formation is still based on the analysis of main-sequence,
    solar-type stars. 
    Conversely, detailed chemical studies of large samples of M-dwarf planet hosts are still missing.  
    }
   {
   We aim to test whether the correlations between the metallicity, individual chemical abundances, and mass of the star
   and the presence of different type of planets found for FGK stars
   still holds for the less massive M dwarf stars.  
   Methods to determine in a consistent way stellar abundances of M dwarfs from high-resolution optical spectra
   are still missing. 
   The present work is a first attempt to fill this gap.
   } 
   {We analyse in a coherent and homogeneous way a large sample of M dwarfs with and without known planetary companions. 
    We develop for the first time a methodology to determine stellar abundances
    of elements others than iron
    for M dwarf stars from high-resolution, optical spectra.
    Our methodology is based on the use of principal component analysis and sparse Bayesian's methods.
    We made use of a set of M dwarfs orbiting around an FGK primary with known abundances to train our methods.
    We applied our methods to derive stellar metalliticies and abundances of a large sample of M dwarfs observed within the framework of current radial
    velocity surveys. 
    We then used a sample of nearby FGK stars to cross-validate our technique by comparing the derived abundance trends
    in the M dwarf sample with those found on the FGK stars. 
    %
    %
   }
   {The metallicity distribution of the different subsamples shows that M dwarfs hosting giant planets show a
   planet-metallicity correlation as well as a correlation with the stellar mass. M dwarfs hosting low-mass planets do not
   seem to follow the planet-metallicity correlation. 
   We also found that the  frequency of low-mass
   planets does not depend on the mass of the stellar host.
   These results seem in agreement
   with previous works.
   However, we note that for giant planet hosts our
   metallicities predict a weaker planet metallicity correlation but a stronger mass-dependency
   than photometric values. 
   We show, for the first time, that there seems to be no differences in the abundance distribution of elements different 
   from iron between M dwarfs with and without known planets. 
   }
   {Our data shows that low-mass stars with planets follow the same metallicity, mass, and abundance trends than their
   FGK counterparts, which are usually explained within the framework of core-accretion models.}


   \maketitle
%
\section{Introduction}\label{introduction}
 
 More than twenty-five years after the first exoplanets discoveries \citep{1992Natur.355..145W,1995Natur.378..355M}
 it is now well-established that planetary systems are found around a wide variety
 of stellar hosts from brown dwarfs and low-mass stars to red giants, pulsars and probably white dwarfs
 \citep[see e.g.][and references therein]{2018exha.book.....P}. 
 However, the vast majority
 of known planets are found to orbit around Sun-like stars
 (see e.g. http://exoplanet.eu/ or http://exoplanets.org/).
 Therefore, our understanding
 of the dependence of planet formation on the stellar mass is still far from being complete. 

 The study of chemical abundances in Sun-like planet host stars has been crucial for our
 understanding of planet formation with the finding that the frequency of gas-giant planets
 is a function of the host star metallicity \citep[e.g.][]{1997MNRAS.285..403G,2004A&A...415.1153S,2005ApJ...622.1102F,2009ApJ...697..544S}
 while stars with orbiting low-mass planets do not seem to be preferentially metal-rich
 \citep[e.g.][]{2010ApJ...720.1290G,2011arXiv1109.2497M,2011A&A...533A.141S,2012Natur.486..375B}.
 This trend is explained within the framework of core-accretion models, which
 assume that the timescale needed to form an icy or rocky core is largely dependent on the metal content
 of the protostellar cloud \citep[e.g.][]{1996Icar..124...62P,2004ApJ...616..567I,2005Icar..179..415H,2009A&A...501.1139M,2012A&A...541A..97M}. 

 Besides the gas-giant planet metallicity correlation any other claim of a chemical trend in planet hosts
 has been controversial or at least disputed. Most studies show that planet hosts have
 abundances similar to those of stars
 without planets \citep[e.g.][]{2003A&A...404..715B,2006A&A...445..633E,2006PASP..118.1494G,2006A&A...449..723G,2011A&A...526A..71D,2012A&A...543A..89A}.
 For instance, it has been suggested that the $\alpha$-elements enhancement found in planet hosts of intermediate metallicity 
 \citep{2012A&A...543A..89A}
 or the small depletion on refractory elements with respect to volatiles found in the Sun and other solar analogues  \citep[e.g.][]{2009ApJ...704L..66M,2009A&A...508L..17R}
 can be effects of galactic chemical evolution \citep[e.g.][]{2010ApJ...720.1592G,2013A&A...552A...6G}
 or related to an inner Galactic origin of the planet hosts 
 \citep[e.g.][]{2014A&A...564L..15A,2016A&A...588A..98M}.

 At the  low end of the stellar mass scale, low-mass stars (e.g. M dwarfs) are promising targets
 in the search for small, rocky planets with the potential capabilities of hosting life
 \citep[e.g.][]{2013ApJ...767...95D}.
 Unlike their FGK counterparts, detailed chemical studies of samples of M dwarf planet hosts
 have focused only on the iron content, or metallicity and are based on a relatively small
 number of planet hosts. Furthermore, metallicity values are often based on photometric values
 which makes difficult a comparison with the spectroscopic results from Sun-like stars. 
 As example, the stellar sample analysed in \cite{2010PASP..122..905J} comprises only five planets around M dwarfs
 while \cite{2013A&A...551A..36N} studied a sample with 13 stars hosting 20 planets (seven stars hosting gas-giant planets, and
 six stars with only low-mass planet hosts). 
 The sample in \cite{2012ApJ...748...93R} includes a total of 113 M dwarfs, but only 11 planet hosts.
 \cite{2014ApJ...781...28M} analysed a sample of 111 M dwarfs with only eight planet hosts. 
 These numbers contrast dramatically with the large number of FGK planet hosts for which 
 detailed chemical abundances have been performed \citep[e.g.][]{2003A&A...404..715B,2014A&A...564L..15A,2015A&A...579A..20M}. 

 Despite the small number statistics, several interesting trends regarding the planet-metallicity correlation
 in M dwarfs have already been
 discussed.  
 It is known that there is a systematically lower fraction of Jovian planets around M dwarfs than 
 around FGK stars 
 \citep{2003AJ....126.3099E,2006ApJ...649..436E,2006PASP..118.1685B,2007A&A...474..293B,2008PASP..120..531C,2010PASP..122..149J}.
 On the other hand, low-mass planets seem to be common around M dwarfs 
 \citep{2013A&A...549A.109B,2012ApJS..201...15H,2015ApJ...798..112M,2015ApJ...814..130M}.
 Previous works also suggest that the gas-giant planet metallicity correlation is also apparent in the M dwarf sample
 but the correlation is not present when Neptunian and smaller planets are considered 
 \citep{2007A&A...474..293B,2009ApJ...699..933J,2010A&A...519A.105S,2012ApJ...748...93R,2012ApJ...747L..38T,2013A&A...551A..36N,2016MNRAS.461.1841C}.
 More recently, \cite{2019A&A...625A.126P} found a moderate-to-weak dependence of the planetary minimum mass on the stellar metallicity for
 M dwarfs.

Regarding other chemical abundances besides iron, and to the very best of our knowledge, a detailed analysis 
has not been performed yet. 
This is because the accurate determination of the stellar parameters and abundances of
M dwarfs is a difficult task as these stars are faint at optical wavelengths and their
optical spectra are largely covered by molecular bands that blend or hide most of the atomic lines. 
Whilst spectral synthesis has been tested in several works, it is computationally expensive and requires
a good knowledge of the atomic and molecular data. It has been tested usually on small
number of stars, focusing on strong atomic lines, and on spectral windows known to be less affected by molecular lines
or in the near infrared \citep[e.g.][]{2005MNRAS.356..963W,2006ApJ...653L..65B,2012A&A...542A..33O,2017ApJ...835..239S},
that is, at wavelengths redder than the spectral coverage of many spectrographs
especially designed to achieve accurate radial velocities and used on current planetary surveys of M dwarfs.


We believe that the analysis of a homogeneous and large sample of M dwarfs hosting planets is needed
before a confirmation or rejection of preliminary trends or a proper comparison with FGK stars is invoked.
Several radial velocity projects have been working on different methodologies to derive stellar
abundances of M dwarfs using the same spectra that are used for the radial velocity determinations.
Within the framework of the CARMENES \citep{2018SPIE10702E..0WQ} collaboration,
\cite{2018A&A...615A...6P,2019A&A...627A.161P,2019A&A...625A..68S} derived stellar parameters for almost 300
M dwarfs by applying spectral synthesis to both visible and near-infrared spectra.
For the HARPS GTO M dwarf survey \citep{2013A&A...549A.109B}, a technique based on the use of pseudo-equivalent widths of spectral features identified
in optical high-resolution spectra was developed by \cite{2014A&A...568A.121N}.
 In a recent work, machine learning methods applied to pseudo-equivalent widths have been developed by \cite{2020A&A...636A...9A}.
A different technique, although also based on pseudo-equivalent widths, was presented in \cite{2015A&A...577A.132M} within the framework
of the HADES survey \citep{2016A&A...593A.117A}.

In this paper we present a completely different methodology. Unlike previous works it has the advantage
that it can be used to determine the elemental abundances 
of different elements besides iron. 
Our approach is based on the use of principal component analysis and sparse Bayesian's fitting methods.
A set of M dwarfs in binary systems orbiting around an FGK primary was observed and is used to train our method.



We use our derived abundances to revisit the correlation between the presence of planets around M dwarfs and
 the stellar properties, namely the stellar mass and the chemical composition.
 One of the motivations is the increase with respect to
 previous works of the number of M dwarfs known to host a planet.
  In particular we analyse the properties of at least five times more M dwarfs 
 hosting low-mass planets (m$\sin i$ $\lesssim$ 30 M$_{\oplus}$) than previous works.
 We made use of the available high-resolution HARPS and HARPS-N \'echelle spectra 
 as well as our spectroscopic tools specifically designed for the analysis of M dwarfs spectra
 to homogeneously determine stellar properties and chemical abundances. This allows us to increase consistently
 the stellar sample analysed in this work. 

 The paper is organised as follows. Sect.~\ref{observations} describes the stellar sample analysed in this work
 and how stellar parameters are obtained.
 We describe our technique to measure stellar abundances for M dwarf stars from optical spectra in Sect.~\ref{abundance_analysis}.
 Detection limits for our target stars are discussed in Sect.~\ref{detection_limits}.
 The mass, metallicity, and individual abundance distributions are presented in Sect.~\ref{results}.
Our conclusions follow in Sect.~\ref{conclusions}.

\section{Spectroscopic data}\label{observations}

 A list of late-K and M dwarf stars observed within the framework of radial velocity surveys was compiled
 by carefully checking the stars within
 the California Planet Survey (CPS) late-K and M-type dwarf sample \citep{2006PASP..118..617R},
 the HARPS GTO M dwarf sample \citep{2013A&A...549A.109B},
 and the HADES radial velocity program sample of M dwarfs \citep{2016A&A...593A.117A}. 

 In order to derive homogeneous stellar properties, only stars with HARPS \citep{2003Msngr.114...20M} and HARPS-N \citep{2012SPIE.8446E..1VC}
 data were considered.
 HARPS data were obtained from the ESO Science Data Products Archive\footnote{http://archive.eso.org/wdb/wdb/adp/phase3\_spectral/form?}, while HARPS-N data were taken from the
 public archive of the Telescopio Nazionale Galileo (TNG)\footnote{http://archives.ia2.inaf.it/tng/faces/search.xhtml?dswid=4905}. 
 The instrumental setup of HARPS and HARPS-N is almost identical. The spectra cover the range 383-693 nm (HARPS-N) and 378-691 nm (HARPS). 
 Both instruments provide a resolving power of R $\sim$ 115000. The spectra are provided already reduced using HARPS/HARPS-N standard calibration pipelines.
 For each star a coadded spectrum combining all the available observations was made. 
 Typical values of the signal-to-noise ratio (measured around 605 nm) for the combined spectra are from 45 to 99 with a median value of 67. 

 Stellar effective temperatures and metallicities were determined for each star by using the code msdlines\footnote{https://github.com/jesusmaldonadoprado/mdslines} \citep{2015A&A...577A.132M}
 which is based on the use of spectral features and their ratios. 
  The iron abundance values given by the code msdlines are based on the photometric M$_{\rm K}$-[Fe/H] relationship by \citet[][hereafter N12]{2012A&A...538A..25N}
   while the effective temperatures are based on the revised scale by \cite{2013ApJ...779..188M}.
 Physical parameters namely, surface gravity, stellar mass, stellar radius, and luminosity are derived
  from the derived  stellar effective temperature and metallicity values by using the empirical calibrations
 provided in the msdlines code.
 The stellar masses provided by this code are based on near-infrared photometry \citep{1993AJ....106..773H}
 and have typical uncertainties of the order of 13\%. 
 We note that for eight targets our methodology provides unrealistic low stellar masses (below 0.10 M$_{\odot}$). In these particular cases, photometric masses were considered  using the calibration by \cite{1993AJ....106..773H}. 
These calibrations are provided in the CIT photometric system. We therefore converted the 2MASS \citep{2003yCat.2246....0C}
magnitudes into CIT magnitudes before applying these calibrations, following the transformations provided by 
\cite{2001AJ....121.2851C}.
 Stellar radius is computed using the mass-radius relationship provided in \cite{2015A&A...577A.132M},
 surface gravities are derived from the stellar masses and radii, and luminosities by applying the Stefan-Boltzmann law.
 We note that this is exactly the procedure followed in \cite{2015A&A...577A.132M} to derive the empirical
 calibrations of the physical parameters as a function of T$_{\rm eff}$ and metallicity.

 Galactic spatial-velocity components $(U,V,W)$ were computed using {\it Gaia} DR2 \citep{2018A&A...616A...1G}
 proper motions and parallaxes, together with radial velocities from \cite{2018A&A...616A...7S} and {\it Gaia} DR2. 
 When no values from these sources were available, data from the  {\it Simbad}\footnote{http://simbad.u-strasbg.fr/simbad/sim-fid}  database were
 considered. To compute $(U, V, W)$ the procedure of \cite{2001MNRAS.328...45M} was followed. This procedure
 updates the original algorithm \citep{1987AJ.....93..864J} to epoch J2000 in the International Celestial Reference System (ICRS).
 When possible, the full covariance matrix was used in computing the uncertainties in order to take
 the possible correlations between the astrometric parameters into account. 
 Finally, stars were classified as belonging to the thin/thick disk applying the methodology described in 
 \cite{2003A&A...410..527B,2005A&A...433..185B}.

 Stellar age is one of the parameters more difficult to determine in an accurate way, especially when dealing with
  low-mass stars. An estimate of the age of our stars was obtained by interpolating 
  parallaxes and stellar parameters within a grid of Yonsei-Yale isochrones \citep[e.g.][]{2001ApJS..136..417Y,2002ApJS..143..499K}.
  The code {\it q2}\footnote{https://github.com/astroChasqui/q2} was used for the interpolation \citep{2014A&A...572A..48R}.
  However, we caution that the use of isochrones 
  for M dwarf stars shows important limitations. Indeed, for roughly $\sim$ 20\% of the stars we were not able to recover
  a reliable age estimate.

\begin{figure}[htb]
\centering
\begin{minipage}{\linewidth}
\includegraphics[scale=0.5]{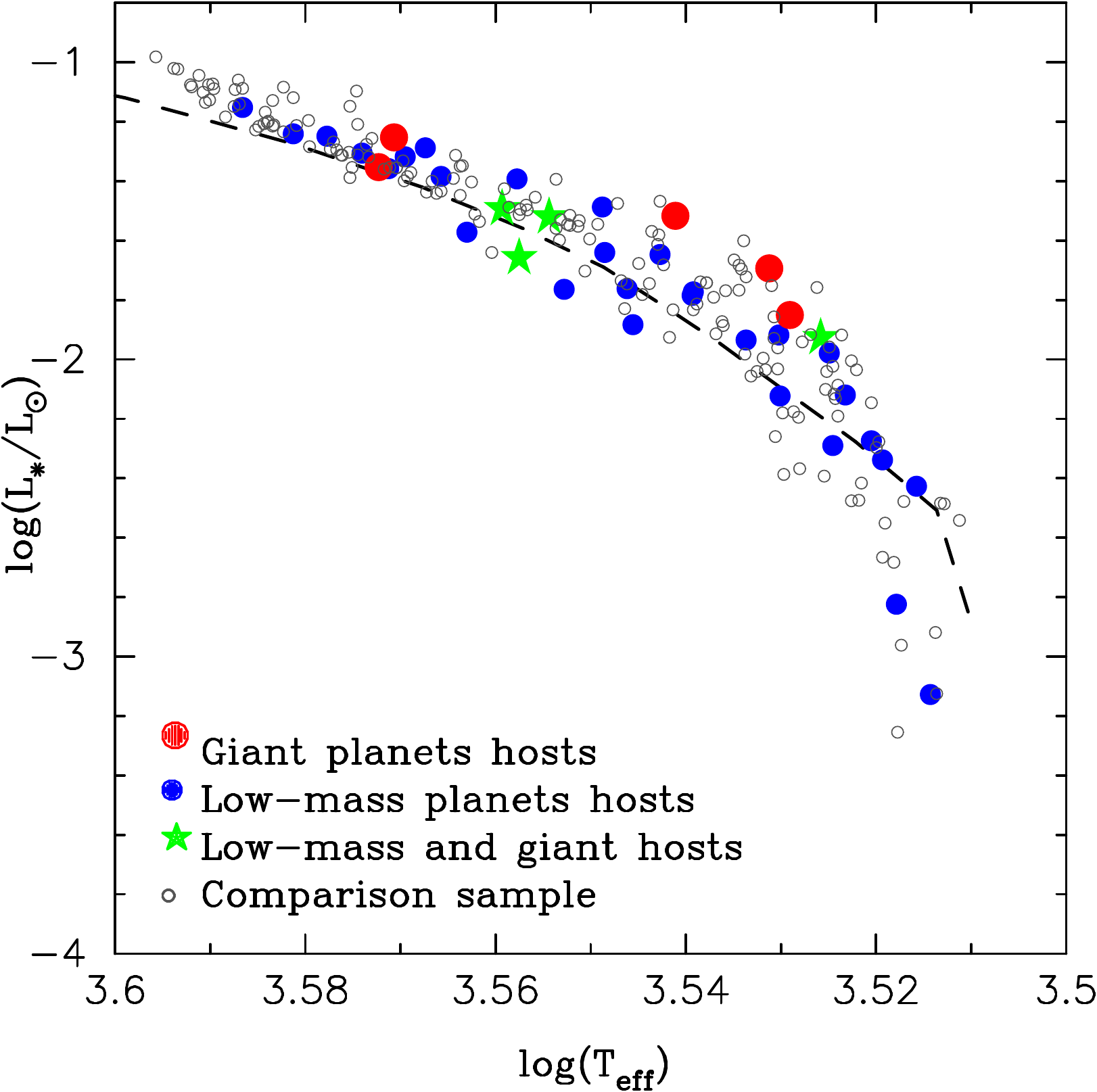}
\end{minipage}
\caption{ Luminosity versus T$_{\rm eff}$ diagram for the stars analysed. 
Stars with gas-giant planets are shown with red filled circles, while low-mass planet hosts
are shown with blue filled circles. Stars harbouring simultaneously gas-giant and low-mass planets are shown
with green stars. 
A 5 Gyr isochrone from \cite{2013ApJ...776...87S} is also shown for comparison.}
\label{diagrama_hr}
\end{figure}

 The final number of stars with available spectra is 204.
 They are listed in Table~\ref{parameters_table_full} while the corresponding HR diagram is shown in Figure~\ref{diagrama_hr}.
 The kinematic properties of the stars are also listed in Table~\ref{parameters_table_kinematics}.
 In order to identify those stars hosting planets the available information in the 
 NASA exoplanets archive\footnote{https://exoplanetarchive.ipac.caltech.edu/} 
 was carefully checked (up to June 2020). Five stars host at least one giant planet, one (or more) low-mass planets are found orbiting around 29 stars,
 while four stars host simultaneously both giant and low-mass planets. 
  Table~\ref{table_planets} shows the planet hosts, number of planets, and planetary properties 
 taken from the NASA exoplanets archive.

\section{Abundance determination in M dwarfs}
\label{abundance_analysis}
\subsection{Spectroscopic observations}\label{spec_analysis}

 The common procedure used in the literature to derive chemical abundances of M dwarfs is to search
 for M stars in common proper motion pairs orbiting around solar-type stars for which the
 accurate determination of spectroscopic abundances is possible. The list provided by
 \cite{2013AJ....145...52M}  was used as a starting reference.
 Several additional M dwarfs in binary systems were selected from
 \cite{2018MNRAS.479.1332M}.
 After searching the lists of possible calibrators available in the
 ESO and TNG archives, HARPS/HARPS-N data were found for only
 six stars. As these data were not sufficient to try 
 kind of calibration, additional observations were performed.
 
 Spectroscopic observations of 14 M dwarfs (and 12 FGK primaries)
 were performed in an observing run between the 9th and 13rd of November, 2018.
 Observations were performed using the HARPS spectrograph \citep{2003Msngr.114...20M} at La Silla ESO observatory.
 By adding the data already public in the ESO/TNG archives the total number of M dwarfs amounts to 
 20. Typical values of the signal-to-noise ratio (S/N measured at $\sim$ 605 nm)
 for the FGK primaries are between 40 and 130. The sample of M dwarfs is significantly fainter and
 even using relatively long integration times (up to two hours for some targets) the achieved S/N ratio is rather modest
 between 30 and 70 in the best cases. The star
  NLTT11500 (primary HIP17076)
   was excluded as a calibrator
    as the primary is a spectroscopic binary with both components clearly visible in our spectra.

%
%

\subsection{Stellar parameters and chemical abundances of the FGK primaries}

 Basic stellar parameters (T$_{\rm eff}$, $\log$g, microturbulent velocity, and [Fe/H]) for the primary stars
 were determined by using the code TGVIT\footnote{http://www2.nao.ac.jp/\textasciitilde{}takedayi/tgv/}
 \citep{2005PASJ...57...27T} which applies
 the iron ionisation and excitation equilibrium conditions 
 to a set
 of 302 Fe~{\sc i} and 28 Fe~{\sc ii} lines.
 The input stellar equivalent widths (EWs) were measured automatically using the code ARES\footnote{http://www.astro.up.pt/\textasciitilde{}sousasag/ares/}
 \citep{2007A&A...469..783S,2015A&A...577A..67S}.
 The {\it reject} parameter was adjusted according to the signal-to-noise ratio of the spectra as described in
 \cite{2008A&A...487..373S}.
 Uncertainties in the stellar parameters are computed by progressively changing each stellar parameter from the converged solution to a value in which any of the conditions (excitation equilibrium, match of the curve of growth, ionisation equilibrium) are no longer fulfilled. 
 We note that this procedure only evaluates statistical errors \citep[see for details][]{2002PASJ...54..451T,2002PASJ...54.1041T}.
 The uncertainties due to the errors in the measurement of the EWs were also computed
 and added in quadrature to the ones derived by the TGVIT code.
 Other sources of uncertainties, such as the choice of the model atmosphere,
 the list lines used, or the adopted atomic parameters are not taken into account.
 The derived parameters are provided in Table~\ref{table_fgk_parameters}. 

\begin{table*}
\centering
\caption{
 Spectroscopic properties of the primaries FGK stars.}
\label{table_fgk_parameters}
\begin{tabular}{lllccccc}
\hline\noalign{\smallskip}
 Primary & Secondary        & SpType & V      & T$_{\rm eff}$ & $\log g$ & v$_{\rm mic}$      & [Fe/H]   \\
           &                &        & (mag)  &    (K)        &  (cm s$^{\rm -2}$)   &  (km s$^{\rm -1}$) &  (dex)  \\
\hline 
HIP 3540	&	NLTT 2478	&	F8	&	7.0	&	6185	$\pm$	28	&	4.22	$\pm$	0.06	&	1.13	$\pm$	0.17	&	0.15	$\pm$	0.07	\\
HIP 5110	&	NLTT 3598	&	K2V	&	8.7	&	4715	$\pm$	23	&	4.78	$\pm$	0.07	&	0.49	$\pm$	0.29	&	-0.07	$\pm$	0.05	\\
HIP 6130	&	Gl 56.3	&	K1V	&	8.0	&	5288	$\pm$	12	&	4.70	$\pm$	0.02	&	0.77	$\pm$	0.10	&	-0.07	$\pm$	0.05	\\
HIP 6431	&	NLTT 4568	&	K5	&	9.5	&	4926	$\pm$	34	&	4.66	$\pm$	0.09	&	0.52	$\pm$	0.34	&	0.21	$\pm$	0.07	\\
HIP 6456	&	NLTT 4599	&	K0IV	&	7.9	&	5281	$\pm$	23	&	4.48	$\pm$	0.06	&	0.81	$\pm$	0.19	&	0.42	$\pm$	0.05	\\
HIP 9094	&	Gl 81.1	&	G5	&	6.4	&	5298	$\pm$	9	&	3.94	$\pm$	0.03	&	1.01	$\pm$	0.05	&	0.04	$\pm$	0.03	\\
HIP 11565	&	Gl 100	&	K4V	&	8.8	&	4594	$\pm$	37	&	4.84	$\pm$	0.09	&	0.50	$\pm$	0.60	&	-0.06	$\pm$	0.07	\\
HIP 11572	&	BD +22353B	&	K0	&	9.2	&	5105	$\pm$	22	&	4.53	$\pm$	0.06	&	0.49	$\pm$	0.31	&	0.03	$\pm$	0.05	\\
HIP 12114	&	Gl 105	&	K3V	&	5.8	&	4832	$\pm$	24	&	4.72	$\pm$	0.04	&	0.12	$\pm$	0.26	&	-0.09	$\pm$	0.04	\\
HIP 21710	&	Gl 173.1	&	K3V	&	9.2	&	4836	$\pm$	65	&	4.62	$\pm$	0.13	&	0.17	$\pm$	0.51	&	-0.02	$\pm$	0.11	\\
HIP 26907	&	NLTT 15511	&	K1V	&	8.6	&	5079	$\pm$	19	&	4.61	$\pm$	0.05	&	0.55	$\pm$	0.29	&	0.10	$\pm$	0.06	\\
HIP 27253	&	NLTT 15601	&	G4V	&	6.0	&	5588	$\pm$	20	&	3.86	$\pm$	0.06	&	1.23	$\pm$	0.08	&	0.33	$\pm$	0.05	\\
HIP 28671	&	NLTT 15974	&	G0V	&	9.3	&	5761	$\pm$	46	&	4.74	$\pm$	0.11	&	2.03	$\pm$	0.37	&	-1.05	$\pm$	0.07	\\
HIP 32984	&	Gl 250	&	K3V	&	6.6	&	4760	$\pm$	46	&	4.79	$\pm$	0.13	&	0.41	$\pm$	0.41	&	0.09	$\pm$	0.07	\\
HIP 40035	&	Gl 297.2	&	F7V	&	5.5	&	6310	$\pm$	42	&	4.27	$\pm$	0.07	&	1.57	$\pm$	0.15	&	-0.06	$\pm$	0.04	\\
HIP 62471	&	HIP 62471B	&	K4/K5V	&	8.9	&	4743	$\pm$	122	&	4.69	$\pm$	0.37	&	1.08	$\pm$	0.59	&	-0.33	$\pm$	0.12	\\
HIP 83591	&	Gl 654	&	K5V	&	7.7	&	4666	$\pm$	46	&	4.43	$\pm$	0.38	&	0.28	$\pm$	1.10	&	-0.58	$\pm$	0.16	\\
HIP 116906	&	NLTT 57675	&	G5	&	7.7	&	5760	$\pm$	11	&	4.36	$\pm$	0.03	&	1.05	$\pm$	0.05	&	-0.02	$\pm$	0.03	\\
HD 24916	&	HD 24916B	&	K4V	&	8.1	&	4635	$\pm$	43	&	4.91	$\pm$	0.17	&	0.40	$\pm$	0.42	&	0.06	$\pm$	0.07	\\
\hline
\end{tabular}
\end{table*}

 Chemical abundance of individual elements, C, O, Na, Mg, Al, Si, Ca, Sc, Ti, V, Cr, Mn, Co, Ni, and Zn are obtained using the
 2017 version of the code MOOG\footnote{http://www.as.utexas.edu/\textasciitilde{}chris/moog.html} \citep{1973PhDT.......180S}
 together with ATLAS9 atmosphere models \citep{1993KurCD..13.....K}. 
 The used line list is given in  \citet[][and references therein]{2015A&A...579A..20M}.
 Hyperfine structure (HFS) was taken into
 account for V~{\sc i}, and Co~{\sc i}, using the MOOG driver {\it blends}.
 Our derived abundances are provided in Table~\ref{table_fgk_abd}. They are expressed relative to the solar values derived  in 
 \cite{2015A&A...579A..20M},
 using the same methodology and similar spectra.
 We do not consider HFS effects on Mn abundances, as  \cite{2015A&A...579A..20M}  found
  an offset between the HFS abundances of Mn when using different lines.
 The uncertainties take into account the line-to-line scatter errors as well as the uncertainties due to the propagation
 of the errors in the stellar parameters and the equivalent widths (computed using the star HIP116906 as reference).

 Carbon abundances were derived by spectral synthesis of the CH molecular band at 430 nm using the CH line list provided in 
 \cite{2014A&A...571A..47M}. They are expressed relative to the solar value obtained in \cite{2020A&A...640A.123B}.
 Oxygen abundances were derived by spectral fitting of the [O~{\sc  i}] 630 nm line. However, reliable abundances
 were obtained only for seven stars. Carbon and oxygen uncertainties are those due to the fitting procedure. 

%
%
%
\begin{sidewaystable*}
\centering
\caption{
 Derived abundances and associated uncertainties ([X/H] in dex) for the primaries FGK stars.}
\label{table_fgk_abd}
 \begin{tabular}{lrrrrrrrrrrrrrrr}
 \hline\noalign{\smallskip}
  Star & [C/H] &  [O/H] & [Na/H] & [Mg/H] & [Al/H] & [Si/H] & [Ca/H] & [Sc~\sc ii/H] & [Ti/H] & [V/H] & [Cr/H] & [Mn/H] & [Co/H] & [Ni/H] &  [Zn/H] \\ 
\hline
HIP 3540 &        0.08 &           &  -0.01 &   0.20 &           &  0.08 &    0.12 &    0.13 &    0.07 &    0.29 &    0.16 &    0.09 &    0.00 &    0.10 &    0.32 \\  
    &  0.15 &               &    0.05 &       0.06 &               &    0.05 &       0.09 &       0.08 &       0.06 &       0.10 &       0.07 &       0.09 &       0.23 &       0.05 &       0.09 \\     
HIP 5110 &        -0.09 &          &  -0.21 &   -0.27 &   -0.11 &   -0.11 &   -0.26 &   -0.20 &   -0.09 &   0.10 &    -0.09 &   -0.12 &   -0.07 &   -0.13 &   -0.07 \\ 
    &  0.15 &               &    0.04 &       0.06 &       0.04 &       0.04 &       0.11 &       0.08 &       0.06 &       0.06 &       0.06 &       0.08 &       0.06 &       0.05 &       0.11 \\     
HIP 6130 &        -0.15 &          &  -0.15 &   -0.08 &   -0.03 &   -0.10 &   -0.12 &   -0.12 &   -0.05 &   -0.03 &   -0.07 &   -0.06 &   -0.09 &   -0.12 &   -0.14 \\ 
    &  0.17 &               &    0.04 &       0.07 &       0.04 &       0.03 &       0.07 &       0.07 &       0.05 &       0.06 &       0.06 &       0.08 &       0.06 &       0.05 &       0.10 \\     
HIP 6431 &        0.16 &           &  0.15 &    0.01 &    0.26 &    0.14 &    -0.03 &   0.13 &    0.24 &    0.37 &    0.17 &    0.32 &    0.22 &    0.16 &    0.19 \\  
    &  0.12 &               &    0.05 &       0.06 &       0.06 &       0.04 &       0.09 &       0.08 &       0.05 &       0.08 &       0.06 &       0.09 &       0.06 &       0.05 &       0.09 \\     
HIP 6456 &        0.38 &    0.22 &    0.61 &    0.49 &    0.58 &    0.44 &    0.27 &    0.43 &    0.43 &    0.55 &    0.43 &    0.78 &    0.50 &    0.48 &    0.90 \\  
    &  0.15 &       0.05 &       0.04 &       0.06 &       0.08 &       0.04 &       0.08 &       0.08 &       0.05 &       0.06 &       0.06 &       0.11 &       0.06 &       0.05 &       0.09 \\     
HIP 9094 &        -0.01 &   -0.06 &   -0.02 &   0.16 &    0.13 &    0.08 &    0.03 &    0.02 &    0.04 &    0.03 &    0.05 &    0.15 &    0.02 &    0.01 &    0.19 \\  
    &  0.12 &       0.07 &       0.04 &       0.06 &       0.04 &       0.03 &       0.07 &       0.07 &       0.05 &       0.06 &       0.06 &       0.08 &       0.06 &       0.05 &       0.26 \\     
HIP 11565 &       -0.18 &          &  -0.21 &   -0.37 &   -0.16 &   -0.16 &   -0.41 &   -0.21 &   -0.09 &   0.18 &    -0.13 &   -0.24 &   -0.09 &   -0.21 &   0.01 \\  
    &  0.20 &               &    0.05 &       0.06 &       0.04 &       0.07 &       0.15 &       0.10 &       0.06 &       0.06 &       0.06 &       0.08 &       0.08 &       0.05 &       0.30 \\     
HIP 11572 &       -0.04 &          &  0.18 &    0.09 &    0.21 &    0.10 &    -0.03 &   0.09 &    0.14 &    0.11 &    -0.02 &   0.05 &    0.08 &    0.02 &    0.23 \\  
    &  0.20 &               &    0.27 &       0.06 &       0.04 &       0.04 &       0.08 &       0.08 &       0.05 &       0.06 &       0.06 &       0.08 &       0.07 &       0.05 &       0.11 \\     
HIP 12114 &       -0.16 &   0.01 &    -0.05 &   -0.13 &   0.10 &    -0.03 &   -0.16 &   -0.05 &   0.08 &    0.21 &    -0.08 &   -0.06 &   -0.01 &   -0.09 &   -0.08 \\ 
    &  0.20 &       0.10 &       0.05 &       0.06 &       0.03 &       0.04 &       0.08 &       0.08 &       0.06 &       0.07 &       0.06 &       0.08 &       0.06 &       0.05 &       0.13 \\     
HIP 21710 &       -0.14 &          &  -0.06 &   -0.14 &   -0.06 &   -0.10 &   -0.23 &   -0.07 &   0.07 &    0.20 &    -0.04 &   -0.08 &   0.01 &    -0.15 &   -0.09 \\ 
    &  0.20 &               &    0.05 &       0.06 &       0.04 &       0.03 &       0.12 &       0.13 &       0.06 &       0.07 &       0.08 &       0.08 &       0.06 &       0.06 &       0.09 \\     
HIP 26907 &       0.08 &    0.10 &    0.05 &    0.07 &    0.20 &    0.10 &    -0.02 &   -0.03 &   0.16 &    0.18 &    0.05 &    0.17 &    0.10 &    0.05 &    0.00 \\  
    &  0.15 &       0.2	 &       0.04 &       0.06 &       0.03 &       0.04 &       0.08 &       0.10 &       0.05 &       0.06 &       0.06 &       0.08 &       0.06 &       0.05 &       0.15 \\     
HIP 27253 &       0.26 &    0.35 &    0.48 &    0.34 &    0.41 &    0.36 &    0.23 &    0.38 &    0.29 &    0.34 &    0.33 &    0.58 &    0.35 &    0.36 &    0.29 \\  
    &  0.10 &       0.10 &       0.09 &       0.11 &       0.04 &       0.03 &       0.07 &       0.08 &       0.05 &       0.06 &       0.06 &       0.09 &       0.06 &       0.05 &       0.11 \\     
HIP 28671 &       -1.12 &          &  -1.27 &   -0.68 &          &  -0.67 &   -0.96 &   -0.84 &   -0.70 &   -0.67 &   -1.13 &   -0.85 &   -0.01 &   -0.88 &   -1.07 \\ 
    &  0.15 &               &    0.05 &       0.06 &               &    0.05 &       0.07 &       0.24 &       0.08 &       0.18 &       0.10 &       0.16 &       0.05 &       0.11 &       0.09 \\     
HIP 32984 &       0.07 &           &  -0.05 &   -0.20 &   -0.01 &   0.03 &    -0.09 &   -0.08 &   0.06 &    0.26 &    0.05 &    0.06 &    0.05 &    0.02 &    0.04 \\  
    &  0.18 &               &    0.04 &       0.06 &       0.04 &       0.04 &       0.13 &       0.08 &       0.06 &       0.06 &       0.07 &       0.09 &       0.06 &       0.05 &       0.09 \\     
HIP 40035 &       -0.08 &          &  0.03 &    -0.06 &          &  0.00 &    -0.13 &   -0.22 &   -0.09 &   -0.05 &   -0.11 &   -0.18 &   -0.09 &   -0.11 &   -0.33 \\ 
    &  0.10 &               &    0.09 &       0.06 &               &    0.07 &       0.09 &       0.08 &       0.06 &       0.07 &       0.07 &       0.08 &       0.05 &       0.05 &       0.09 \\     
HIP 62471 &       -0.60 &          &  -0.19 &   -0.59 &   -0.13 &   -0.28 &   -0.36 &   -0.56 &   -0.03 &   0.18 &    -0.08 &   -0.41 &   -0.30 &   -0.42 &   0.00 \\  
    &  0.20 &               &    0.09 &       0.06 &       0.07 &       0.11 &       0.15 &       0.10 &       0.06 &       0.06 &       0.09 &       0.10 &       0.10 &       0.05 &       0.09 \\     
HIP 83591 &              &         &  -0.22 &   -0.46 &   -0.03 &   -0.61 &   -0.46 &   -0.71 &   0.05 &    0.32 &    -0.33 &   -0.70 &   -0.39 &   -0.64 &   -0.80 \\ 
   &          &            &    0.07 &       0.13 &       0.04 &       0.12 &       0.16 &       0.08 &       0.07 &       0.06 &       0.06 &       0.09 &       0.08 &       0.06 &       0.09 \\     
HIP 11690 &       -0.04 &   -0.02 &   0.01 &    0.09 &    0.06 &    0.00 &    -0.01 &   -0.04 &   -0.02 &   -0.02 &   -0.01 &   -0.01 &   -0.05 &   -0.03 &   -0.05 \\ 
          &  0.10 &       0.10 &       0.04 &       0.06 &       0.03 &       0.03 &       0.07 &       0.08 &       0.05 &       0.05 &       0.06 &       0.08 &       0.06 &       0.05 &       0.12 \\     
HD 24916 &        0.04 &    0.06 &    -0.20 &   -0.27 &   -0.03 &   0.02 &    -0.37 &   -0.10 &   0.00 &    0.31 &    0.04 &    0.01 &    0.06 &    -0.03 &   0.08 \\  
         &  0.20 &       0.07 &       0.07 &       0.06 &       0.04 &       0.04 &       0.18 &       0.10 &       0.06 &       0.06 &       0.06 &       0.09 &       0.06 &       0.05 &       0.13 \\ 
\hline
\end{tabular}
\end{sidewaystable*}

\subsection{M dwarf abundances: Methodology}

 In order to derive the abundances of the M stars
 we initially tried to proceed as in \cite{2015A&A...577A.132M}
 where a large dataset of spectral features and ratios of features were identified as
 temperature and metallicity diagnostics.
 However, in spite of the fact that many of these features are found to 
 correlate with the elemental abundances,
 a quick inspection of the plots of 
 pseudo-equivalent width measurements versus [X/H] reveals flat curves or
 complex patterns difficult to fit. Therefore, we conclude that even if the pseudo-equivalent width of a given spectroscopic feature might show a
 significant correlation with a specific ion abundance, the values of pseudo-equivalent widths depend
 on other parameters such as effective temperature, surface gravity, the global metallicity content, or the abundance of alkali metals. In addition, many
 of the features are likely to be a blend of different atomic or molecular lines.
 In other words, pseudo-equivalent widths values contain too much information to deal with.

 A common technique to reduce the dimensionality of the data and find the variables in which the spread or variance
 of the data is larger is the so-called  principal component (PCA) analysis.
 The PCA technique \citep[e.g.][]{1999ASPC..162..363F} is used to extract information of correlated data sets and find
 a new basis in which  the largest amount of variance is explained with the least number of basis vectors.
 This methodology has been successfully
 applied to samples of FGK stars even with spectroscopic data at moderate resolution
 \citep[e.g.][]{2013A&A...553A..95M,2017MNRAS.464.3657X,2019A&A...629A..33G}.

 We initially tried to identify spectral regions sensitive to the elemental abundance of different ions.
 For this 
 purpose we made use of atmospheric models together with the lists of lines used in the chemical analysis
 of solar-type stars and already identified spectral indexes sensitive to metallicity and other elements
 \citep[e.g.][]{2014AJ....148..105G}. 
 We made use of ATLAS9 models together with the spectral synthesis code SYNTHE
 \citep{1993KurCD..13.....K,1993KurCD..18.....K} as 
 it is easy to change the abundances of different elements and compute spectra with the desired abundances. We are aware that
 other sets of models (PHOENIX, MARCS) are usually considered to better reproduce the atmospheres
 of low-mass stars
 \citep[e.g.][]{2008A&A...485..823B,2010MNRAS.409L..49S,j_maldonado_2012_3738204}.
  However, we found that better results were obtained if the full
   spectral range was considered (from 534 nm, to avoid the bluest region of the spectra which suffers
    from a lower S/N and the gap between the two CCDs in HARPS spectra).


 Before applying  the PCA, each coadded spectrum was rebinned to a common wavelength grid and smoothed
 to a lower resolution using a Gaussian filter of a 12 nm width.
 To estimate how this smoothing reduces the spectral resolution we measured the ratio $\lambda$/($\delta$$\lambda$)
 on several lines in a ThAr spectra after and before the smoothing
 finding that it goes from $\sim$ 115000 to $\sim$ 1000-2000.
  While it is true that smoothing and resampling the data might destroy information,
  it also reduces the impact of high-frequency distortion in the data, and 
  previous works have found better results at lower resolution \citep{2013A&A...553A..95M}.
 Finally, 
 all the spectra were set to a common flux scale. 
 In order to do this, we consider the spectral flux at the R band centred at 609 nm with only 2 nm width
 to avoid the inclusion of strong molecular bands.
 We use one of the stellar spectra as reference and perform a linear fit between the ``reference''
 and ``problem'' flux. 
 
 Then, a flux matrix, $F(n,j)$, was obtained were $j$
 indexes the stars (including both the training and the problem datasets)
 and $n$ is the index corresponding to the wavelength bin. 
 Finally,  PCs were computed 
 using the available routines in the
 {\sc scikit-learn}
 python package \citep{scikit-learn}. 



  Our next step is to use the training dataset (that is, the M dwarfs in binary systems
  around an FGK primary star) to find a relationship between the  PCs and the stellar
  abundances. While some authors have explored the possibility of using a large number of
   PCs \citep[e.g.][]{2017MNRAS.464.3657X}, our training dataset is composed of
  only 19 stars, so the use of a large number of  PCs lead us to a 
  reduced number of degrees of freedom. For example,
  a fit of the stellar abundance as a liner combination of the  PCs using
  17  PCs  would leave us
  with only one degree of freedom. 
  In order to avoid over-fitting, the use of sparse Bayesian learning algorithms have been proposed 
  in the literature \citep[e.g.][]{2013A&A...553A..95M}, like the automatic relevance
  determination regression (ARDR).

 In brief, the target value is expected to be a linear combination
 of the features,

 \begin{equation}
 y(\mathbf{x,w})=\sum_{i=1}^{p}w_{i}x_{i}
 \end{equation}

 \noindent each coefficient $w_{i}$ is drawn from a Gaussian distribution centred at zero
 and a standard deviation $\alpha_{\rm i}$ so that:

 \begin{equation}
 p(\mathbf{w}|\mathbf{\alpha})=\mathcal{N}(w|0,A^{-1})
 \end{equation}

 \noindent with $A=diag(\alpha_{1}\ldots\alpha_{p})$.
 The aim of sparse Bayesian's fitting is to use the available data to compute the posterior
 distribution for the vector of weights ${\mathbf w}$ and the noise variance, $\sigma^{2}$. 
 ARDR is based on defining
 the inverse variances of these Gaussian distributions, $\alpha$, as
 variables, and to infer their values as well
 so 
 if the $\alpha_{i}$ parameter of the $i$-th feature tends to infinity, then its weight is very likely close to zero
 and
 is therefore pruned. In this way the solution that contains the least number of non-zero elements in ${\bf w}$
 is favoured. In this way, over-fitting is avoided.  

 Bayesian inference proceeds by applying the Bayes's rule to compute
 the posterior distribution over all unknowns given the data ${\mathbf x}$:
 
 \begin{equation}\label{posterior}
 P(\mathbf{w},\mathbf{\alpha},\sigma^{2}|\mathbf{x})=\frac{P(\mathbf{x}|\mathbf{w},\mathbf{\alpha},\sigma^{2})P(\mathbf{w},\mathbf{\alpha},\sigma^{2})}{P(\mathbf{x})}
 \end{equation}

 \noindent where the data, $\mathbf{x}$, are in our case the  PCs  and
 the stellar abundances, $P(\mathbf{x}|\mathbf{w},\mathbf{\alpha},\sigma^{2})$
 is the likelihood function and gives a measure of how well the model
 fits the data, and $P(\mathbf{w},\mathbf{\alpha},\sigma^{2})$ 
 is the prior distribution for the parameters.

 The posterior in Eqn.~\ref{posterior} cannot be computed directly. It is common
 to decompose it, as:

 \begin{equation}\label{posterior2}
 P(\mathbf{w},\mathbf{\alpha},\sigma^{2}|\mathbf{x})=P(\mathbf{w}|\mathbf{x},\alpha,\sigma^{2})P(\mathbf{\alpha},\sigma^{2}|\mathbf{x})
 \end{equation}

 \noindent where the first term in Eqn.~\ref{posterior2} is the posterior over the weights which
 can be computed analytically. Therefore, the optimisation of the evidence
 or the learning process reduces to the maximisation of:

 \begin{equation}
 P(\mathbf{\alpha},\sigma^{2}|\mathbf{x}) \propto P(\mathbf{x}|\mathbf{\alpha},\sigma^{2})P(\mathbf{\alpha})P(\sigma^{2})
 \end{equation}

  \noindent with respect to $\mathbf{\alpha}$ and $\sigma^{2}$.
  All the ARDR analysis was done with the {\sc scikit-learn} python package \citep{scikit-learn}
  where the $\alpha$ and $\sigma^{2}$ hyper-parameters are assumed to follow a Gamma distribution.
  More details about the Bayesian's inference and the ARDR technique can be found in 
  \cite{mackay92,10.1162/15324430152748236} as well as on the {\sc scikit-learn} tutorials.

  Initially, all the 19 training stars and a total of 17 PCs were used.
  Stellar abundances were computed in this way for our sample of M dwarfs. 
  The derived M dwarfs metallicities were compared with a sample of nearby solar-type FGK stars 
  coming from our previous works \citep{2015A&A...579A..20M}. This sample was selected as comparison since the 
  stars were analysed using similar spectra and the same methodology than the abundances of the FGK
  primaries of our training M dwarfs\footnote{
  With the only exception of the carbon abundances that in \cite{2015A&A...579A..20M} were 
  derived from atomic lines.}.  
  We found that the
  derived M dwarfs metallicities were shifted towards higher values when compared with the metallicities of 
  the nearby FGK stars. This could be related
  to the fact that most of our training stars have metallicities larger than -0.10 dex, while only three training
  stars have metallicities below this value.
  We will discuss this issue with more detail in Sect.~\ref{previous_works_sub}.
  To overcome this difficulty we performed a series of simulations
  in which 17 out of the 19 stars in the training dataset were randomly selected as effectively training stars.
  For each simulation we compared the metallicity distribution derived for our M dwarfs with the known metallicity
  distribution of the nearby solar-type FGK stars by using a two-sample Kolmogorov-Smirnov test (hereafter KS test). 
  We selected as the ``optimal'' training dataset the one that provides the most similar metallicity distribution
  to the one of the FGK stars. 
  We note that to force the metallicity distribution of M dwarfs to be similar to
  the one of nearby FGK stars is a common practice and has been used before to derive empirical calibrations
  for M dwarfs metallicities \citep[e.g.][]{2009ApJ...699..933J,2010A&A...519A.105S}.
  It is also worth noticing that instead of using all the stars included in \cite{2015A&A...579A..20M}
  we restricted the comparison to FGK stars within $\sim$ 70 pc (as our training stars are located within this distance) and with
  galactic-spatial velocity components ($U$, $V$, $W$) similar to the ones of the training data set.
  In this way we ensure that the comparison is not biased by stars located at different distances or belonging
  to different kinematic populations. 
  Further, the star NLTT15601 was discarded from the training dataset as its  PCs  deviate in a clear way from the values
  of the rest of the stars in the training dataset. We note that this star is the one with the lowest S/N among the whole training dataset.

\begin{figure*}[htb]
\centering
\includegraphics[scale=0.75]{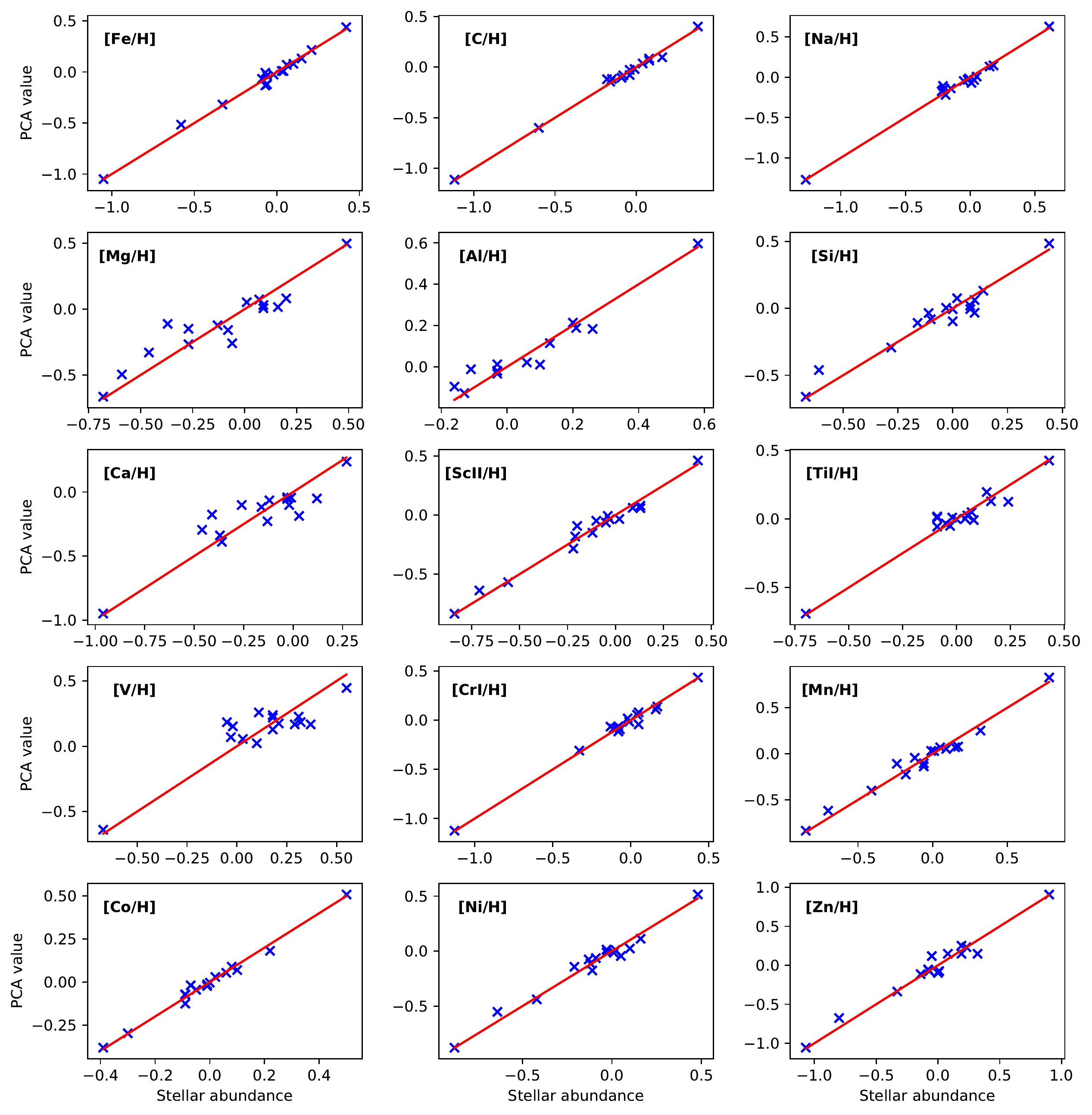}
\caption{  
Stellar abundances, [X/H], derived for the M dwarfs included in the training dataset using our PCA/Bayes technique versus the abundances measured in the corresponding primary stars. The red line denotes the one-to-one relationship. 
}
\label{pca_results}
\end{figure*}

   The final number of training stars used in the computations amounts to 16 while the number of  PCs 
  is 14. In this way we use the maximum number of  PCs while letting one degree of freedom. 
  Figure~\ref{pca_results} shows the comparison of the abundances derived for the training stars using our  PCA technique
  with the abundances derived from the primaries, while the standard deviation of the differences for
  each element are shown in Table~\ref{standard_deviation}. We also show the number of training stars and  PCs 
  used as well as the residual mean square (RMS), the root-mean squared error (RMSE) and the coefficient of determination
  (R$^{\rm 2}$), \citep[see e.g. Appendix in][]{2012ApJ...748...93R}.
  We note that for Ti, and Cr, whose abundances can be derived from lines of the neutral atom
  as well as from lines from the single ionised atom we consider the abundances derived from
  lines of the neutral atom as these are more abundant than the single ionised lines. 
  On the contrary for Sc, abundances from lines of the single ionised atom
  were considered as they are more abundant.  
  In general, we find that the differences between our derived abundances and those
  measured in the FGK primaries have standard deviation lower than $\sim$ 0.10 dex,
  the RMS and RMSE values are close to zero and R$^{\rm 2}$ close to one as expected when
  a model fit is useful for prediction. 
  Slightly higher dispersions are found for elements like Mg, Ca and V for which
  we note that their abundances are more difficult to measure even in solar-type stars.
  Indeed, for some stars we were not able to derive the abundances of 
  C, and Al. Therefore, for these elements we were forced to use less training stars and
  a lower number of  PCs.
  As far as oxygen abundances are concerned, we were able to measure the FGK primaries'
  abundances in only six stars and the methodology (using four PCs) fails to
  derive reliable values. 
  We will discuss the effect on the results of the number of training stars
  and  PCs used in the next subsection. 

  The derived abundances for our sample of M dwarfs are given in Table~\ref{abundances_table_full}.

\begin{table}
\centering
\caption{ 
 Standard deviation of the differences, residual mean square, mean squared error, and coefficient of determination
 between the PCA-derived abundances of the training stars and those 
 measured in the corresponding primaries.}
\label{standard_deviation}
\begin{tabular}{lcccccc}
\hline\noalign{\smallskip}
 [X/H]  &   n$_{\rm training}$ & n$(PCs)$ &    $\sigma$ &      RMS      &      RMSE     &       R$^{\rm 2}$ \\
\hline
Fe	&	16	&	14	&	0.04	&	0.03	&	0.00	&	0.99	\\
C       &       15      &       13      &       0.03    &       0.03    &       0.00    &       0.99    \\
Na	&	16	&	14	&	0.05	&	0.05	&	0.00	&	0.98	\\
Mg	&	16	&	14	&	0.12	&	0.11	&	0.01	&	0.86	\\
Al	&	13	&	11	&	0.05	&	0.05	&	0.00	&	0.93	\\
Si	&	16	&	14	&	0.07	&	0.07	&	0.00	&	0.93	\\
Ca	&	16	&	14	&	0.12	&	0.11	&	0.01	&	0.83	\\
ScII	&	16	&	14	&	0.05	&	0.05	&	0.00	&	0.98	\\
TiI	&	16	&	14	&	0.06	&	0.06	&	0.00	&	0.94	\\
V	&	16	&	14	&	0.12	&	0.12	&	0.01	&	0.79	\\
CrI	&	16	&	14	&	0.04	&	0.04	&	0.00	&	0.99	\\
Mn	&	16	&	14	&	0.07	&	0.06	&	0.00	&	0.97	\\
Co	&	16	&	14	&	0.02	&	0.02	&	0.00	&	0.99	\\
Ni	&	16	&	14	&	0.05	&	0.05	&	0.00	&	0.97	\\
Zn	&	16	&	14	&	0.09	&	0.08	&	0.01	&	0.96	\\
\hline
\end{tabular}
\end{table}

\subsection{Validation of the methodology I: Comparison with FGK stars}

 As mentioned we applied the  PCA plus ARDR methodology to our sample of ``problem'' M dwarfs 
 described in Sect.~\ref{observations}. Unfortunately,
 we do not have a sample of M dwarfs with known abundances determined from spectroscopy to validate our methodology.
 Some efforts have been done to apply the spectral synthesis
 methods to M dwarfs but, to the very best of our understanding, they are mainly focused on a small number of stars and mainly 
 analyse strong atomic lines at wavelengths redder than the HARPS/HARPS-N spectral coverage
 \citep[e.g.][]{2005MNRAS.356..963W,2006ApJ...653L..65B,2020arXiv200900876A}.
 Abundance analysis of M dwarfs in the near infrared has also been performed but on a small number of
 stars \citep[e.g.][]{2012A&A...542A..33O,2017ApJ...835..239S}.

\begin{figure*}[htb]
\centering
\includegraphics[scale=0.625]{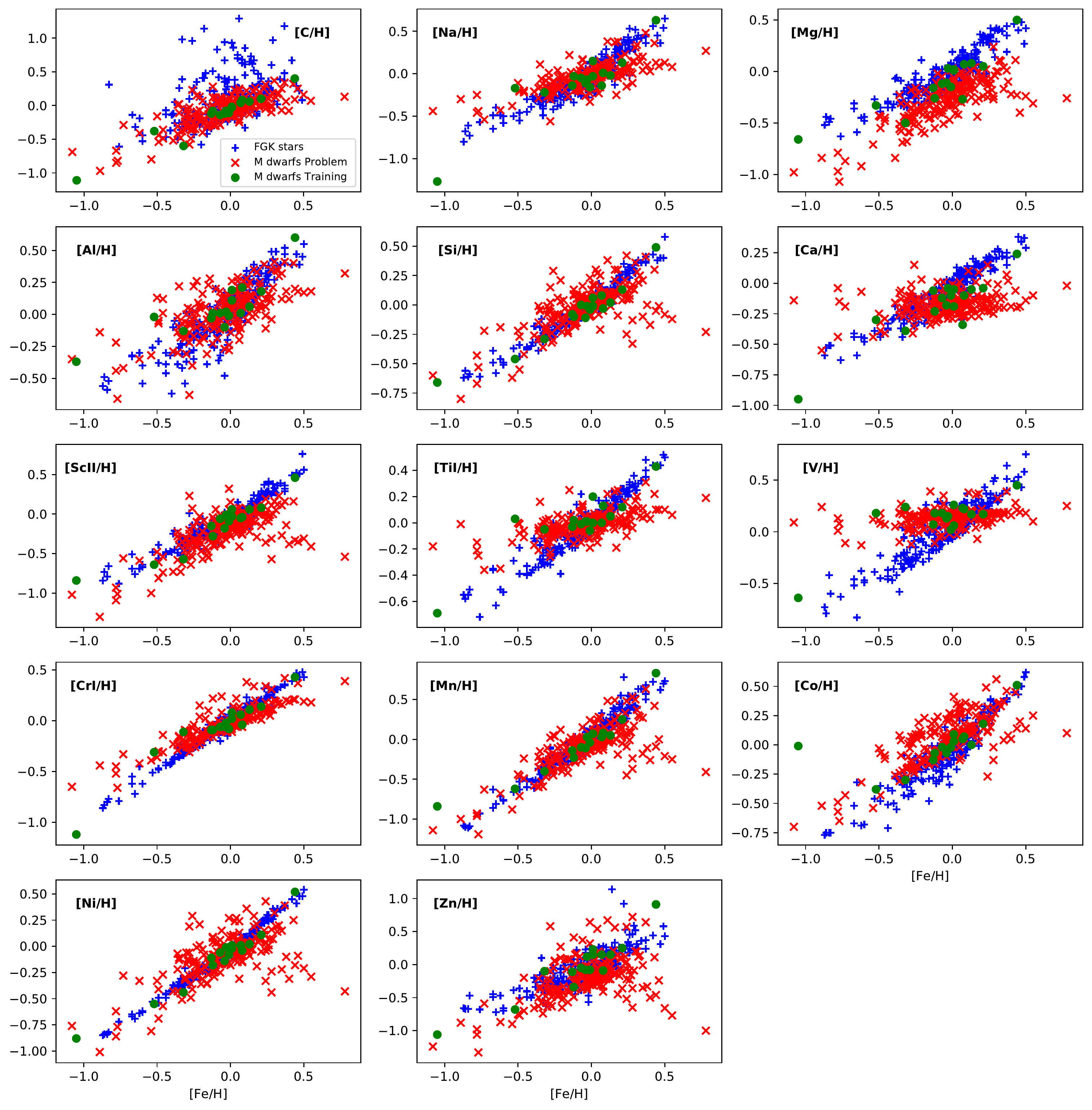}
\caption{   
[X/H] versus [Fe/H] plots for FGK stars (blue plus symbols), the M dwarf training stars (green circles) and
 our sample of ``problem'' M stars (red crosses).}
\label{pca_fgk_comparison}
\end{figure*}

 Therefore, we need to find alternative ways of testing the reliability of our methods.
 As a first step to validate our methodology we compare the [X/H] vs [Fe/H] trends derived
 from our M dwarfs with those known for FGK stars, as it is reasonable to expect similar trends
 for both types of stars. 
 As before, the  abundance for the FGK stars came from our previous works \citep{2015A&A...579A..20M}, although
 we now use the full dataset. 
 The corresponding plots are shown in Figure~\ref{pca_fgk_comparison}, where FGK stars are shown as
 blue plus symbols, the M dwarf training dataset as green circles, and the ``problem'' M stars as red crosses.
 Several conclusions can be drawn from this figure:

 {\it i)} A good agreement between FGK and M stars is found for the abundances of C, Na, Si, Ti, Cr, Mn, and Ni although
 some outliers can be seen specially at low and high abundance values.

 {\it ii)} The [X/H] versus [Fe/H] relationship shows a larger spread when the abundances of Al and Zn are considered,
 having the M dwarfs slightly lower Zn abundances.
 We note that for these elements, even for solar-type
  stars, their abundances are based on a relatively few number of lines.

{\it iii)} For V and Ca the general tendency for FGK and M dwarfs are slightly different.
 For Mg and Sc, the tendency is similar but M dwarfs seem to have lower abundances. 
 The opposite happens for Co, with M dwarfs having higher abundances.


\subsection{Validation of the methodology II: Cross-validation with FGK stars} 
 
 While  PCA and Bayesian methods have been already used to determine stellar parameters and abundances,
 it is clear that the small number of training stars used in this work might raise some doubts on the
 applicability of these techniques. 
 In other words, we need to be sure that sparse Bayesian methods do effectively avoid over-fitting
 and can be safely applied even with a small number of training stars.

 Therefore, we performed a cross-validation of the technique using it to derive the stellar abundances
 of a sample of FGK stars with HARPS/HARPS-N spectra. 
 We use a total of 37 HARPS spectra from our previous works.
  Giant stars were excluded and only
 stars with measured abundances for all elements were considered.

\begin{figure*}[!htb]
\centering
\includegraphics[scale=0.875]{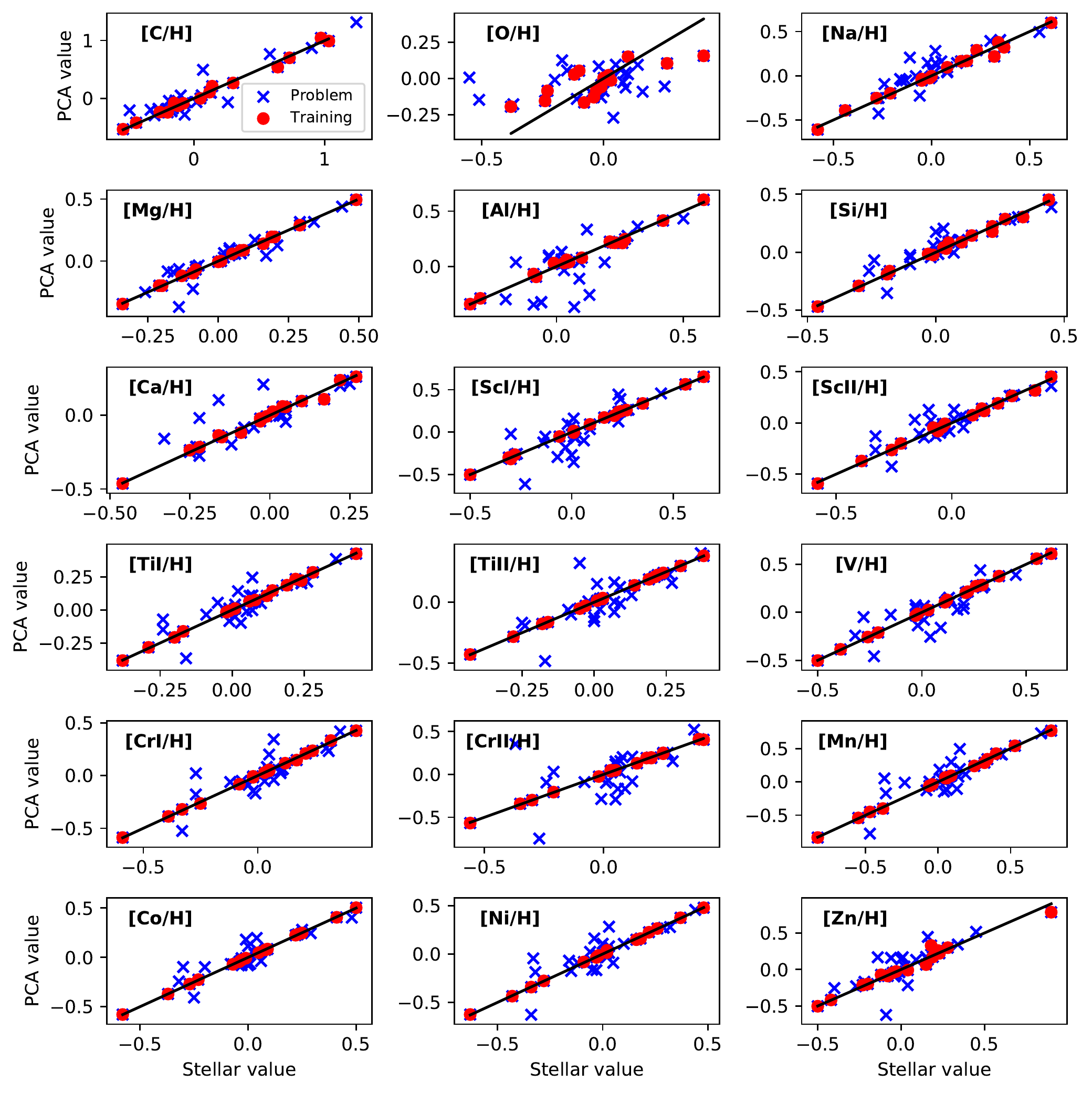}
\caption{
 PCA-derived abundances versus stellar abundances for our sample of FGK stars with
HARPS/HARPS-N spectra from one of the performed simulations. The training dataset is shown in red circles
while problem stars are shown in blue crosses.}
\label{simu_fgk}
\end{figure*}

 A total of 100 simulations were performed. In each simulation 16 stars were randomly selected
 as the training dataset and exactly the same procedure used for the M dwarfs was applied.
 In particular, we note that we keep the number of  PCs used to 14 (as in the M dwarfs case).

 Figure~\ref{simu_fgk} shows the results for  one of the simulations, where the abundances derived from 
 the  PCA methodology are compared with the measured abundances for both 
 the training (red circles) and problem stars (blue crosses).

\begin{figure*}[!htb]
\centering
\includegraphics{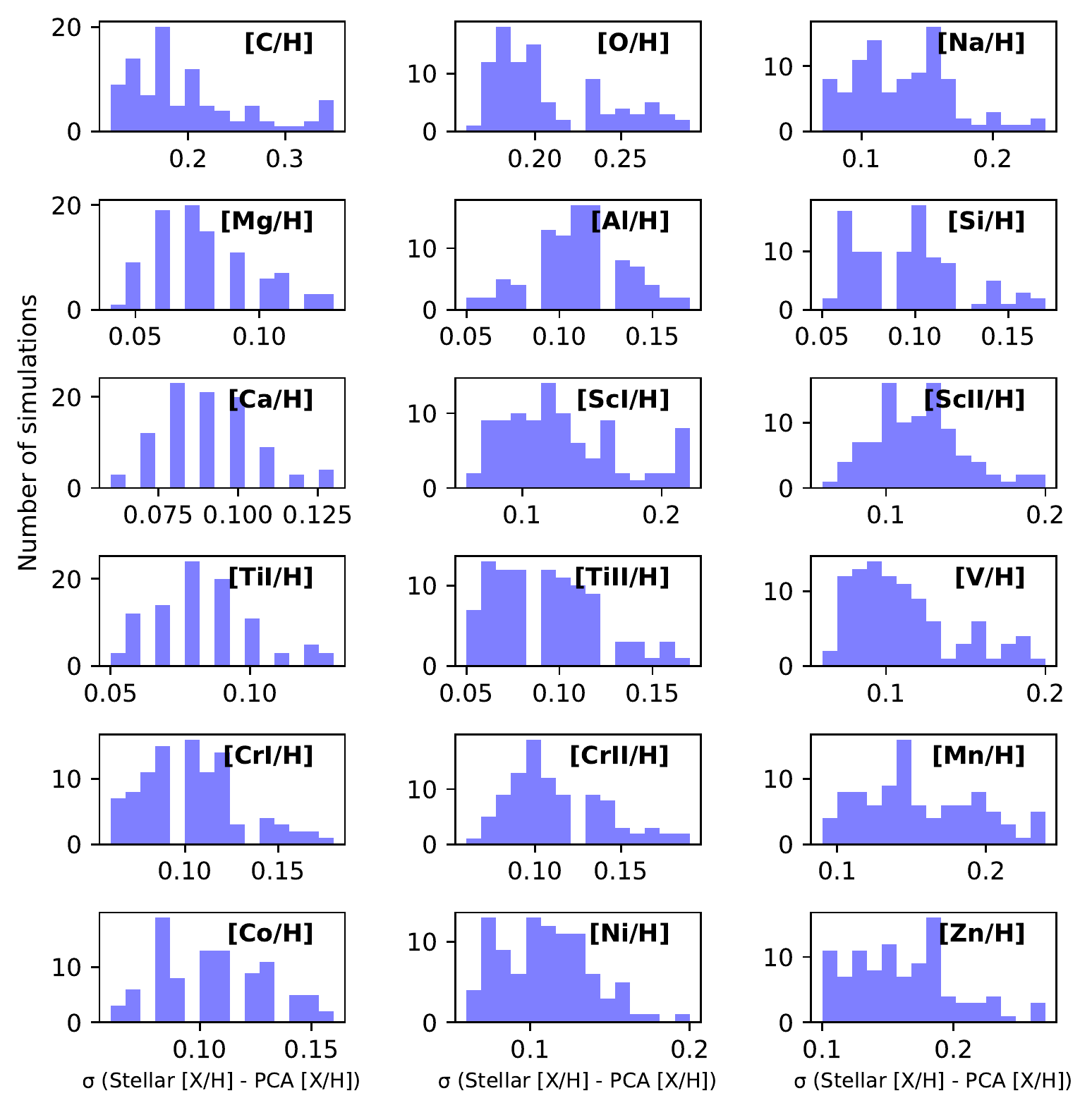}
\caption{ 
Histograms of the obtained standard deviations of the differences between the
PCA-derived abundances and the measured stellar abundances for our sample of FGK stars with
HARPS/-N spectra. The histograms show the results from 100 simulations.  Outliers are excluded from the plot. }
\label{histograms}
\end{figure*}

 We consider the standard deviation of the differences between the stellar abundances and the 
  PCA-derived values as a measure of the goodness of the technique. 
 For each simulation and for each element, we compute the standard deviation of the differences between
 the  PCA/Bayes derived abundances and those derived from a curve-of-growth approach from 
 measured EWs of selected lines in the stellar spectra (as described in Sect.~\ref{spec_analysis}).
 We plot the distributions of the 100 standard deviations from the 100 simulation in Figure~\ref{histograms}. 
 Excluding some outliers,
 for most elements, the agreement between the  PCA-abundances and those measured in the usual way is
 better than 0.10/0.15 dex. 
 An agreement between 0.10 and 0.20 dex is found for some elements like Na, Mn, and Zn, whose abundances
 are based only on a few number of lines. 


Our results show that the PCA-Bayesian fit technique can provide reliable results even if a low number of
 stars (as it is our case) is considered in the training dataset. Certainly, a larger number of M dwarfs
  in the training dataset, especially at low-abundances values would be desirable. We will discuss about
   this possibility in Sect.~\ref{conclusions}.

\subsection{On the metallicity scale of M dwarfs. Comparison with previous works.}\label{previous_works_sub} 

\begin{figure*}[!htb]
\centering
\includegraphics[scale=0.75]{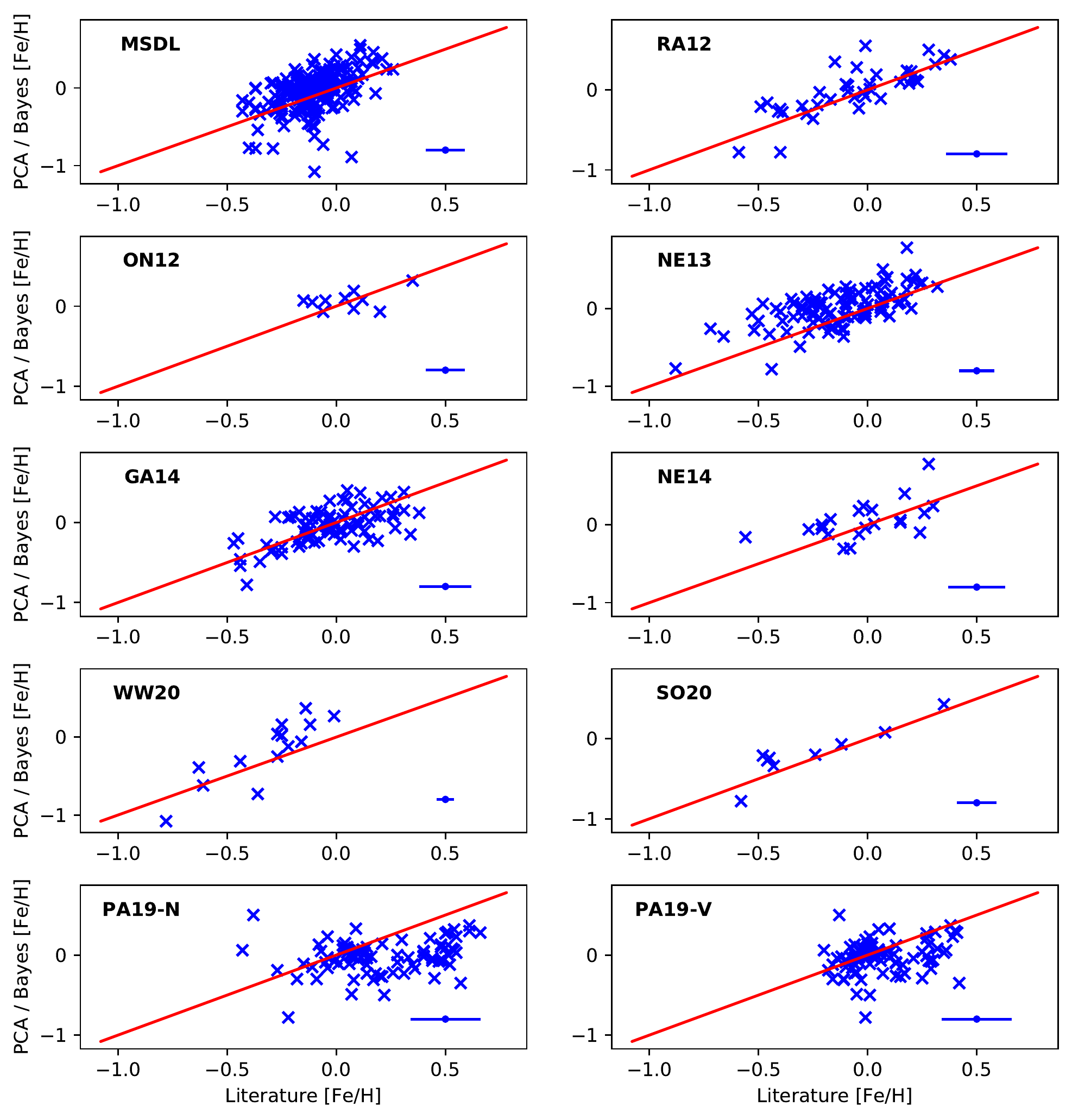}
\caption{ 
Comparison between the iron abundances derived in this work and those previously reported in the literature.
The red line denotes the one-to-one relationship.
}
\label{metal_comparison}
\end{figure*}

  We compare the metallicity values derived from the new  PCA technique with those 
  previously reported in the literature. 
  Values for the comparison include: those values computed by us using our previous msdlines technique (see Sect.~\ref{observations}, MSDL);
  the metallicities provided by \citet[][hereafter RA12]{2012ApJ...748...93R} which are derived from spectral indexes in the near infrared domain;
  the values by \citet[][hereafter ON12]{2012A&A...542A..33O} which perform spectral synthesis in the infrared J band; 
  those derived by \citet[][NE13]{2013A&A...551A..36N} using a technique based on pseudo-equivalent widths of spectral features measured in
  optical high-resolution spectra;
  the values by \citet[][GA14]{2014MNRAS.443.2561G} computed from metal-sensitive atomic and molecular features;
  the work by \citet[][NE14]{2014AJ....147...20N} derived from spectral lines and indexes in moderate-resolution near infrared spectra;
  \citet[][WW20]{2020MNRAS.494.2718W} which perform spectral synthesis in the spectral range 5700 - 10000\AA;
   \citet[][SO20]{2020ApJ...890..133S} which perform spectral synthesis on APOGEE spectra in the H band; 
   and \citet[][]{2019A&A...627A.161P} 
   which derive stellar parameters from high-resolution spectra by performing spectral synthesis
   in both the visible wavelength range (PA19-V) and the near infrared  (PA19-N).
   
  The comparison for iron abundance is shown in Figure~\ref{metal_comparison} while Table~\ref{rmse_previous} shows the main statistics of the comparisons. 
  The agreement is in all cases overall good with RMSE values lower than 0.10 dex (that is, the typical uncertainties
  reported in M dwarf abundances). 

\begin{figure}[!htb]
\centering
\includegraphics[scale=0.60]{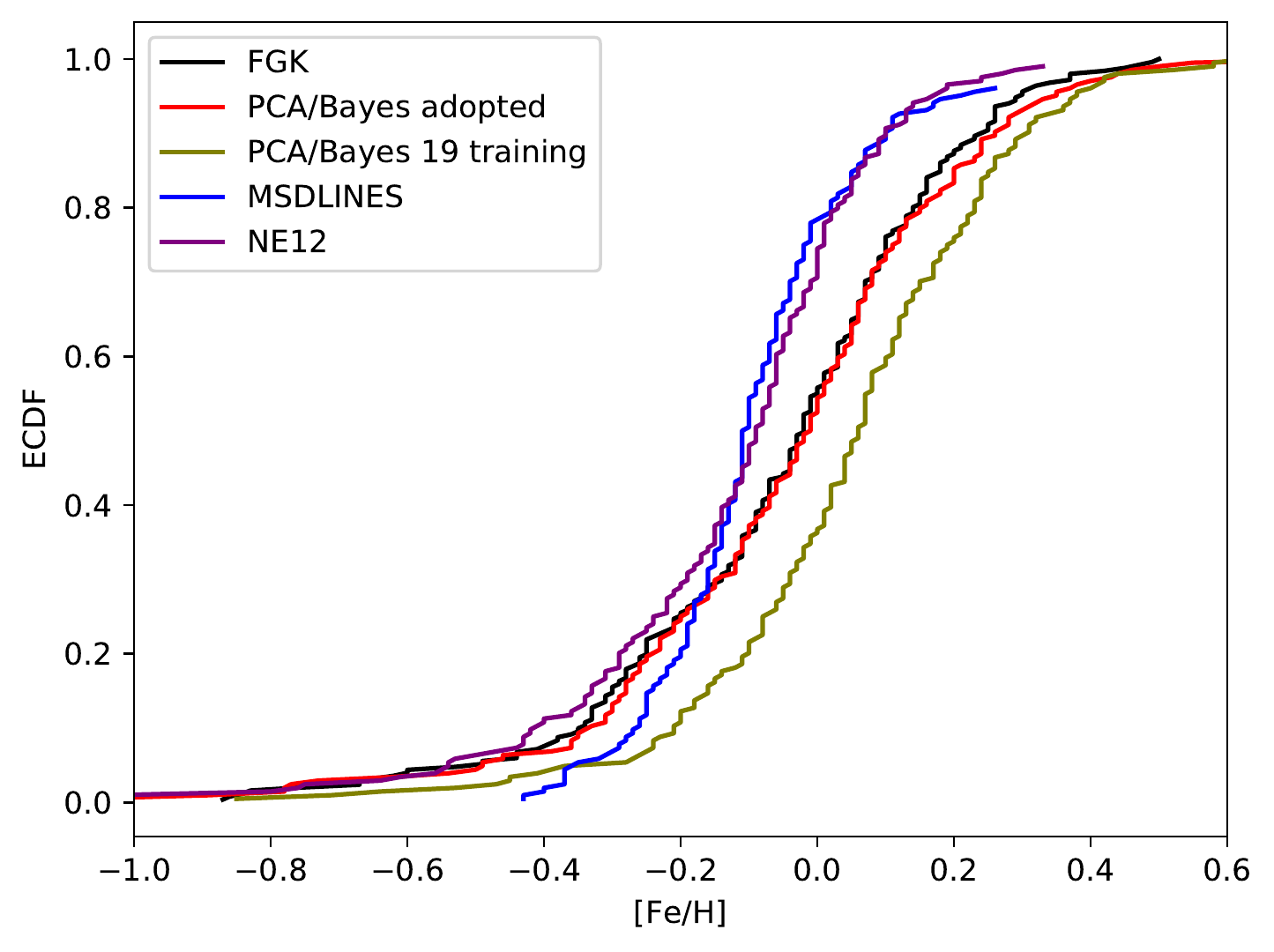}
\caption{ 
[Fe/H] empirical cumulative distribution function for the derived metallicities with those obtained with other methods. See text for details.}
\label{metal_ECDF}
\end{figure}

 As an additional test, in 
 Fig.~\ref{metal_ECDF} we show the empirical cumulative distribution function (ECDF) of the derived metallicities
 compared to the metallicities of FGK stars. We also show the results if all the 19 training stars are
 considered as well as the ECDF obtained using the msdlines metallicities and the NE12 calibration.
 Several conclusions can be drawn from this figure:

 {\it i)} Metallicity values derived from photometry (we recall that the msdlines method was calibrated using NE12
 due to the lack of high-resolution spectroscopic observations of M dwarfs in binary systems around
 solar-type stars) provide a ECDF shifted towards lower metallicities in comparison with 
 FGK stars for metallicity values higher than -0.20 dex. 

{\it ii)} If no selection of the training dataset is performed,  PCA-Bayes derived metallicities are shifted
 towards higher values when compared with FGK stars. As mentioned before, this is likely a dataset shift from selection bias as
 most of our training stars have metallicities higher than +0.00 dex.
 Although several methods are available to deal with this problem (e.g. density ratio estimator) they have provided
 unsuccessful results, probably due to our reduced number of training stars.

{\it iii)} Our finally adopted metallicity values for M dwarfs do reproduce the metallicity behaviour of nearby solar-type stars. 
 A KS test between our derived M dwarf metallicities and the FGK metallicities returns 
 the values $D$=0.04, $p$-value=0.99, with n$_{\rm eff}$=112.5, meaning that both samples show statistically identical metallicity distributions.

  As a final test we checked whether our derived metallicities show any correlation with the stellar effective temperatures
  and masses finding no significant correlation between these quantities.
  For metallicity and temperature the Spearman's rank $\rho$ is -0.0959 $\pm$ 0.0696 with a $z$-score =  -0.578 $\pm$ 0.422, while   
  for metallicity and stellar mass we obtain $\rho$ = 0.0548 $\pm$ 0.0700 and
  $z$-score = 0.329 $\pm$ 0.422. The statistical tests were performed  by a bootstrap Monte Carlo (MC) simulation plus a Gaussian random shift of each data-point within its error bars
  \citep{2014arXiv1411.3816C}\footnote{https://github.com/PACurran/MCSpearman/}.
  This result is expected as we derive these quantities using
  independent methodologies. Note that, unlike photometric calibrations, our  PCA/Bayes metallicities are not based on stellar evolution models. 
  
\begin{figure*}[!htb]
\centering
\begin{minipage}{0.49\linewidth}
\includegraphics[scale=0.50]{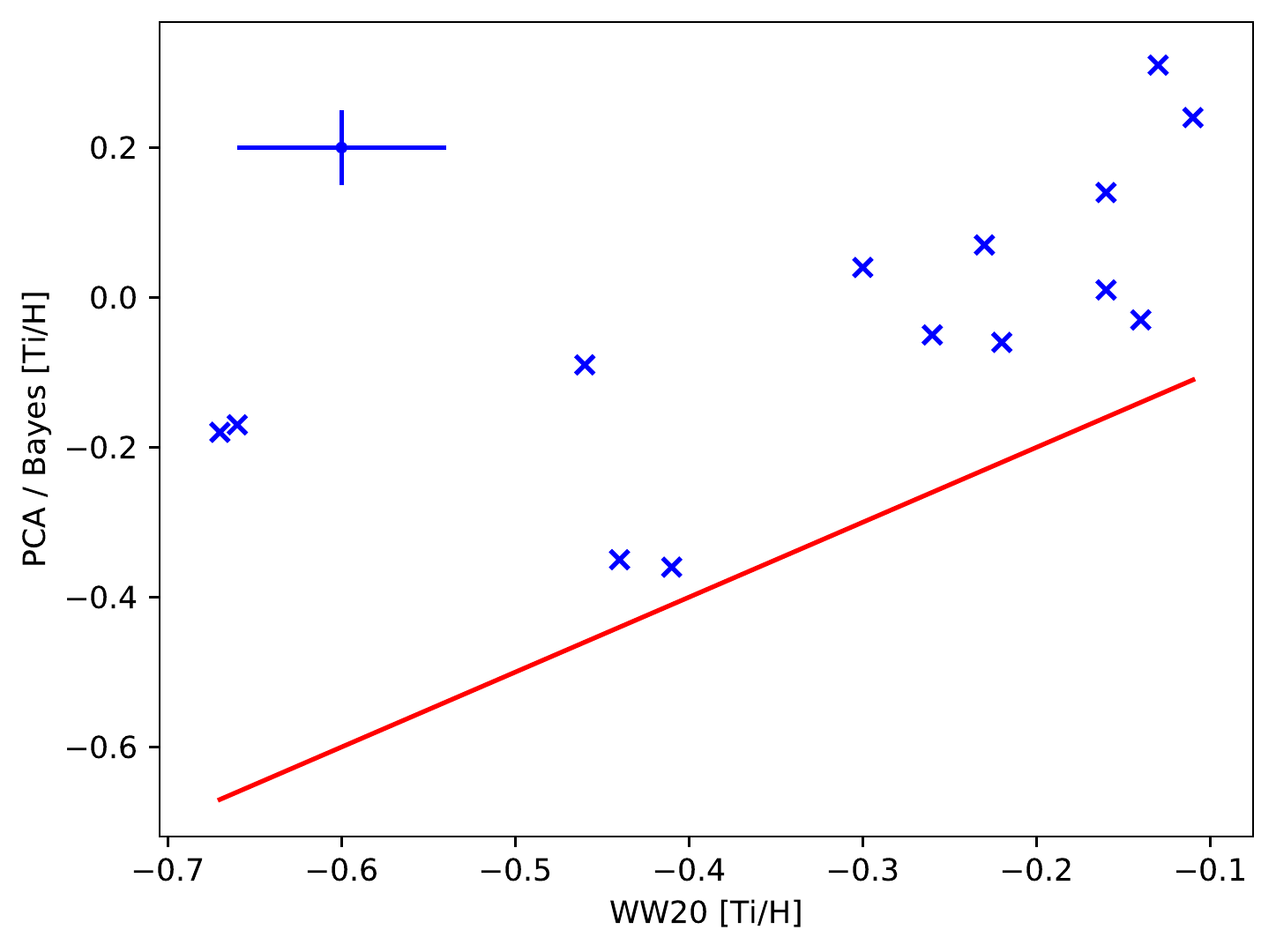}
\end{minipage}
\begin{minipage}{0.49\linewidth}
\includegraphics[scale=0.50]{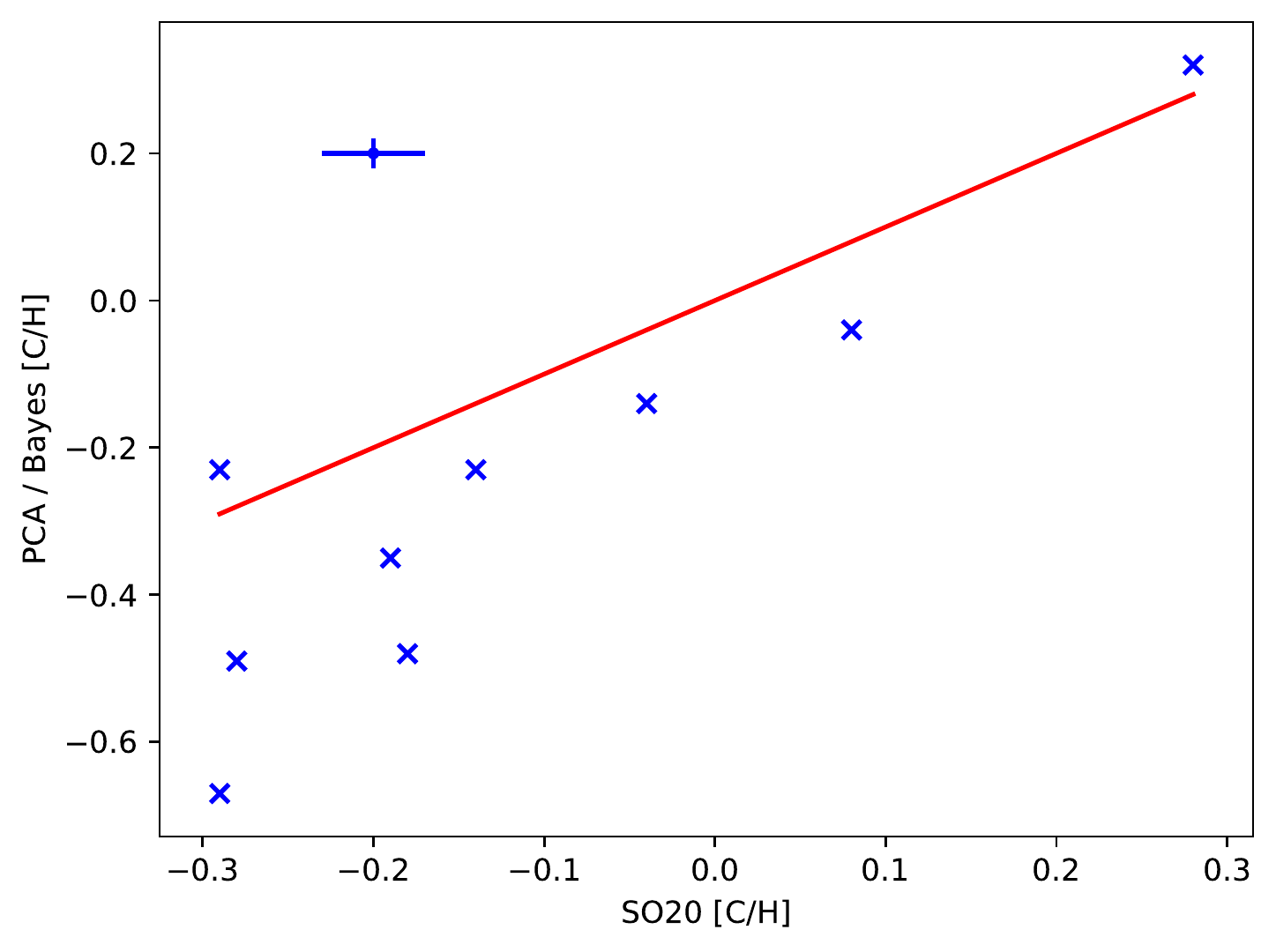}
\end{minipage}\\
\begin{minipage}{0.49\linewidth}
\includegraphics[scale=0.50]{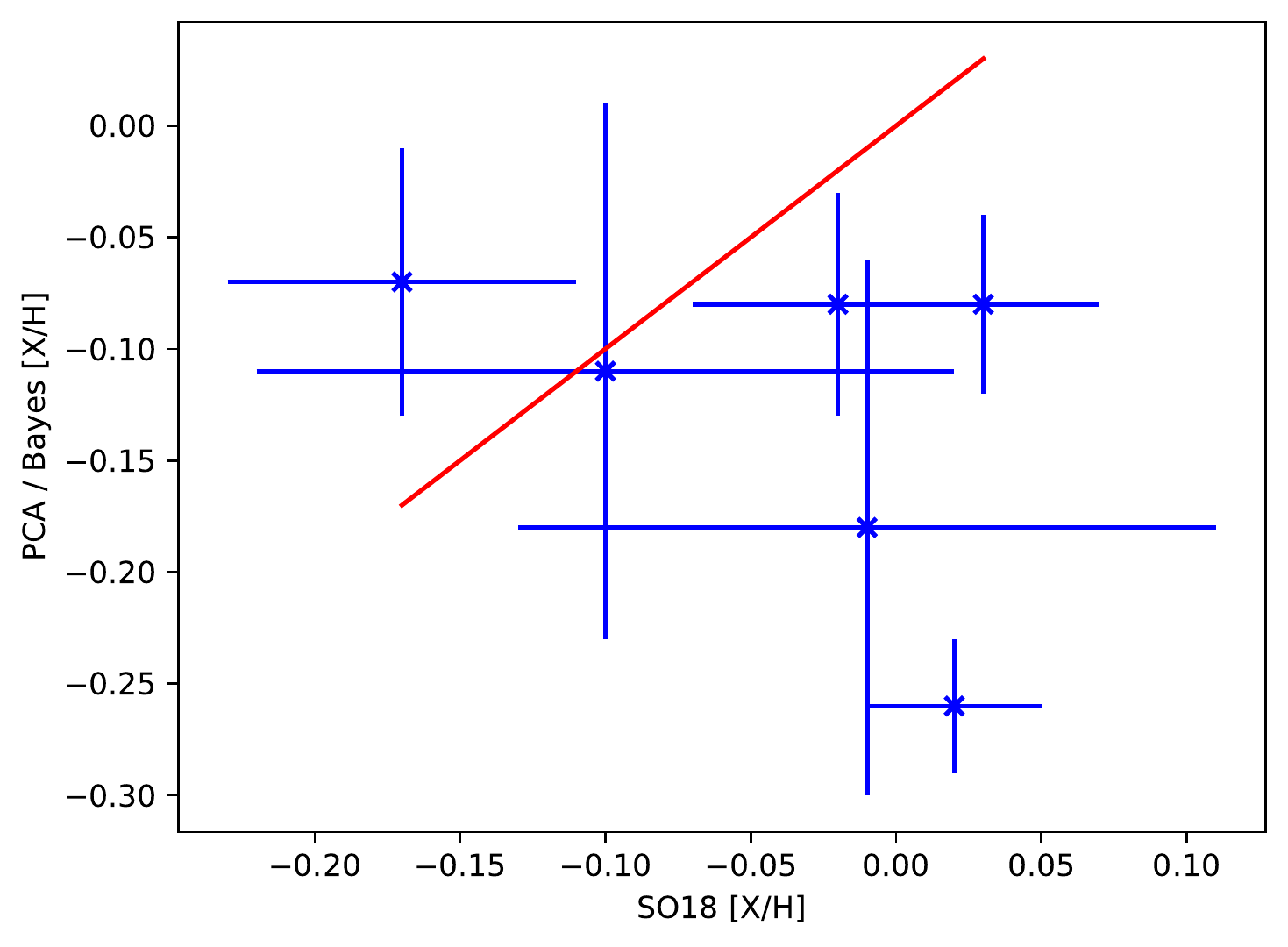}
\end{minipage}
\caption{ 
 Comparison between our derived  PCA/Bayes abundances and those reported by WW20 (Ti, top left panel), 
 SO20 (C, top right panel), and SO18 (Fe, C, Mg, Al, Ca, and Ti for the star Gl 447, bottom panel).
 The red line denotes the one-to-one relationship.
 } 
\label{other_abundances}
\end{figure*}

  As far as other abundances besides iron are considered,
  only few elemental abundances for few stars are available in the literature.
  WW20 also compute abundances of Ti, while SO20 derive abundances of C.
  The corresponding comparisons are shown in Figure~\ref{other_abundances}.
  It is clear that our derived Ti abundances are higher than those previously reported and there seems to be an offset
  between our derived abundances and those reported in WW20. 
  The reason for this offset is not clear,
  given the good agreement between WW20 
  and our results for the iron abundance. 
  For carbon,  SO20 abundances are higher but the agreement is overall good 
  and there seems to be a linear relationship between our abundances and those given in SO20. 
  Finally, \citet[][SO18]{2018ApJ...860L..15S} derived abundances of several elements (namely Fe, C, Mg, Al, Ca, and Ti) from near infrared spectra for the star
  Gl 447. 

\begin{table}
\centering
\caption{ 
 Standard deviation of the differences, residual mean square, mean squared error, and coefficient of determination
 between the   PCA-derived abundances of the training stars and those
 provided in previous works. Note that for SO18 we compare the abundances of six different species for one single star.}
   \label{rmse_previous}
   \begin{tabular}{lcccccc}
   \hline\noalign{\smallskip}
            &   [X/H] & n$_{\rm stars}$  &  $\sigma$ &      RMS      &      RMSE     &       R$^{\rm 2}$ \\
    \hline
MSDL	&	Fe	&	196	&	0.22	&	0.23	&	0.05	&	-2.04	\\
RA12	&	Fe	&	38	&	0.19	&	0.19	&	0.04	&	0.42	\\
ON12	&	Fe	&	10	&	0.14	&	0.14	&	0.02	&	0.07	\\
NE13	&	Fe	&	100	&	0.19	&	0.24	&	0.06	&	-0.16	\\
GA14	&	Fe	&	81	&	0.19	&	0.19	&	0.03	&	0.10	\\
NE14	&	Fe	&	21	&	0.22	&	0.22	&	0.05	&	-0.12	\\
WW20	&	Fe	&	14	&	0.25	&	0.28	&	0.08	&	-0.72	\\
SO20	&	Fe	&	9	&	0.14	&	0.15	&	0.02	&	0.72	\\
PA19-N  &       Fe      &       84      &       0.28    &       0.36    &       0.13    &       -1.17   \\ 
PA19-V  &       Fe      &       85      &       0.23    &       0.24    &       0.06    &       -1.31   \\ 
\hline
WW20	&	Ti	&	14	&	0.15	&	0.31	&	0.10	&	-1.90	\\
SO20	&	C	&	9	&	0.14	&	0.19	&	0.04	&	-0.15	\\
\hline
SO18	&		&	6	&	0.13	&	0.15	&	0.02	&	-3.40	\\
\hline
\end{tabular}
\end{table}


\subsection{Stellar abundance and kinematics}\label{abkin} 

 Stars of the thin and thick disc populations are known to differ in terms of age, chemical composition, spatial distribution,
 and kinematics.  
 Thin disc stars rotate faster than the local standard of rest and show solar $\alpha$-element abundances. On the other hand,
 the thick disc is enriched in $\alpha$ elements and lags the local standard of rest
 \citep[e.g.][]{2003MNRAS.340..304R,2006MNRAS.367.1329R,2014A&A...562A..71B}.
 Most of our targets have a kinematics compatible with 
 the thin disc population.
 If we consider the metallicity values derived from our   PCA/Bayes analysis, we note that
 the mean metallicity of these stars is +0.01 dex. 
 On the other hand, for the  stars possible members of the thick disc population our new technique 
 gives a mean metallicity of -0.39 dex. 
 As it is common in the literature we consider Mg, Si, Ca, and Ti as $\alpha$ elements. 
 Whilst thin disc stars in our sample show a mean [$\alpha$/Fe] value of 
 -0.11 dex, thick discs have larger $\alpha$ abundances with a mean value of +0.10 dex.

\begin{figure*}[!htb]
\centering
\begin{minipage}{0.49\linewidth}
\includegraphics[scale=0.75]{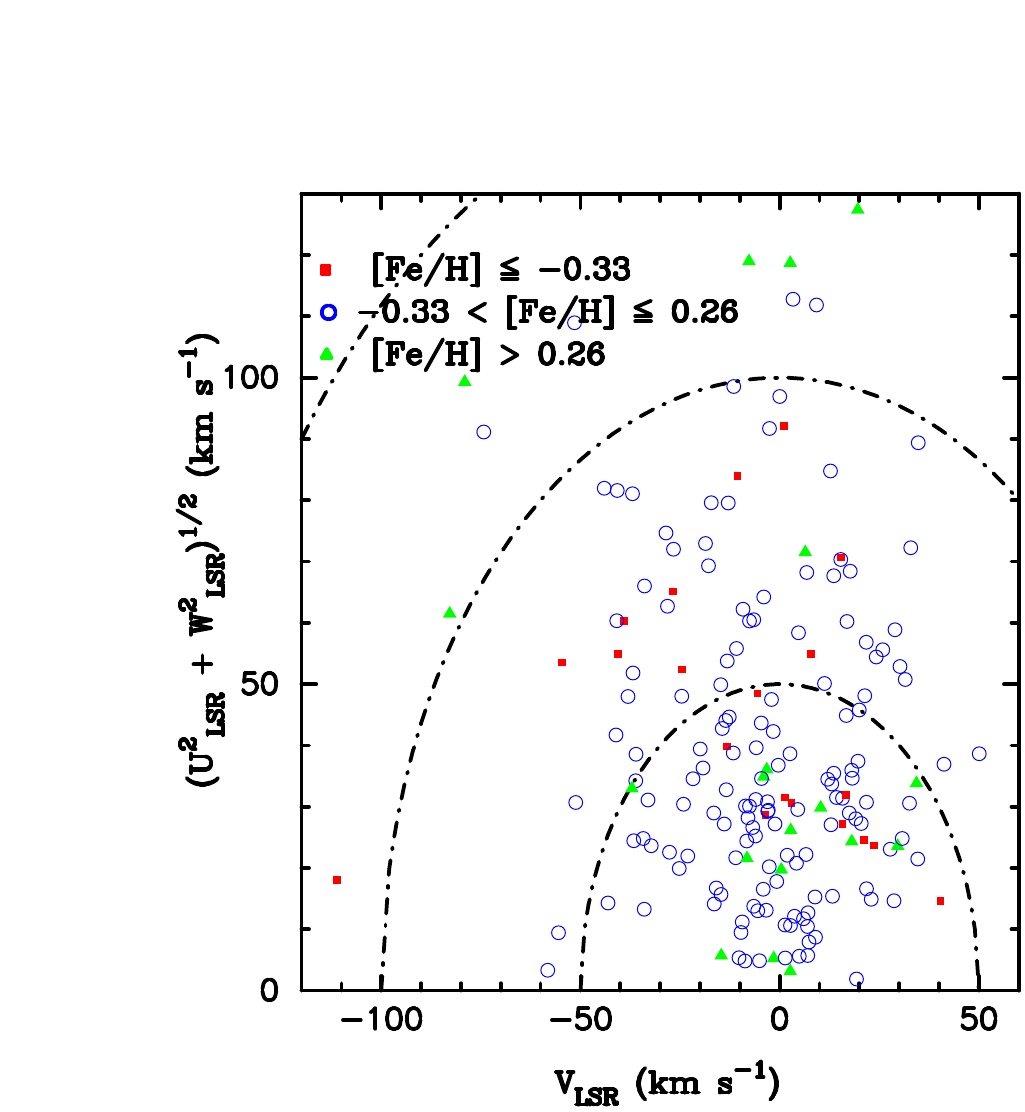}
\end{minipage}
\begin{minipage}{0.49\linewidth}
\includegraphics[scale=0.75]{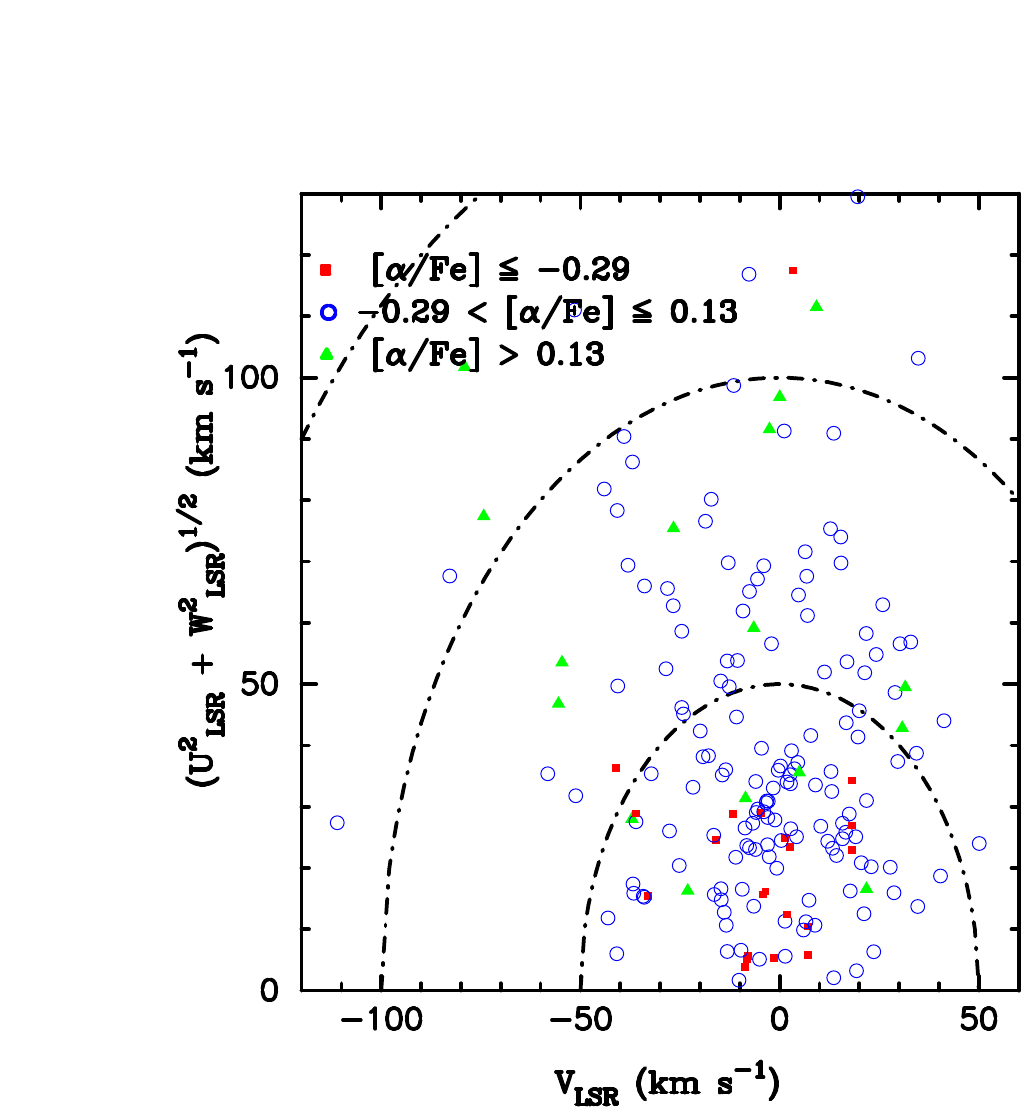}
\end{minipage}
\caption{
Toomre diagram for the stars analysed in this work. Stars are shown with different colours
and symbols according to their iron abundances (left) and [$\alpha$/Fe] abundances (right).
Intervals in [Fe/H] and [$\alpha$/Fe] were selected to show the stars in the 10 and 90
quartiles of the distribution. 
}
\label{toomre_diagram}
\end{figure*}

 This can also be seen in Fig.~\ref{toomre_diagram} where we show the Toomre diagram of the observed stars.
 Stars are plotted with different colours and symbols according to their metallicities (left)
 and [$\alpha$/Fe] abundances (right). The dashed lines indicate values of constant total velocities.
 The figure shows that low-metallicity stars span a much larger range of total velocities than
 the metal-rich stars. On the other hand, stars with higher [$\alpha$/Fe] abundances show 
 larger total velocities. We conclude that our derived abundances reproduce the known 
 kinematic trends from the different stellar populations.

\subsection{Stellar abundances and activity.}

 In cool stars with convective outer-layers  
 chromospheric activity and rotation are linked by the stellar dynamo
 \citep[e.g.][]{1967ApJ...150..551K,1984ApJ...279..763N,2001MNRAS.326..877M}
  and both (activity and rotation) are known to decrease with time as a star loses angular momentum with
 stellar winds via magnetic braking \citep{1967ApJ...148..217W,1993MNRAS.261..766J}.
 On the other hand, stellar metallicity reflects the enrichment history of the interstellar medium
 \citep[e.g.][]{1995ApJS...98..617T} 
 and for a given spectral type young stars are expected to show higher metallicity values. 
 In addition, stars with a greater metallicity may be more active because their convection zones tend to be
 deeper than those of less-metallic stars with the same parameters and because the chromospheric emission
 in metal spectral lines may be enhanced \citep[cf.][]{2018ApJ...852...46K}.
  We caution that this result has been observed in only one solar twin star so similar analysis on other stars
 are needed to test it. In particular it is unclear whether it would also hold for low-mass stars,
 and specially for fully convective late-M dwarfs.

\begin{figure*}[!htb]
\centering
\begin{minipage}{0.49\linewidth}
\includegraphics[scale=0.45]{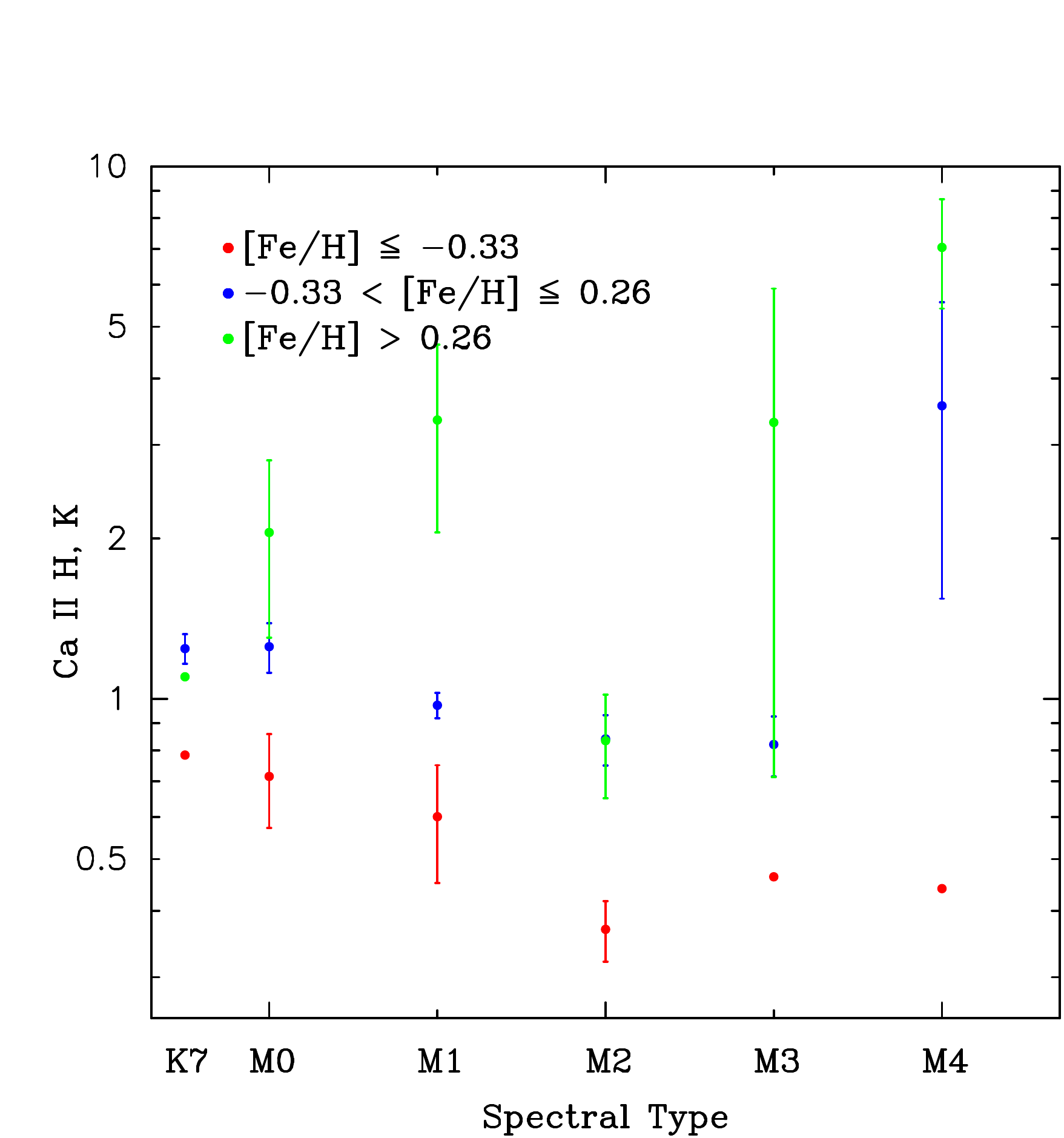}
\end{minipage}
\begin{minipage}{0.49\linewidth}
\includegraphics[scale=0.45]{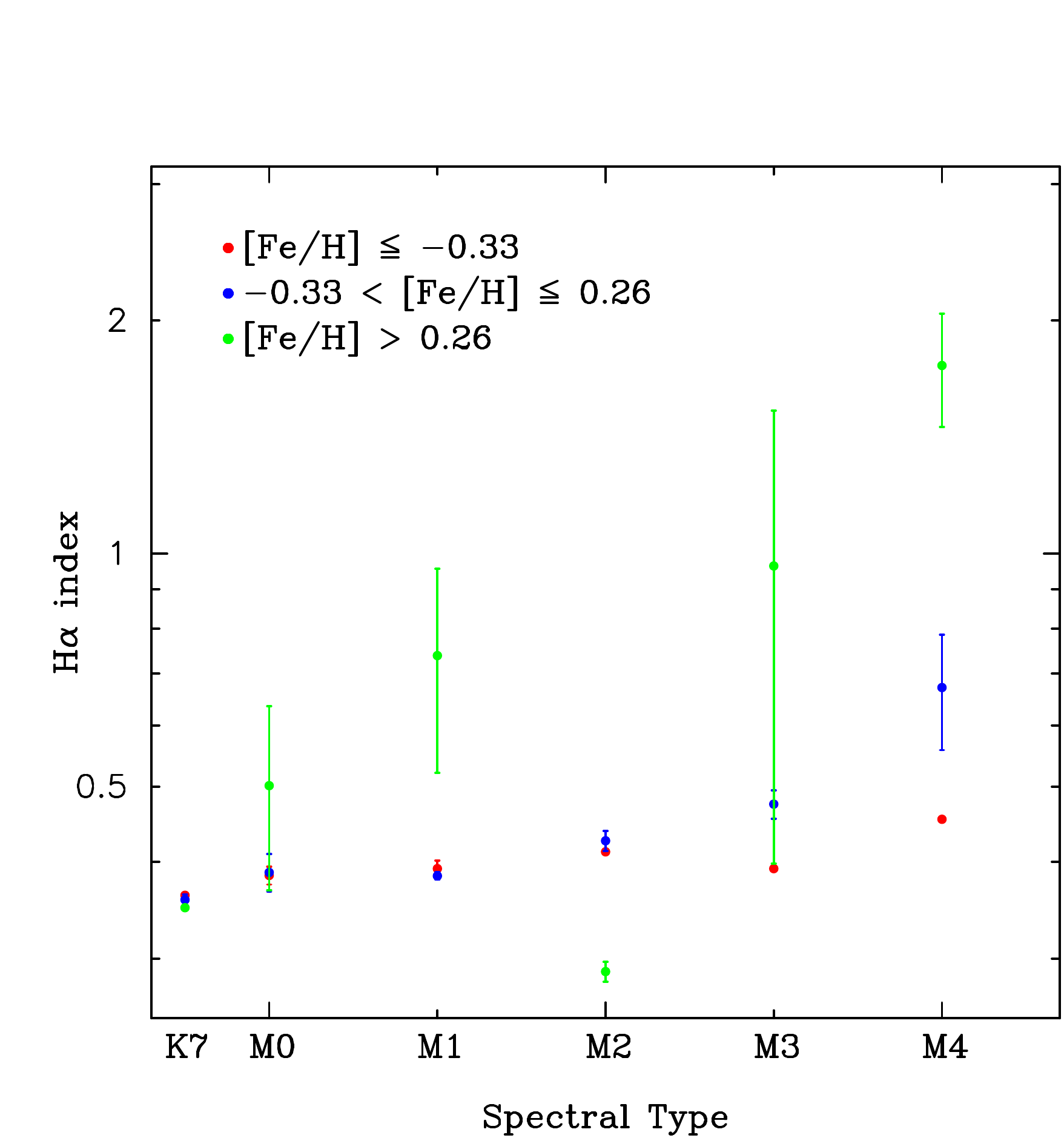}
\end{minipage}
\caption{ 
Ca~{\sc ii} H \& K (left), and H$\alpha$ (right) activity indexes as a function of the spectral sub-type.
Mean values for stars into three bins of metallicity ranges are shown with different colours.}
\label{activity_diagram}
\end{figure*}

 We tested whether this is the case. Activity indexes in the main optical indicators, Ca~{\sc ii} H \& K,
 Balmer lines (from H$\alpha$ to H$\epsilon$), Na~{\sc i} D$_{\rm 1}$  D$_{\rm 2}$ and He~{\sc i} D$_{\rm 3}$
 were computed following the bandapasses defined in \cite{2019A&A...627A.118M}. 
 Figure~\ref{activity_diagram} shows the
  mean Ca~{\sc ii} (left) and H$\alpha$ (right) activity indexes
 values of the stars divided into three metallicity bins
 as a function of
 the  spectral type.  
 When considering the calcium index the figure clearly shows a tendency of higher levels of activity at higher metallicities
 for a given spectral type,
  which
 suggests that more active stars have higher metallicities. 
 For H$\alpha$, the tendency is  even more clear.
 Similar results are obtained for the other activity indexes (not shown). 
 Our conclusion is that our derived abundances reproduce the expected trends with stellar activity. 

\section{Completeness of the planet host sample}\label{detection_limits}

 Our aim is to understand planet formation and evolution as a function of the main stellar properties.
 It is therefore fundamental to select stars for which planets of a given mass and period can be detected
 in a uniform way.
 In order to estimate the detectability limits of our planet hosts, we proceeded as in \cite{2019A&A...624A..94M}.
 For each star we collected all the available HARPS and HARPS-N data. We are aware that for many stars there should
 be radial velocities measured with other instruments. However, given that radial velocity time series are usually
 non available for non-planet detections, we decided to focus only on HARPS and HARPS-/N data. 
 This is certainly a ``conservative'' approach, as it might overestimate the detection limits for some periods.
 However, it is a robust and homogeneous method and allows us to obtain an efficient determination of the detection limits.
  More detailed detection limits around samples of M dwarfs, in particular for the stars included in the  HADES
 survey will be addressed in forthcoming works.

 For each star, radial velocities were computed from the available data using the 
 TERRA pipeline \citep{2012ApJS..200...15A}, which provides a better radial velocity accuracy when applied to M dwarfs
 \citep{2017A&A...598A..26P}.
 For those stars with known planets, we used the code rvlin\footnote{http://exoplanets.org/code/} \citep{2009ApJS..182..205W}
  to subtract the contribution of the planets to the radial velocity by using a multiplanet Keplerian fit
 with the planetary periods fixed to the published values.

 We then computed the expected radial velocity semi-amplitude due to the presence of different types of planets by
 sampling in logarithmic space the planetary mass-period space.
  In particular, planetary masses from 0.005 to 80 M$_{\rm Jup}$ and orbital periods from one to 10$^{\rm 4}$ days
 were considered.
 Circular orbits were considered as it has been shown that planet detection's limits are not
 strongly dependent on the eccentricity \citep{2002A&A...392..671E,2010MNRAS.401.1029C}.
 For each planet, the expected radial velocities were computed keeping the same time as in the original observations.
 Several realisations of the radial velocity were performed, each one corresponding to a different phase offset.
 Following previous works \citep{2005A&A...443..337G,2012A&A...542A..18L,2012A&A...545A..87M}
 a planet is considered to be detectable around a given star if
 the root mean square (rms) of the planet's expected radial velocity is larger
 than the rms of the residuals of the stellar radial velocity (that is, after subtracting the
 signals of the already known planets) in each of the simulated phases.

\begin{figure}[htb]
\centering
\begin{minipage}{\linewidth}
\includegraphics[scale=0.55]{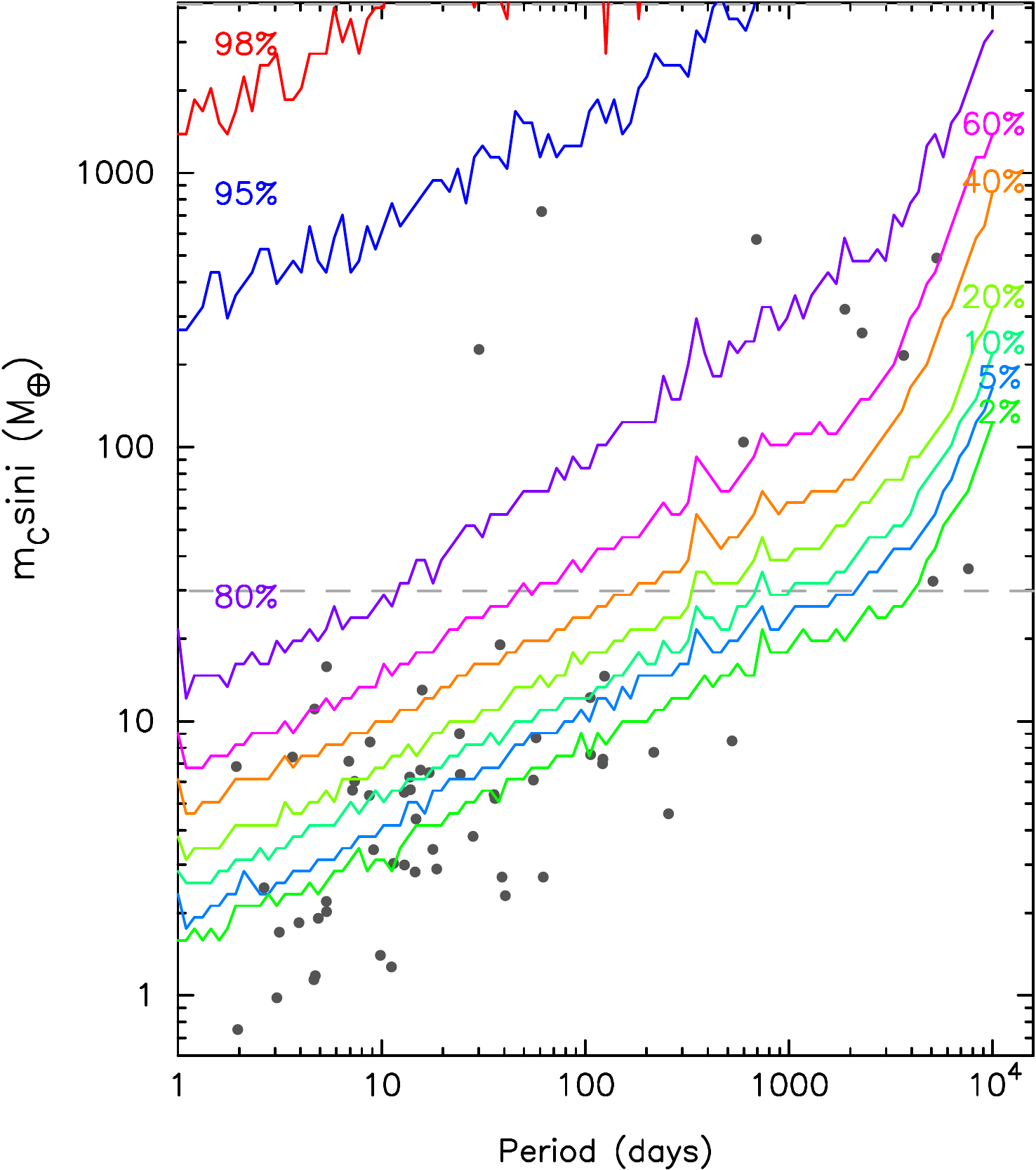}
\end{minipage}
\caption{  
Minimum mass versus planetary period diagram. Substellar companions known around M dwarfs are shown in grey circles.
Detection probability curves are superimposed with different colours. 
The horizontal dashed line indicates the standard mass loci of low-mass planets, and gas-giant planet companions.}
\label{detection_limits_plot}
\end{figure} 

  The derived detection probability curves are shown in Figure~\ref{detection_limits_plot}. The curves show for each period
  the percentage of stars from our sample for which planets with corresponding minimum mass can be detected.
  The results show that for $\sim$ 80\% of the stars in our sample, low-mass planets within a 10 days period might be
  detected. Gaseous planets more massive than 100 M$_{\oplus}$ and $\sim$ 200 M$_{\oplus}$ with periods within 100 and 1000 days, respectively
  can also be excluded for $\sim$ 80\% of the stars. 
  It can be seen that several known planets are actually located under the 2\% probability curve. This is not surprising. As mentioned before
  our approach is quite conservative.

\section{Results}\label{results}

  Our sample is composed of a total of 204 stars with homogeneous mass and metallicity measurements.
  Five stars host at least one giant planet (hereafter giant host subsample), while 29 stars harbour at least one
  low-mass planet (hereafter low-mass host subsample).
  Two stars namely GJ 15A and Gl 433 host one (two) low-mass planets plus an additional planet somehow in the boundary 
  between low-mass and giant (minimum masses 36 and 32 M$_{\oplus}$, respectively). These two stars were considered in the
  low-mass planet subsample. 
  On the other hand, Gl 832 and Gl 876 harbour simultaneously low-mass companions as well as planets more massive than 200 M$_{\oplus}$ so
  we add these stars to the giant host subsample.
  As seen before, given the available data, it is unlikely that stars in the low-mass host and comparison subsamples host
  a non-detected gas-giant planet at short periods
   (with the data at hand they should have been already detected).
  However, we cannot rule out the possibility that stars in the comparison subsample
  host non-detected low-mass planets. In the following, unless otherwise noticed, we consider the stellar metallicity values derived
  with the  PCA/Bayes technique. 

\subsection{Biases} 
\begin{figure*}[!htb]
\centering
\begin{minipage}{0.33\linewidth}
\includegraphics[scale=0.375]{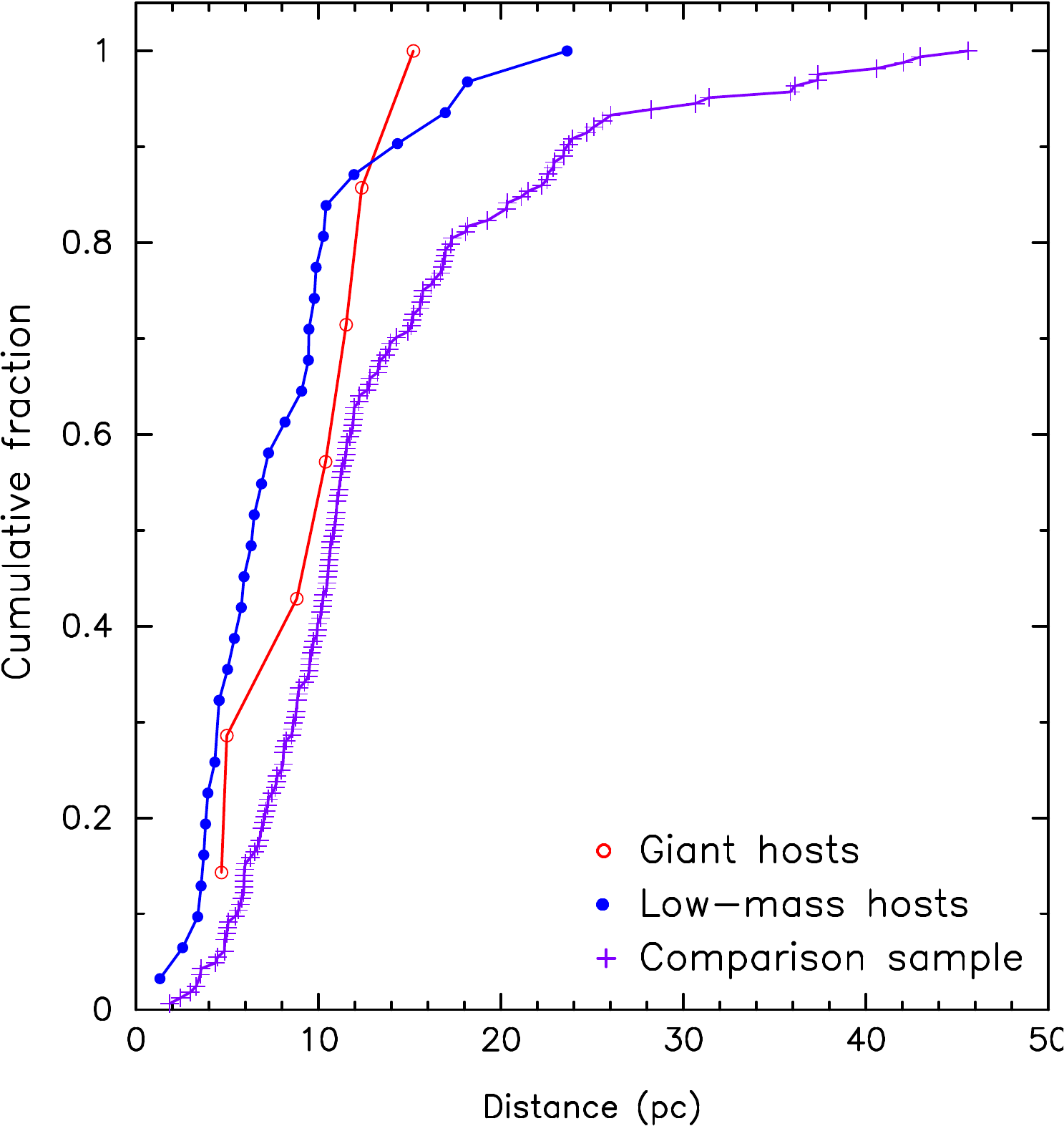}
\end{minipage}
\begin{minipage}{0.33\linewidth}
\includegraphics[scale=0.375]{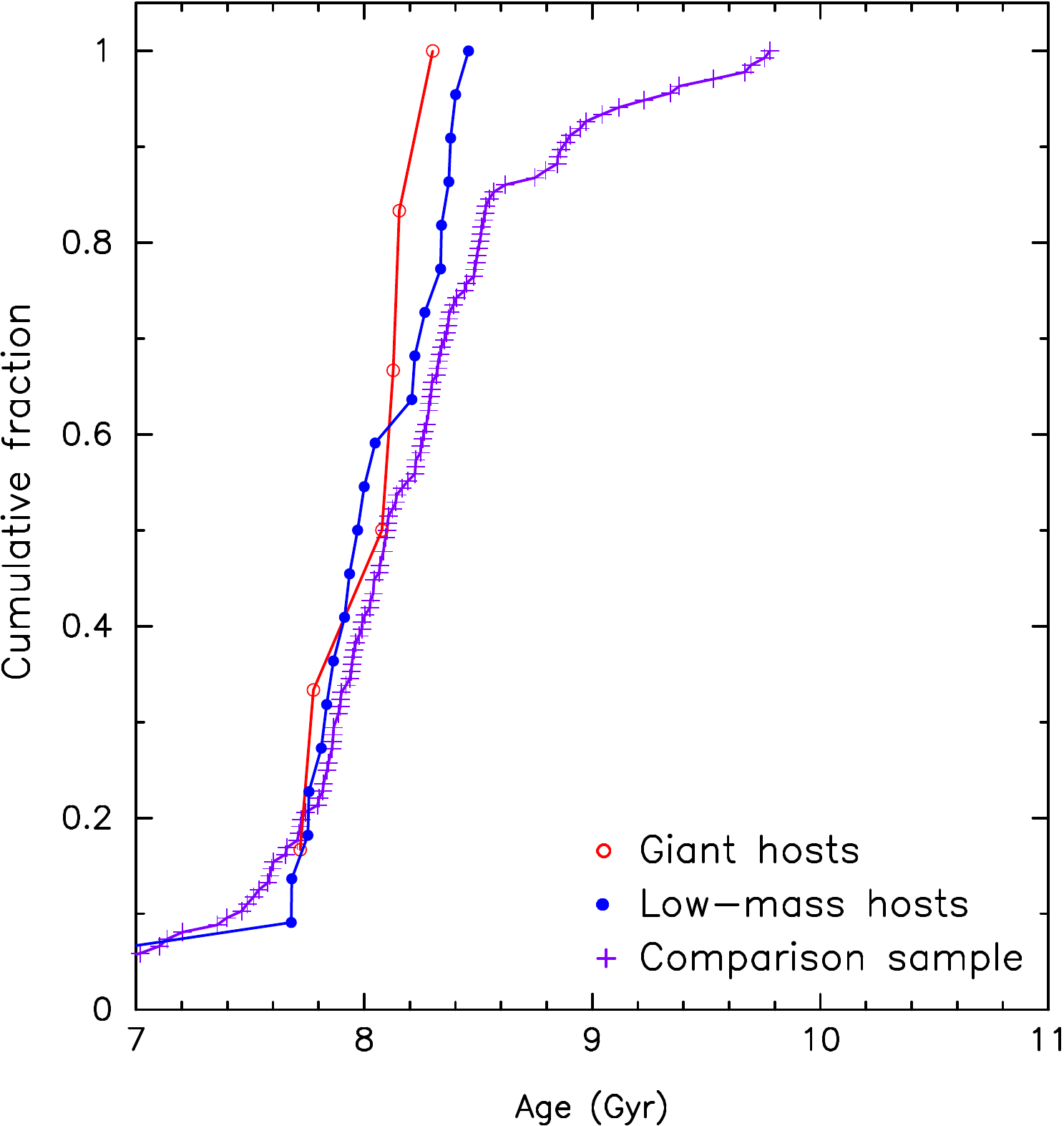}
\end{minipage}
\begin{minipage}{0.33\linewidth}
\includegraphics[scale=0.375]{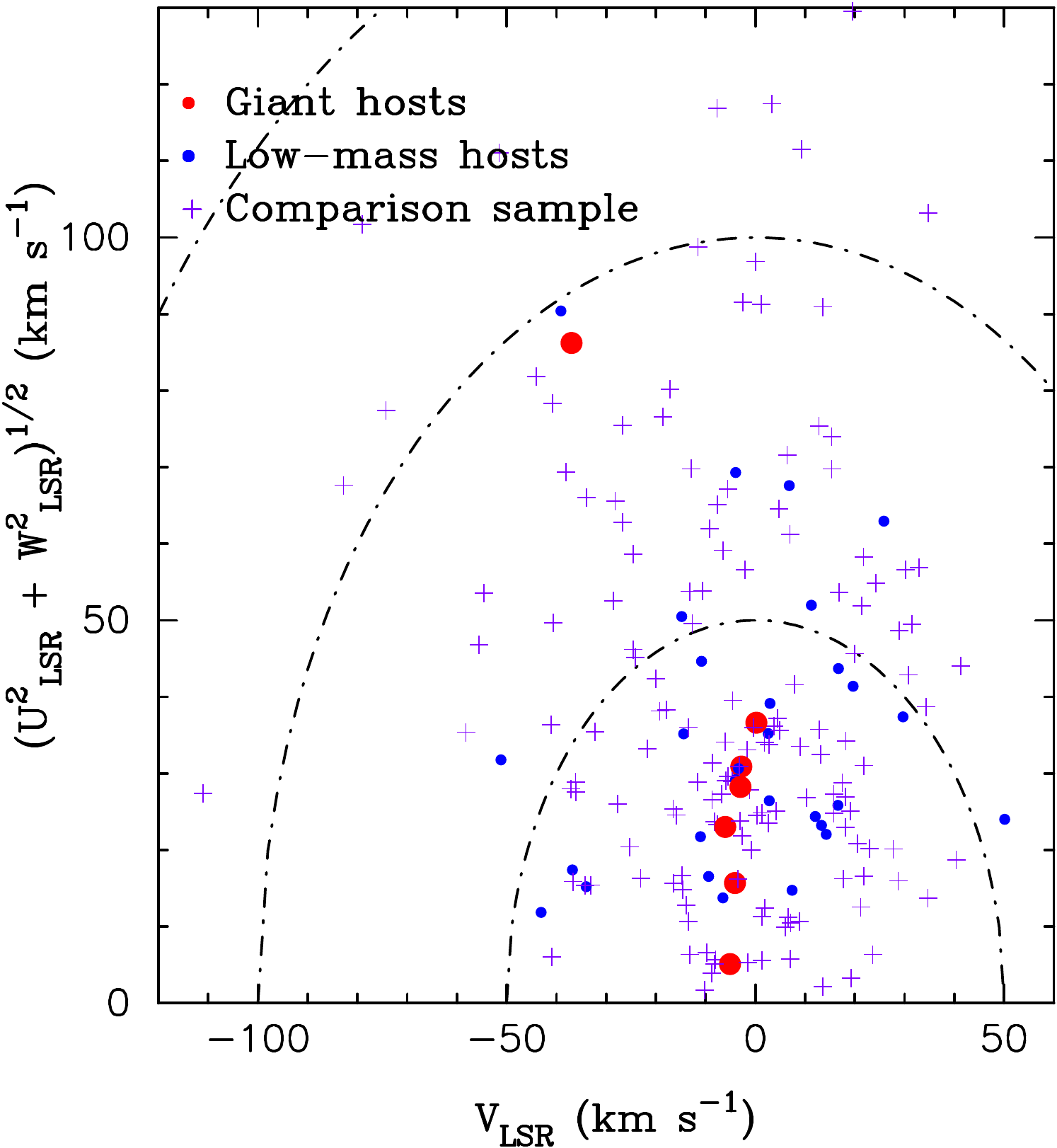}
\end{minipage}
\caption{  
Cumulative distribution function for the distance (left), age (centre),
and Toomre diagram (right) for the stars analysed in this work.}
\label{fwhm_bis_vrad}
\end{figure*}

  Before we proceed further in the comparison between the different subsamples, an exploration of the
  possible sources of bias that could mimic metallicity differences is called for.
  We therefore compared the different subsamples in terms of distance, age, and kinematics,
  which are the parameters most likely to affect the metal content of a star. 
  The comparison is given in Table~\ref{bias_statistics}, while Figure~\ref{fwhm_bis_vrad} shows the corresponding cumulative
  distribution functions for the distance (left), age (centre), and the Toomre diagram (right).

\begin{table*}
\centering
\caption{ Comparison between the properties of the different samples analysed in this work.}
\label{bias_statistics}
\begin{tabular}{lccccccccc}
\hline
       & \multicolumn{3}{c}{\bf{Comparison}}        & \multicolumn{3}{c}{\bf Giant hosts} & \multicolumn{3}{c}{\bf Low-mass hosts} \\ 
\hline
              &  Range      & Mean   & Median  &  Range      &  Mean  & Median   &  Range      &   Mean   &   Median  \\
\hline
Distance (pc) & 1.83/45.60  & 13.24  & 10.90   &  4.68/15.20 &  9.70  &   10.38   &  1.30/23.62 &   7.94   &  6.47     \\
Age (Gyr)     & 0.65/9.78   & 8.00   &  8.10   &  7.72/8.30  &  8.03  &   8.13   &  6.40/8.46  &   7.98    &  8.00     \\ 
\hline
D/TD$^{\dag}$ (\%) & \multicolumn{3}{c}{77 (D);  4 (TD); 17 (R); 2 (H)} & \multicolumn{3}{c}{86 (D);  14 (R)} & \multicolumn{3}{c}{80 ( D);  3.3 (TD); 13.3 (R); 3.3 (H)} \\ 
\hline
\end{tabular}
\tablefoot{$^{\dag}$ D: thin disc, TD: thick disc, R: transition, H: halo}
\end{table*}

  The comparison shows that while there seems to be no differences in terms of age
  or kinematics between planet hosts and the comparison stars, stars with planets
  tend to be systematically located at shorter distances.
  Note that typical uncertainties in the stellar age is around 4 Gyr, while errors
  in the distance are around 0.03 pc.
  A KS test confirms that there are no differences in terms of age
  ($D$ = 0.24, $p$-value = 0.18, n$_{\rm eff}$ = 18.94 for comparison/low-mass hosts;
   and $D$ = 0.36, $p$-value = 0.43, n$_{\rm eff}$ = 5.75 for comparison/giant hosts)
   while low-mass planet hosts are located at shorter distances
   ($D$ = 0.41, $p$-value = 0.0002, n$_{\rm eff}$ = 26.1 for comparison/low-mass hosts;
      and $D$ = 0.28, $p$-value = 0.59, n$_{\rm eff}$ = 6.7 for comparison/giant hosts).
  This trend could reflect a bias in the exoplanet surveys, as closer stars are usually brighter
  and thus higher signal-to-noise observations can be obtained with shorter integration times.
  Therefore, closer and brighter stars can be observed with a more dense temporal cadence and
  higher signal-to-noise values thus favouring
  the detection of planets.

\subsection{Metallicity distribution} 

\begin{figure*}[htb]
\centering
\begin{minipage}{0.45\linewidth}
\includegraphics[scale=0.45]{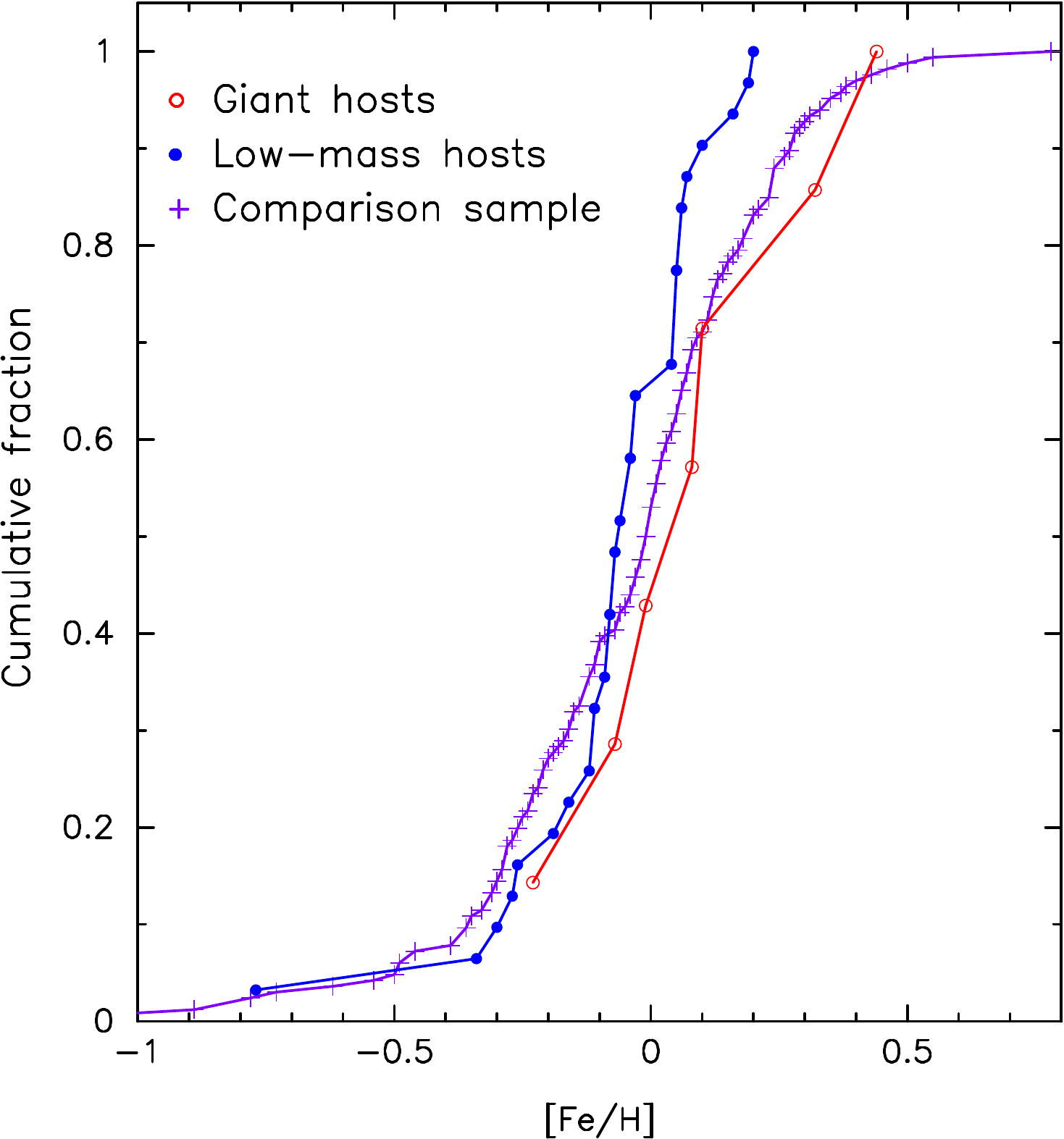}
\end{minipage}
\begin{minipage}{0.45\linewidth}
\includegraphics[scale=0.45]{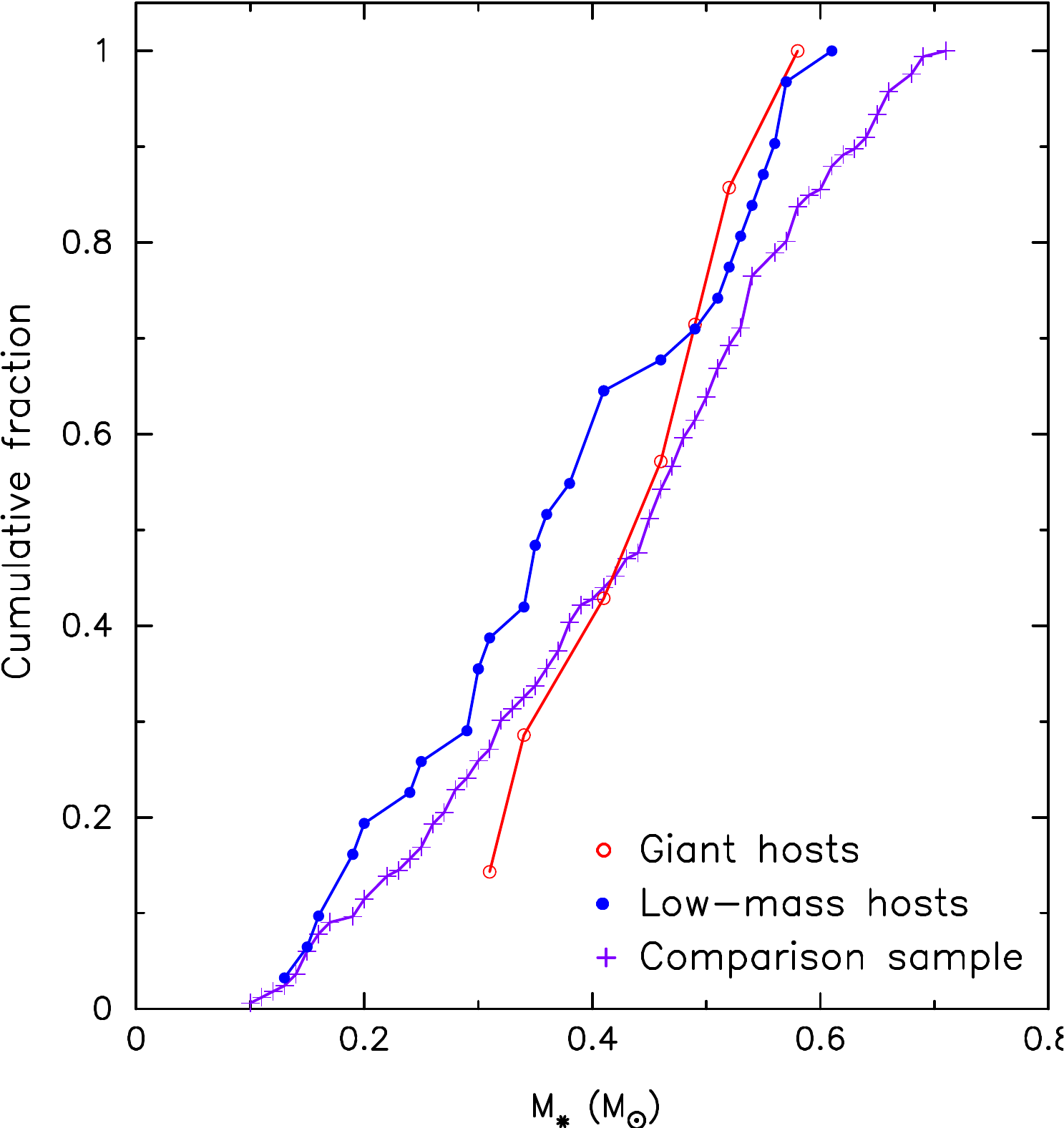}
\end{minipage}
\caption{ [Fe/H] (left) and stellar mass (right) cumulative distributions for the different samples studied in this work.}
\label{metal_distributions}
\end{figure*}

 The cumulative distribution function of the metallicity for the different samples analysed in this
 work is presented in Fig.~\ref{metal_distributions}, left panel. For guidance some statistical diagnostics are also given in Table~\ref{metal_statistics}.
 The figure shows a tendency of giant hosts to show high metallicities although their cumulative distribution does not look
 very different from those of the comparison sample.
 On the other hand the low-mass hosts also show a similar metallicity distribution than the comparison sample.
 A KS test shows that there are no statistically significant differences between
 the metallicity distribution of the three subsamples
 ($D$ = 0.21, $p$-value = 0.16, n$_{\rm eff}$ = 26.1 for comparison/low-mass hosts; 
 and $D$ = 0.25, $p$-value = 0.70, n$_{\rm eff}$ = 6.7 for comparison/giant hosts).

 Figure~\ref{metal_distributions} shows that M dwarfs with gas-giant planets might have high metallicity values
  (suggesting that the gas-giant planet-metallicity correlation
  found on FGK stars might also  hold for M dwarfs)
 although the trend needs statistical confirmation. On the other hand, 
 stars with low-mass planets do not show the metal-rich signature.
 These results seem in line with previous works
 \citep{2007A&A...474..293B,2009ApJ...699..933J,2010A&A...519A.105S,2012ApJ...748...93R,2012ApJ...747L..38T,2013A&A...551A..36N}.

\begin{table*}
\centering
\caption{  [Fe/H] statistics of the stellar samples.}
\label{metal_statistics}
\begin{tabular}{lcccccc}
\hline
 Sample & Mean & Median & Deviation & Min & Max & $N$ \\
\hline
Comparison    &    -0.04 &       +0.00 &        0.27  &      -1.08  &       0.78  & 166 \\
  Low-mass    &    -0.07 &       -0.06 &        0.18  &      -0.77  &       0.20  &  31 \\
      Giants  &     0.09 &        0.08 &        0.21  &      -0.23  &       0.44  &   7 \\
\hline
\end{tabular}
\end{table*}

\begin{figure*}[htb]
\centering
\begin{minipage}{0.45\linewidth}
\includegraphics[scale=0.45]{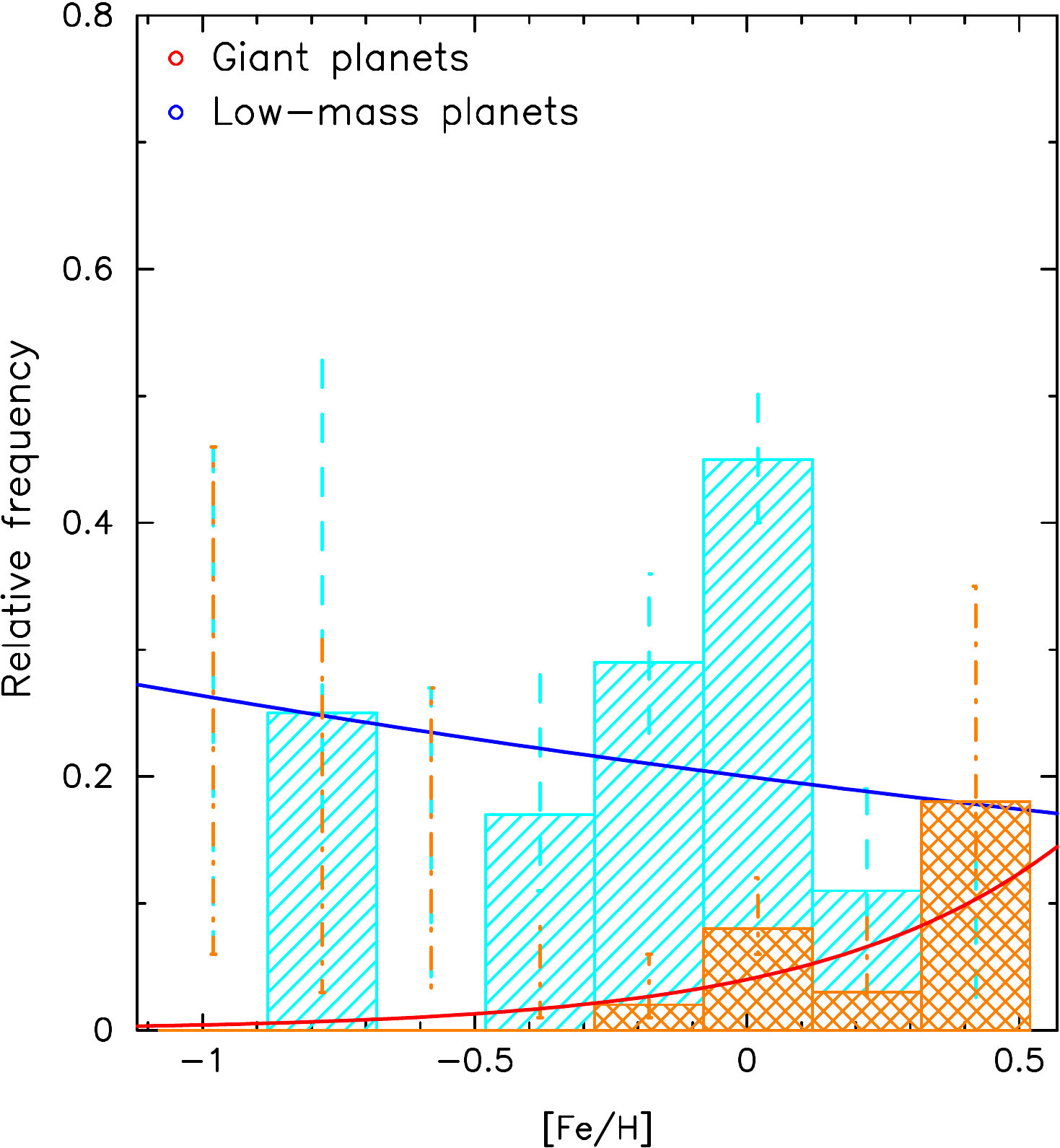}
\end{minipage}
\begin{minipage}{0.45\linewidth}
\includegraphics[scale=0.45]{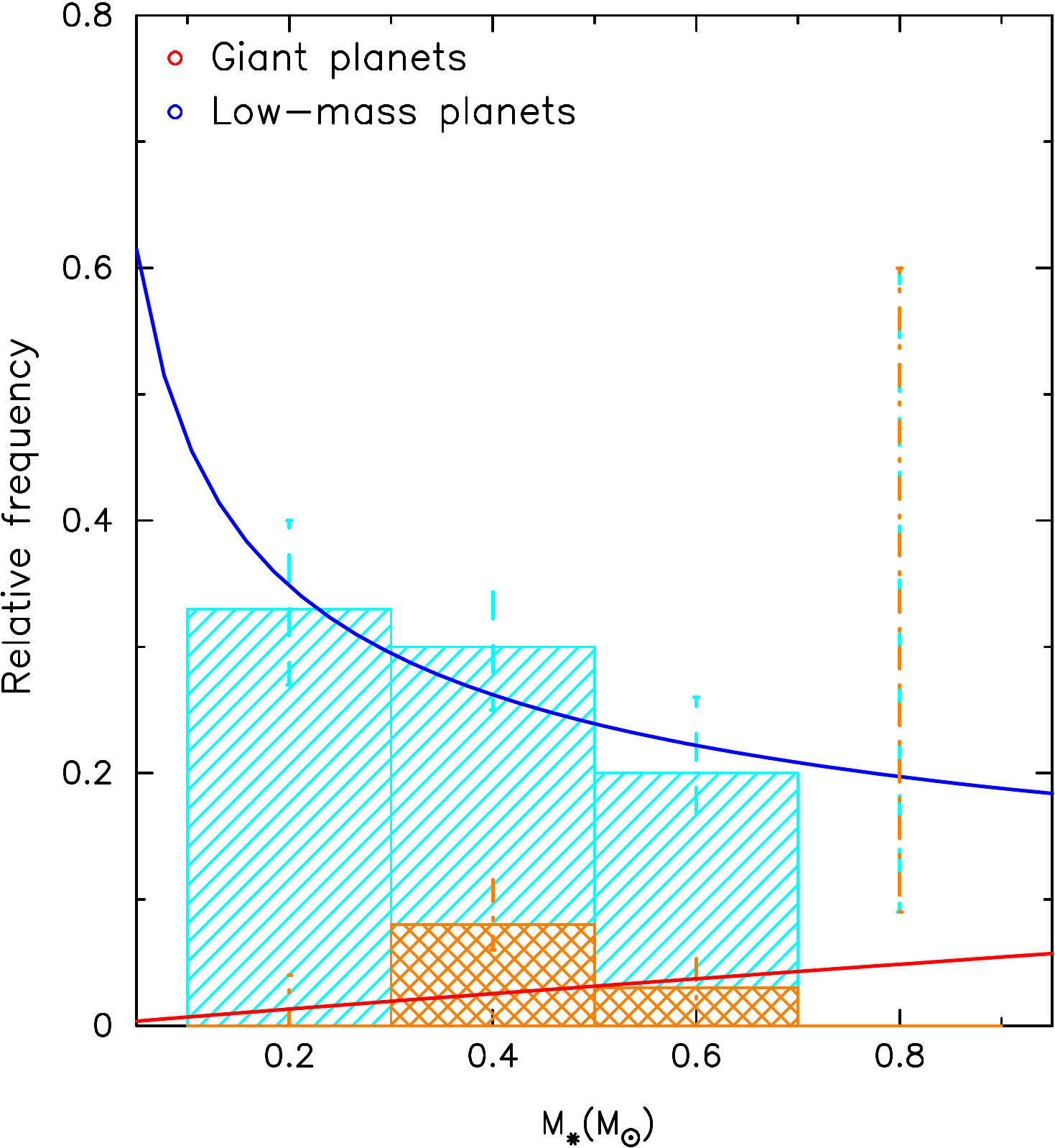}
\end{minipage}
\caption{Fraction of gas-giant planets (light orange) and low-mass planets (light blue) as a function of the stellar [Fe/H] (left) and stellar mass (right).
The best bin fitting are shown in red for gas-giant planets and in blue for low-mass planets.}
\label{metal_histogram}
\end{figure*}

  Figure~\ref{metal_histogram}, left panel, shows the fraction of gas-giant planets (light orange) and
  low-mass planets (light blue) as a function of the stellar metallicity.
  The errors in the frequency of each bin are calculated using
  the binomial distribution, 

  \begin{equation}
  P(f_{p},n,N)=\frac{N!}{n!(N-n)!}f_{p}^{n}(1-f_{p})^{N-n}
  \end{equation}

  \noindent where $P(f_{\rm p}, n, N)$ is the probability of $n$ detections given a sample of size $N$ when the true planetary companion frequency is f$_{\rm p}$.
  Given that the probability distribution is not symmetric about its maximum, a common practice
  is to report the range in planetary fraction that delimits the 68.2\% of the integrated
  probability function, which is equivalent to the 1$\sigma$ limits for a Gaussian distribution
  \citep[e.g.][]{2003ApJ...586..512B,2006ApJ...649..436E,2009ApJ...697..544S,2013A&A...551A..36N}.
  
  The fraction of stars with planets was fitted to a function with the functional form $f = C10^{\alpha[Fe/H]}$, following previous works
  \citep[e.g.][]{2005ApJ...622.1102F,2007ARA&A..45..397U}, although other functional forms have been discussed in the literature \citep[e.g.][]{2013A&A...551A.112M}.
 Our analysis provides $C$ = 0.20 $\pm$ 0.03, $\alpha$ = -0.12 $\pm$ 0.16 for the low-mass planets hosts;
 and $C$ = 0.04 $\pm$ 0.02, $\alpha$ = 0.98 $\pm$ 0.87 for the giant hosts.

  Two conclusions can be drawn from this analysis. First, the frequency of low-mass planets
  does not seem to be a function of the stellar metallicity. Furthermore, as the $\alpha$ value
  is negative there might be an anti correlation, that is, a decreasing fraction of planet frequency with increasing stellar
  metallicity. 
  This result seems to be in line with previous analysis of samples of M dwarfs  \citep{2013A&A...551A..36N}.
  Second, gas-giant planets show a planet-metallicity correlation with an $\alpha$ value
  compatible within errors with the values found for FGK stars, $\sim$ 1.7  \citep[e.g.][]{2005ApJ...622.1102F}.
  However, our value of $\alpha$ is  lower than previous values found in samples of M dwarfs,
  $\alpha \sim$ 1.26-2.94, \citep{2013A&A...551A..36N}.
  Nevertheless, we caution that the uncertainty in $\alpha$ is high, and we have not taken into account
  a possible dependence of the planet fraction on the stellar mass. 
  We address these issues in the following sections. 


\subsection{Mass distribution}

  Planet occurrence is also known to show a dependence on the stellar mass
  \citep[][]{2003AJ....125.2664L,2007A&A...472..657L,2007ApJ...670..833J,2010PASP..122..905J}.
  Figure~\ref{metal_distributions}, right panel, shows the cumulative distribution function of the stellar mass for the samples analysed in this work.
   The figure shows
  a hint of planet hosts to be less massive than the comparison stars although
  this could be an observational bias, as planets are easier to find around less massive stars. 
  Some statistical diagnostics are also provided in Table~\ref{mass_statistics}.
  The results from a KS tests provide
  $D$ = 0.20, $p$-value = 0.19, n$_{\rm eff}$ = 26.1 for comparison/low-mass hosts;
  and $D$ = 0.26, $p$-value = 0.69, n$_{\rm eff}$ = 6.7 for comparison/giant hosts.

\begin{table*}
\centering
\caption{ Stellar mass statistics of the stellar samples.}
\label{mass_statistics}
\begin{tabular}{lcccccc}
\hline
 Sample & Mean & Median & Deviation & Min & Max & $N$ \\
\hline
Comparison     &    0.43  &       0.45   &      0.16   &      0.10  &       0.71 & 166 \\
Low-mass       &    0.38  &       0.36   &      0.14   &      0.13  &       0.61 &  31 \\
Giants         &    0.44  &       0.46   &      0.09   &      0.31  &       0.58 &   7 \\
\hline
\end{tabular}
\end{table*}

  Figure~\ref{metal_histogram}, right panel, shows the fraction of gas-giant planets (light orange) and
  low-mass planets (light blue) as a function of the stellar mass.
  The fraction of stars with planets was fitted to a power-law function $f = C M_{\star}^{\alpha}$
  as previous works have shown that planet occurrence rises monotonically with the stellar mass \citep{2010PASP..122..905J}. 
  The derived values are $C$ = 0.06 $\pm$ 0.06, $\alpha$ = 0.94 $\pm$ 1.23 for the gas-giant planets hosts;
  and $C$ = 0.18 $\pm$ 0.05, $\alpha$ = -0.41 $\pm$ 0.25 for the low-mass planets hosts.

  Our results show that gas-giant planet occurrence increases with the stellar mass in line
  with previous findings \citep{2010PASP..122..905J,2014ApJ...781...28M}, while
  the frequency of low-mass planets tends to decrease as we move towards more massive stars.
  This last fact might be an observational bias due to the fact that small planets are easier to 
  detect around less-massive stars.
  However, we note that a similar trend, that is, a higher frequency of planets towards
  less massive stars has been found in the {\sc Kepler} data \citep{2019AJ....158...75H}.
  

\subsection{Simultaneous fit to stellar mass and metallicity}\label{simultaneous}

  Finally, we test the planet frequency as a function of the stellar mass and metallicity simultaneously.
   For this purpose we follow a Bayesian approach, discussed at length in \cite{2010PASP..122..905J,2013A&A...551A.112M}
   and \cite{2014ApJ...781...28M}.
     
   In brief, we fit the planetary frequency as a function of the mass and metallicity
   through the following relationship:

\begin{equation}
\label{eq1}
f(M,F)=CM^{\alpha}10^{\beta F}
\end{equation}

  \noindent where $M$ is the stellar mass, $F$ is the stellar metallicity and $X$ = $(C, \alpha, \beta)$
  are the parameters to be optimised. 
  Each star represents a Bernoulli trial
  and the probability of finding a planet at a given mass and metallicity is given by the binomial
  distribution.
  For a total of T stars, the probability of a detection around a star $i$ (of H total detections) is
  given by $f(M_{i}, F_{i})$, while the probability of a non-detection around a star $j$ is
  $1 - f(M_{j}, F_{j})$.
  The probability of an specific model $X$, considering our data $d$, is given by the Bayes' theorem:

\begin{equation}
P(X|d)\propto P(X)\prod_{i}^{H}f(M_{i},F_{i})\times\prod_{j}^{T-H}[1-f(M_{j},F_{j})]
\end{equation}
  
  Each star's mass and metallicity is considered to be a Gaussian distribution with means $(M_{i}, F_{i})$
  and standard deviations $(\sigma_{M,i}, \sigma_{F,i})$, so the predicted planet fraction for the $i$-th star is:

\begin{equation}
f(M_{i},f_{i})=\iint p_{obs}(M_{i},F_{i})f(M,F)dMdF
\end{equation}

  The marginal log likelihood is finally obtained by taking logarithms in Eqn.~\ref{eq1}:

\begin{eqnarray} 
\mathcal{L}\equiv\log P(X/d)\propto\sum_{i}^{H}\log f(M_{i},F_{i})+ \sum_{j}^{T-H}\log[1-f(M_{j},F_{j})] \nonumber \\
+\log P(X)
\end{eqnarray} 
  
\begin{figure*}[htb]
\centering
\begin{minipage}{0.45\linewidth}
\includegraphics[scale=0.42]{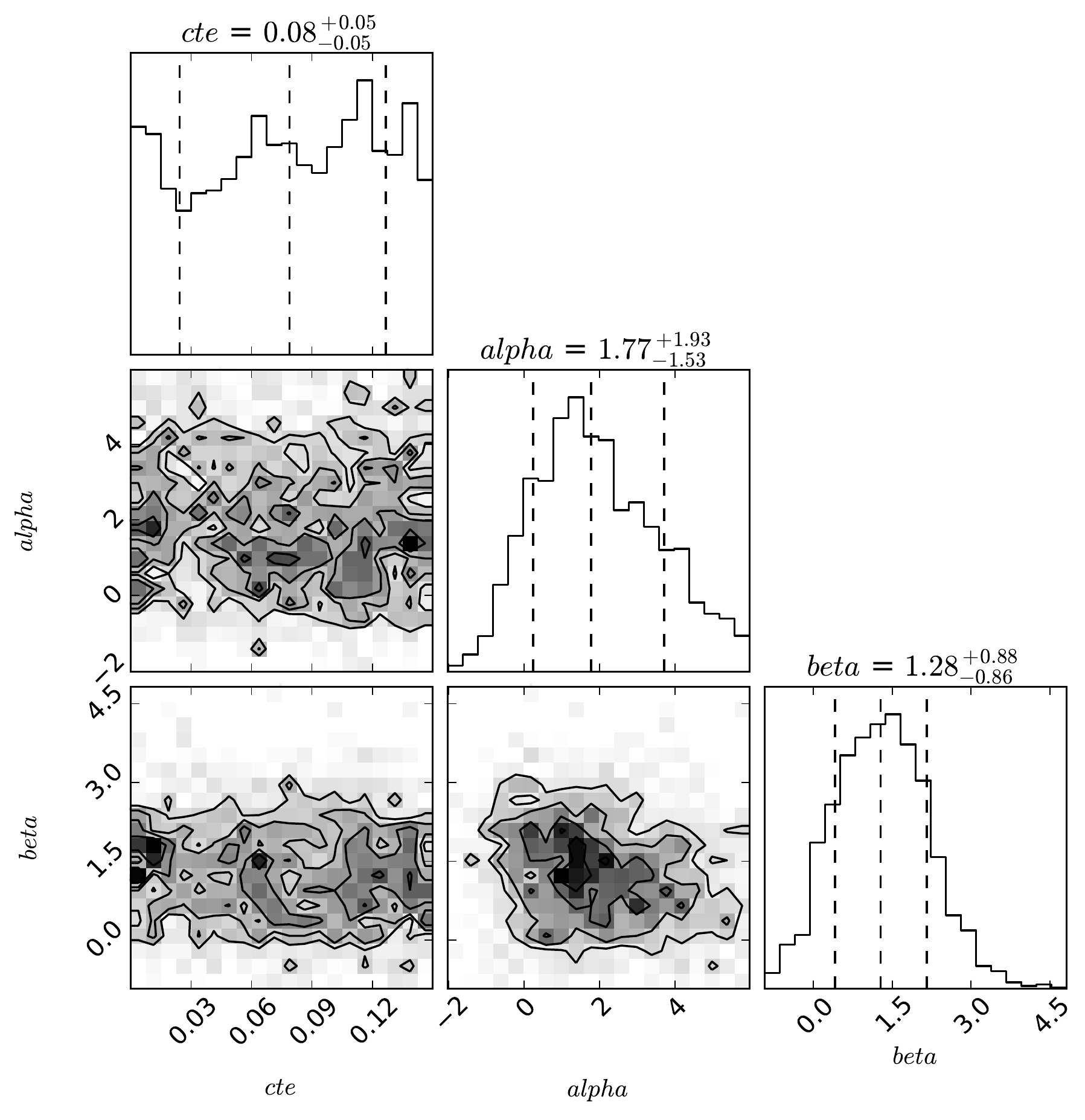}
\end{minipage}
\begin{minipage}{0.45\linewidth}
\includegraphics[scale=0.42]{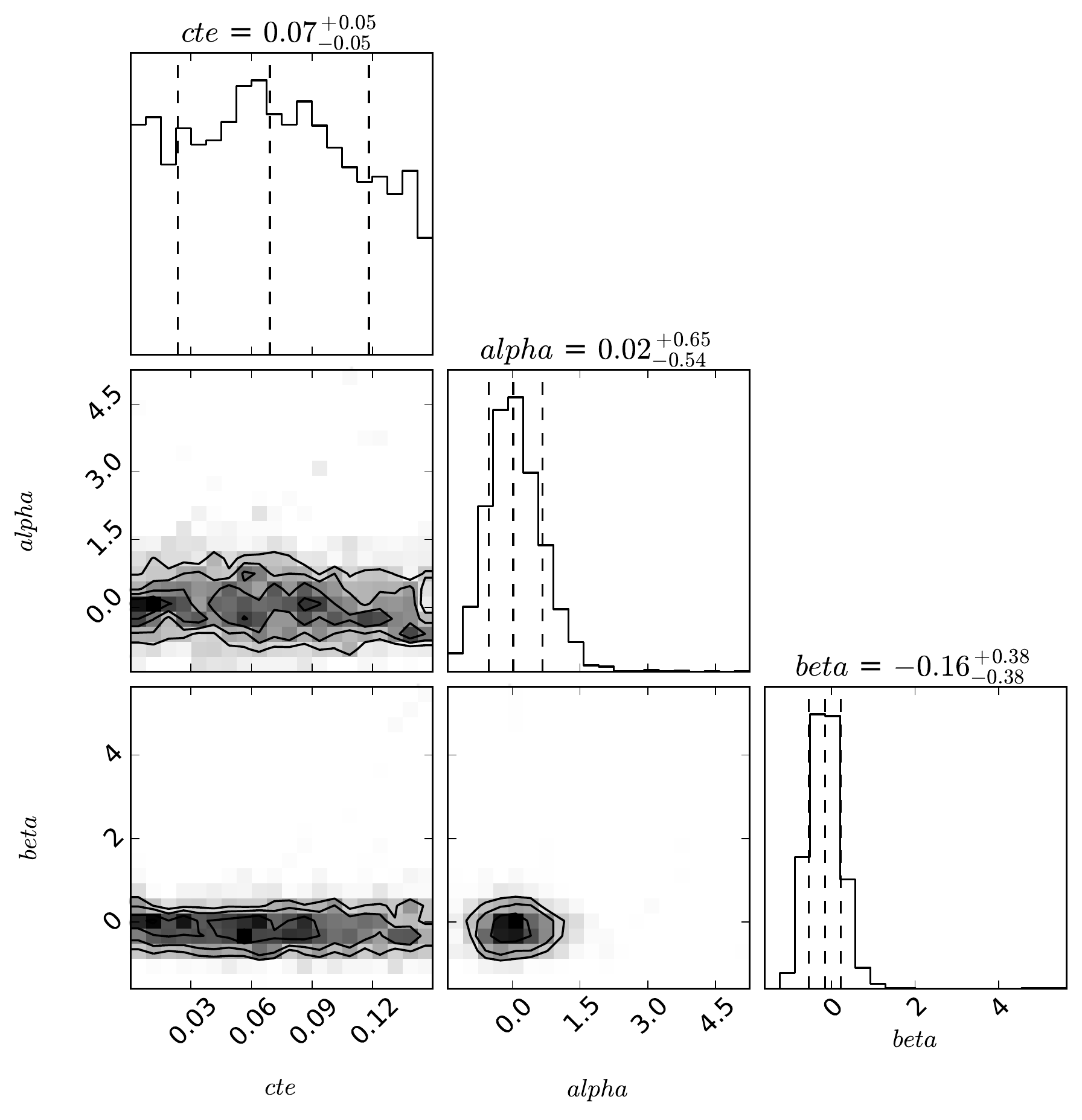}
\end{minipage}
\caption{
Marginal posterior probability distribution functions of the fit to Eqn.~\ref{eq1} for gas-giant planets (left), and for low-mass planets (right).
The metallicity values used in this analysis are those derived with the  PCA/Bayes technique developed in this work.}
\label{corner_plot}
\end{figure*}

 
 Our results are shown in Table~\ref{bayes_results} where the median values and their corresponding
 68.2\% confidence interval are provided
 while Fig.~\ref{corner_plot} shows the marginal posterior
 probability distribution functions for the model parameters for the gas-giant planets (left)
 and for low-mass planets (right). Uniformly distribution priors were chosen for the parameters.

\begin{table}
\centering
\caption{Parameters of the Bayesian fit.}
\label{bayes_results}
\begin{tabular}{cccc}
\hline
           & \multicolumn{3}{c}{\bf Gas-giant hosts}           \\
\hline
Parameter & Uniform       & Value      & 68.2\% confidence  \\
          &  prior        &            & interval         \\
\hline
$C$       & (0.00, 0.15)    & 0.08       &  (+0.05, -0.05)  \\
$\alpha$  & (-3.0, 6.0)     & 1.77       &  (+1.93, -1.53)  \\ 
$\beta$   & (-1.0, 9.0)     & 1.28       &  (+0.88, -0.86)  \\
\hline    
           & \multicolumn{3}{c}{\bf Low-mass hosts}           \\
\hline
Parameter & Uniform       & Value      & 68.2\% confidence  \\
          &  prior        &            & interval         \\
	  \hline
$C$       & (0.00, 0.15)   &  0.07       &  (+0.05, -0.05)  \\
$\alpha$  & (-3.0, 5.0)    &  0.02       &  (+0.65, -0.54)  \\      
$\beta$   & (-3.0, 6.0)    & -0.16       &  (+0.38, -0.38)  \\
\hline
 \end{tabular}
 \end{table}

  Table~\ref{comparison_pworks} shows a comparison between the results from this work and those
  previously reported in the literature.
  Our results are consistent with a planet-metallicity correlation for gas-giant planets.
  We find that the dependence of the gas-giant planet frequency on the stellar metallicity,
  $\beta$ = 1.28, is lower but
  compatible within errors
  with what is known for FGK stars, that is the frequency of gas-giant planets goes as $\sim$ 10$^{\rm 2[Fe/H]}$.
  In particular we note that our result is fully compatible with the findings of \cite{2010PASP..122..905J}, based on an
  analysis of FGKM stars including intermediate-mass subgiants. 
  Our result, however, suggests a weaker gas-giant planet-metallicity correlation 
  than recent works focused on M dwarfs that found a 
  dependency on the metallicity with $\beta$ values between 2 and 4 \citep{2013A&A...551A..36N,2014ApJ...781...28M}.
  We also find a stronger dependence of the gas-giant planet frequency on the stellar mass, $\alpha$ = 1.77, than recent works.
  That is, we find a slightly weaker gas-giant planet-metallicity
  correlation in M dwarfs, but  a stronger planet-stellar mass, so it is possible that
  lower metallicities environments can be compensated by a higher stellar (an therefore disc) mass to allow gas-giant planet formation to occur.
  This is consistent with the view
   that the mass of solids in protoplanetary discs is the main factor controlling the formation of planets and
   can be explained in the framework of core-accretion models \citep[e.g.][]{1996Icar..124...62P,2004ApJ...616..567I,2009A&A...501.1139M,2012A&A...541A..97M}.

  Regarding low-mass planets, their frequency does not seem to depend on the stellar mass
  (in disagreement with what we found in the bin fitting).
  We confirm that the frequency of low-mass planets does not depend on, or might show an
  anti-correlation with, increasing
  stellar metallicity. 



\begin{table*}
\centering
\caption{ Comparison with previous works.}
\label{comparison_pworks}
\begin{tabular}{lcccc}
\hline
       &   [Fe/H]   & M$_{\star}$ & Stellar & Notes \\
       &  parameter & parameter   & sample  &       \\
\hline
 FV05  & 2.0        &             &  FGK    &                 \\
\hline
 US07  &  2.04      &             &  FGK    & [Fe/H] $>$ 0.00 \\
       &  0.00      &             &         & [Fe/H] $<$ 0.00 \\
\hline
 JO10  &  1.2$^{\rm + 1.0}_{\rm - 1.4}$     & 1.0$^{\rm + 0.70}_{\rm - 1.30}$         &  FGKM   &                  \\
\hline
 NE13  &  2.94  $\pm$ 1.03                  &                                         &   M     &  giant hosts     \\
       &  -0.41  $\pm$ 0.77                 &                                         &   M     & low-mass hosts   \\
\hline
M014   &  3.8 $\pm$ 1.2                     & 0.8$^{\rm + 1.1}_{\rm - 0.9}$         &   M     &                  \\
\hline
This work                  &   1.28$^{\rm + 0.88}_{\rm - 0.86}$  &  1.77$^{\rm + 1.93}_{\rm - 1.53}$    &   M     &  giant hosts     \\
( PCA/Bayes)                &   -0.16 $\pm$ 0.38  &  0.02$^{\rm + 0.65}_{\rm - 0.54}$   &   M     & low-mass hosts   \\
\hline
This work                  &  3.94$^{\rm + 2.57}_{\rm - 2.21}$   &   1.65$^{\rm + 2.45}_{\rm - 2.12}$   &   M     &  giant hosts     \\
(msdlines)                 &  -1.13$^{\rm + 1.02}_{\rm - 0.99}$  &   0.22$^{\rm + 0.69}_{\rm - 0.59}$     &   M     & low-mass hosts   \\
\hline
\end{tabular}
\tablefoot{FV05: \cite{2005ApJ...622.1102F}, US07: \cite{2007ARA&A..45..397U}, JO10: \cite{2010PASP..122..905J}, NE13: \cite{2013A&A...551A..36N}, MO14: \cite{2014ApJ...781...28M}.}
\end{table*}

\begin{figure*}[htb]
\centering
\begin{minipage}{0.45\linewidth}
\includegraphics[scale=0.42]{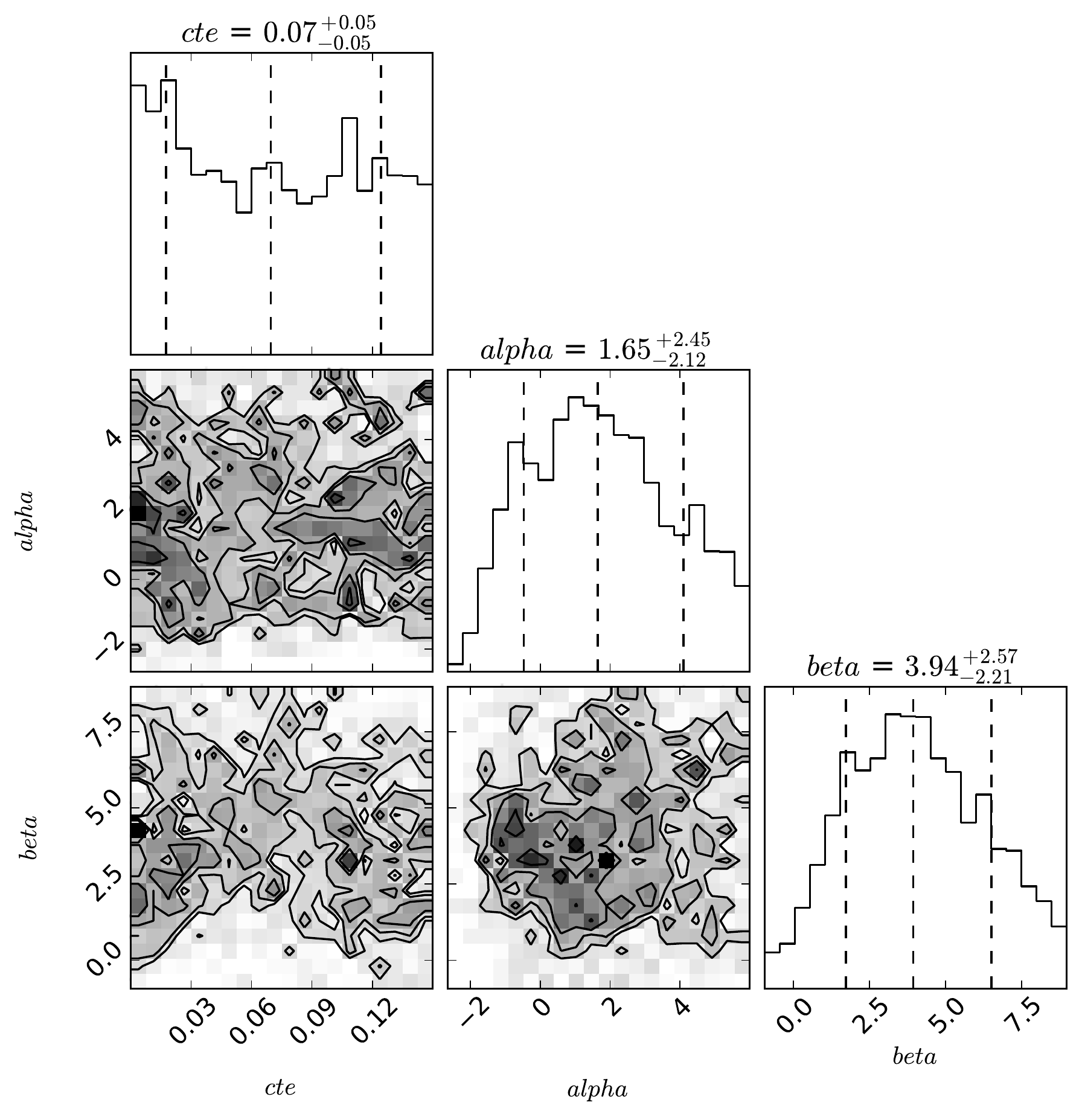}
\end{minipage}
\begin{minipage}{0.45\linewidth}
\includegraphics[scale=0.42]{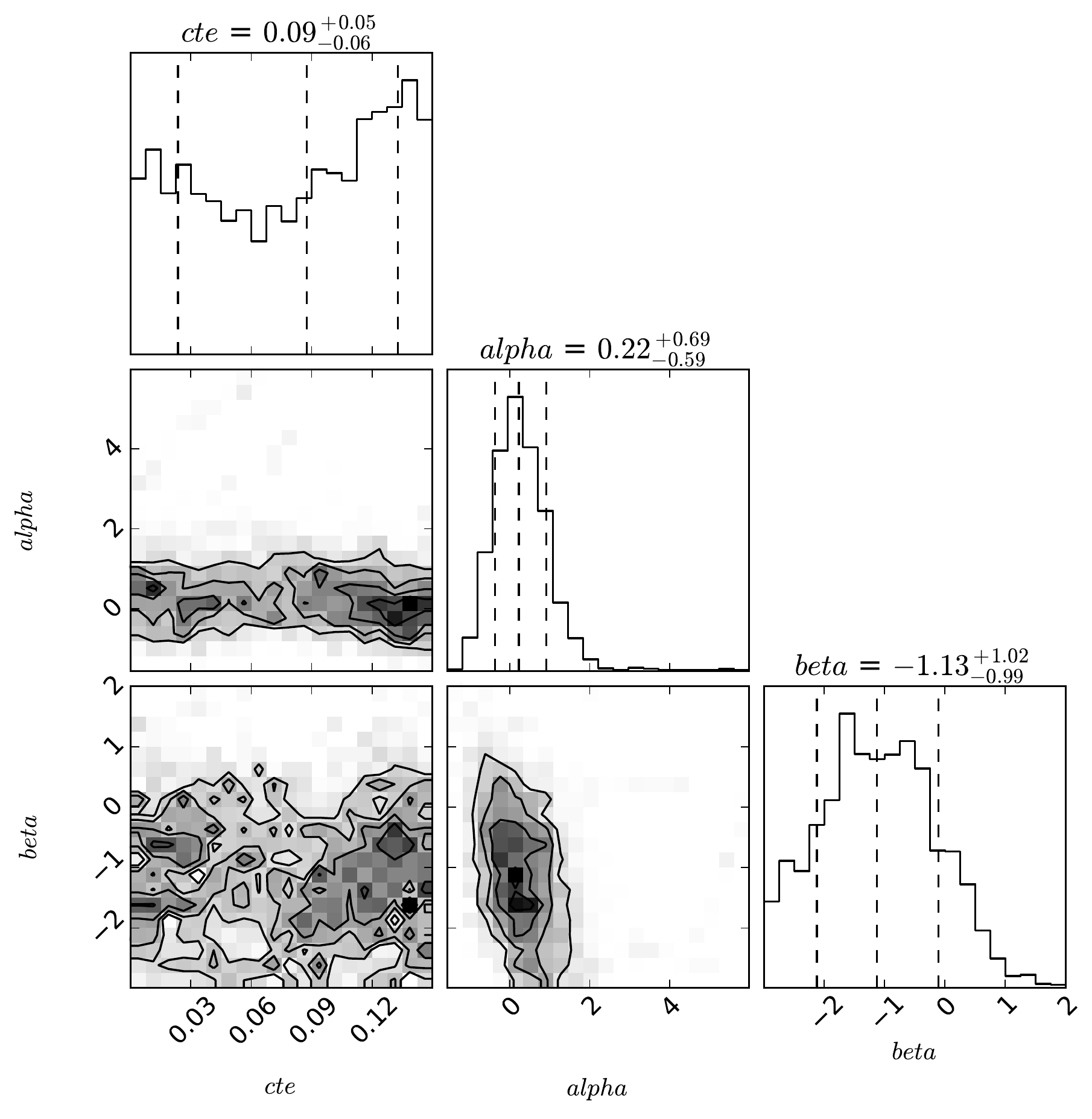}
\end{minipage}
\caption{
Marginal posterior probability distribution functions of the fit to Eqn.~\ref{eq1} for gas-giant planets (left), and for low-mass planets (right).
The metallicity values used in this analysis are those derived with the msdlines code.}
\label{corner_plot2}
\end{figure*}

  Since the value of $\beta$ derived by us is lower than the values obtained by
  other works focused on M dwarfs, we repeated our analysis using, this time, the metallicities obtained
  by the msdlines code.
  The results are C = 0.07$^{\rm +0.05}_{\rm -0.05}$, $\alpha$ = 1.65$^{\rm +2.45}_{\rm -2.12}$, and $\beta$ = 3.94$^{\rm +2.57}_{\rm -2.21}$
  for the gas-giant hosts, and  C = 0.09$^{\rm +0.05}_{\rm -0.06}$, $\alpha$ = 0.22$^{\rm +0.69}_{\rm -0.59}$, and $\beta$ = -1.13$^{\rm +1.02}_{\rm -0.99}$
  when considering the low-mass planet hosts subsample.
  The corresponding posterior distributions are shown in Figure~\ref{corner_plot2}.
  It can be seen that the values of the parameter $\alpha$ obtained using the msdlines
  metallicities are similar to the one obtained by the  PCA/Bayes values for the low-mass planet sample.
  On the contrary, the value of $\beta$ differs. The value obtained by using the  PCA/Bayes metallicities
  is compatible with a rather flat dependency of the low-mass planet frequency on the stellar metallicity. 
  However, the value obtained when using the  msdlines metallicities ($\beta$ = -1.13) 
  might indicate an anti-correlation. 
  
  On the other hand, for the gas-giant sample we obtain a similar mass dependency but, clearly, a stronger 
  correlation between the planet occurrence and the stellar metallicity. This  is in line with the works of
  \cite{2013A&A...551A..36N,2014ApJ...781...28M} who discussed 
  that M dwarfs should show a stronger giant planet-metallicity correlation than their FGK counterparts, given their smaller 
  protoplanetary disc's masses. 
  We caution that the posterior distributions of the parameters $\alpha$, and $\beta$ are very broad
  and their associated uncertainties very high. This could be related to the small size of the giant-hosts subsample. 

  Given that the msdlines metallicities are shifted towards lower values when compared with the nearby
  FGK stars (see Fig.~\ref{metal_ECDF}) we speculate whether the very high $\beta$ values found in previous
  works ($\sim$ 3, 4) and with our msdlines metallicities may be related to the use of a photometric scale. 
  However, further analysis should be done to clarify this issue.

\subsection{Other chemical signatures}

\begin{figure*}[!htb]
\centering
\includegraphics[scale=0.65]{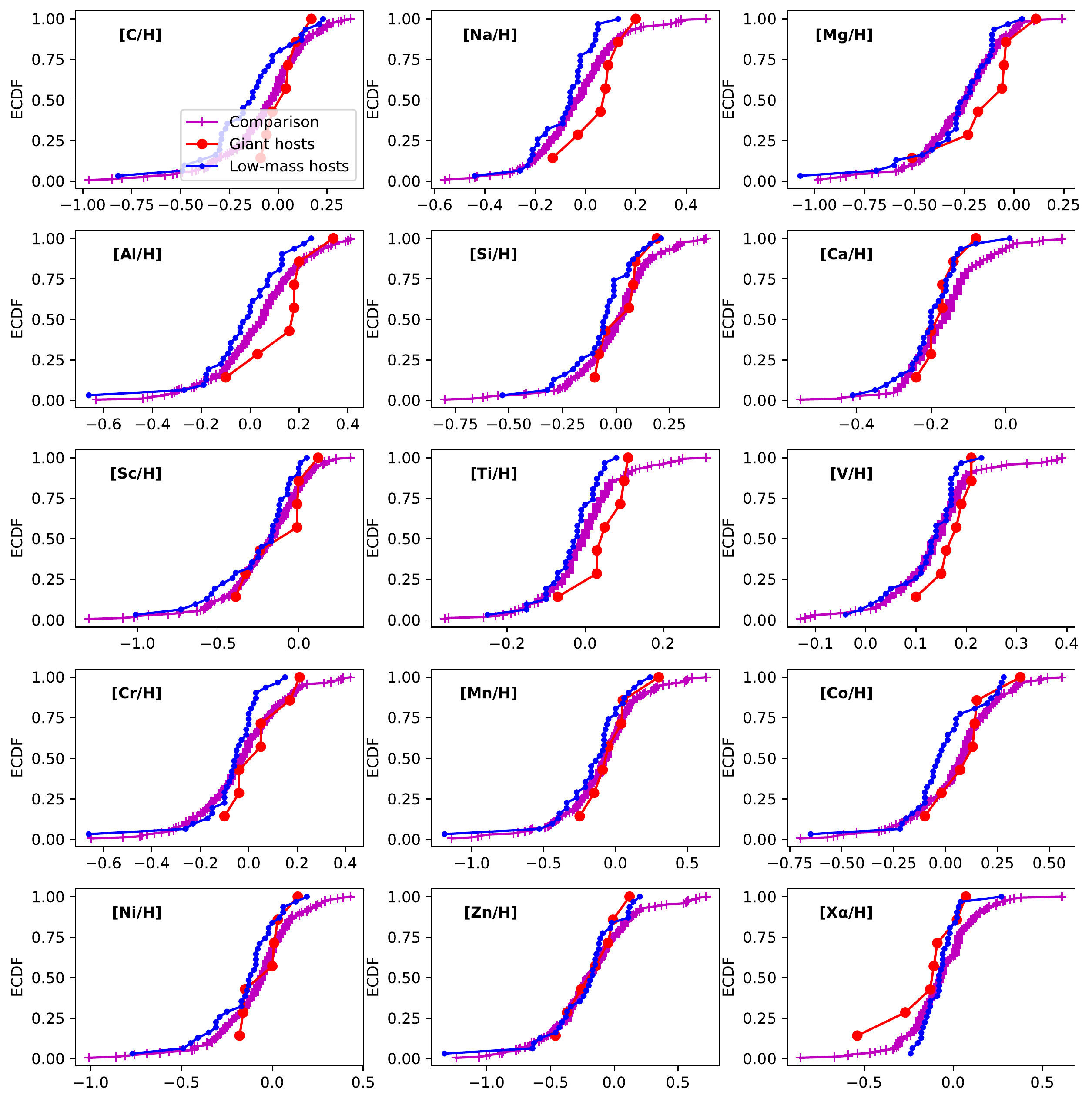}
\caption{ 
[X/H] cumulative fraction of low-mass hosts (blue), gas-giant hosts (red), and
comparison M dwarfs (purple).}
\label{ecdf_mdwarfs}
\end{figure*}

 To try to disclose differences in the abundances of other chemical elements besides iron,
 we show in Fig.~\ref{ecdf_mdwarfs} the cumulative distribution of the abundances  [X/H] 
 of the comparison, the low-mass hosts, and the giant hosts subsamples.
 Some statistic diagnostics are shown in Table~\ref{other_abundances_table}, where the results of a KS test
 for each element are also listed.

 Similar behaviour between planet and non-planet hosts is found in all elements,
 both for gas-giant and low-mass planets.
  However, a small hint of higher abundances of Ti, Na, Mg, and Al can be seen for gas-giant planet hosts.
 Slightly lower abundances of Co for low-mass planet hosts can also be seen. 
 Since these trends are not statistically significant
 we conclude that our results are
 similar to what is found in FGK stars where planet hosts have been shown to have
 indiscernible abundances from stars without planets  
 \citep[e.g.][]{2003A&A...404..715B,2006A&A...445..633E,2006PASP..118.1494G,2006A&A...449..723G,2011A&A...526A..71D,2012A&A...543A..89A}.
 We included the $\alpha$ abundance in our analysis as \cite{2012A&A...547A..36A,2012A&A...543A..89A} 
 showed that stars with planets at low-metallicity values tend to belong to the thick disc and show high $\alpha$ values,
 but besides the general trend of higher $\alpha$ abundances in less-metallic stars discussed 
 in Sect.~\ref{abkin}, 
 no differences in $\alpha$ abundances between planet and
 non-planet hosts were found.
 It is important to note that chemical differences between planet and non planet hosts, when claimed, are
 rather modest. For instance, a depletion of only $\sim$ 0.08 dex on refractory relative to volatile
 elements have been found in the Sun and other solar analogues \citep[e.g.][]{2009ApJ...704L..66M,2009A&A...508L..17R}.
 In Table~\ref{standard_deviation} we list the standard deviation of the differences between the  PCA-derived abundances and those measured in the corresponding primaries for our training dataset. These values can be considered as a lower limit of the uncertainties derived. 
 It is clear that we are still far from achieving the precision
 that can be obtained in the abundance analysis of FGK stars, so
 small chemical differences beyond the limits of our precision can still be present in our sample.

\begin{table*}
\centering
\caption{Comparison between the elemental abundances of the different subsamples.}
\label{other_abundances_table}
\begin{tabular}{lcccccccccc}
\hline
 Ion   & \multicolumn{2}{c}{Comparison}  &  \multicolumn{2}{c}{Giant hosts} &  \multicolumn{2}{c}{Low-mass hosts} & \multicolumn{2}{c}{K-S test comparison/giants} &  \multicolumn{2}{c}{K-S test comparison/low-mass} \\
 &  Mean & $\sigma$ &   Mean & $\sigma$ &  Mean & $\sigma$ & $D$ & $p$-value & $D$ & $p$-value \\
\hline
C & -0.03  &	0.22 &	0.04 &	0.08 &	-0.13 &	0.22 &	0.38 &	0.22 &	0.25 &	0.06 \\
Na	&	-0.03	&	0.17	&	0.08	&	0.10	&	-0.06	&	0.12	&	0.44	&	0.10	&	0.24	&	0.07	\\
Mg	&	-0.25	&	0.21	&	-0.06	&	0.18	&	-0.24	&	0.22	&	0.44	&	0.10	&	0.12	&	0.80	\\
Al	&	0.05	&	0.18	&	0.18	&	0.13	&	-0.01	&	0.17	&	0.48	&	0.06	&	0.19	&	0.28	\\
Si	&	0.02	&	0.19	&	0.06	&	0.10	&	-0.05	&	0.15	&	0.25	&	0.74	&	0.28	&	0.03	\\
Ca	&	-0.16	&	0.10	&	-0.17	&	0.05	&	-0.20	&	0.08	&	0.22	&	0.85	&	0.22	&	0.14	\\
Sc	&	-0.17	&	0.26	&	-0.01	&	0.18	&	-0.17	&	0.24	&	0.35	&	0.30	&	0.15	&	0.58	\\
Ti	&	-0.01	&	0.10	&	0.05	&	0.06	&	-0.02	&	0.07	&	0.51	&	0.04	&	0.18	&	0.35	\\
V	&	0.14	&	0.09	&	0.18	&	0.04	&	0.14	&	0.06	&	0.38	&	0.24	&	0.14	&	0.63	\\
Cr	&	-0.03	&	0.18	&	0.05	&	0.11	&	-0.05	&	0.14	&	0.31	&	0.47	&	0.25	&	0.06	\\
Mn	&	-0.07	&	0.28	&	-0.05	&	0.16	&	-0.11	&	0.26	&	0.22	&	0.85	&	0.18	&	0.35	\\
Co	&	0.08	&	0.21	&	0.13	&	0.14	&	-0.03	&	0.18	&	0.24	&	0.77	&	0.30	&	0.01	\\
Ni	&	-0.06	&	0.23	&	0.00	&	0.11	&	-0.12	&	0.20	&	0.27	&	0.66	&	0.23	&	0.10	\\
Zn	&	-0.22	&	0.34	&	-0.15	&	0.19	&	-0.17	&	0.29	&	0.19	&	0.95	&	0.13	&	0.71	\\
\hline
X$_{\alpha}$ & -0.07 & 0.20 &  -0.11 & 0.19 & -0.08 & 0.10 &  0.25 & 0.73 & 0.20 & 0.23 \\
\hline
\end{tabular} 
\end{table*}

\section{Conclusions}\label{conclusions}

  M dwarfs are considered promising targets in the search for small, rocky,
  potentially habitable planets. They also constitute valuable tools to understand
  planet formation as a function of the stellar mass. 
  However the analysis of their optical spectra has proven to be a difficult task.
  In this work we develop a technique to derive stellar abundances
  of M dwarfs from the same optical, high-resolution spectra that are used
  in exoplanet searches, that is, without the necessity of relying on additional infrared observations
  or atmospheric models.

  Our methodology is based on the use of principal component analysis and Bayesian regression methods.
  The procedure is trained with spectra of M dwarfs orbiting around an FGK primary.
  We found the agreement between the abundances derived from our method and those
  measured in the FGK primary to be better than 0.10 dex in most of the cases.
  While it is true that these dispersions should be regarded as lower limits of the uncertainties
  of our technique, we believe that it is encouraging to see 
  that they are of the same order of magnitude than the dispersion
  found in the individual abundances of FGK stars when measured from different lines of
  the same ion.
  A cross-validation of the technique performed on FGK
  stellar spectra shows that in spite of the small training sample
  used, the methodology seems to give reliable abundances.

  Since we are lacking a suitable comparison sample, we are forced to compare our derived abundances
  for M dwarfs with the known trends for FGK stars. We find similar [X/H] vs. [Fe/H] trends for M dwarfs
  and FGK stars in most of the ions. 
  We also show that our  PCA/Bayes metallicities reproduce the metallicity scale of nearby FGK stars.
  On the contrary, metallicity values derived from photometry provide a ECDF distribution shifted towards
  lower metallicities values in comparison with FGK stars. 
  Nevertheless, we recognise that the methodology can be improved by observing more training stars. This is specially
  evident at low-metallicities. Further observations of M dwarfs in the near infrared domain will help us
  to validate our techniques. 

  To the very best of our knowledge, we performed for the first time a detailed chemical analysis of several elements for a large
  sample of M dwarfs with and without gas-giant and low-mass planets. We note that while it is unlikely that
  giant planets may be hidden in the sample, we can not discard the possibility that there are more non-detected low-mass planets 
  in the different subsamples analysed in this work. A comparison of the properties of the different subsamples reveals no differences in age or kinematics
  between planet and non-planet hosts. We find planet hosts to be located at lower distances than the comparison sample.
  This, likely, reveals the fact that closer stars are brighter and easier to observe at higher S/N.

  Regarding the dependency of the planetary fraction on the stellar metallicity and mass, our results confirm previous works.
  That is, gas-giant planet hosts show a planet-metallicity correlation as well as a dependence with the stellar mass.
  We note, however, that our  PCA/Bayes metallicities predict a weaker planet metallicity correlation and a stronger mass-dependency
  than photometric values for giant-planet hosts. 
  On the other hand the frequency of low-mass planets is not a function of the stellar metallicity, although a weak 
  anti-correlation might be present. 
  We also find an anti-correlation
  between the frequency of low-mass planets and the stellar mass in line with recent results from the {\sc Kepler} mission. 
  However, we caution that this anti-correlation was found only in the bin fitting analysis, but not
  in the Bayesian fit. 
  For other elements besides iron, we do not find differences in the abundance distributions of stars with and without planets.
  These results are in line with what is known from the chemical analysis of solar-type stars and can be explained within the
  framework of core-accretion models.

  We finally note that the homogeneous determination of stellar abundances of M dwarfs might be of interest for many other
  studies dealing with the local properties of the Galaxy, for example the study of stars in clusters and other stellar
  associations. Our codes are public available\footnote{https://github.com/jesusmaldonadoprado/mdwarfs\_abundances}, so the whole community can
  benefit from them.

\begin{acknowledgements}

  This research was supported by the Italian Ministry of Education,
  University, and Research  through the
  \emph{PREMIALE WOW 2013} research project under grant
  \emph{Ricerca di pianeti intorno a stelle di piccola massa}.
  J.M. acknowledges support from the  Accordo Attuativo ASI-INAF n. 2018.22.HH.O,
    \emph {Partecipazione alla fase B1 della missione Ariel (ref. G. Micela).}
  M.P., I.R., J.C.M, and E.H acknowledge support from the Spanish Ministry of Science and Innovation and the European Regional Development Fund through grant PGC2018-098153-B- C33, as well as the support of the Generalitat de Catalunya/CERCA programme.
  J.I.G.H. acknowledges financial support from Spanish Ministry of Science and Innovation (MICINN)  under the 2013 Ram\'on y Cajal program RYC-2013-14875. A.S.M. acknowledges financial support from the Spanish MICINN under the 2019 Juan de la Cierva Programme. B.T.P. acknowledges Fundaci\'on La Caixa for the financial support received in the form of a Ph.D. contract. J.I.G.H., R.R., A.S.M., B.T.P. acknowledge financial support from the Spanish MICINN AYA2017-86389-P. 
  This research has made use of the NASA Exoplanet Archive, which is operated by the California Institute of Technology, under contract with the National Aeronautics
  and Space Administration under the Exoplanet Exploration Program.
  This work has made use of data from the European Space Agency (ESA) mission
  {\it Gaia} (\url{https://www.cosmos.esa.int/gaia}), processed by the {\it Gaia}
  Data Processing and Analysis Consortium (DPAC,
  \url{https://www.cosmos.esa.int/web/gaia/dpac/consortium}). Funding for the DPAC
  has been provided by national institutions, in particular the institutions
  participating in the {\it Gaia} Multilateral Agreement.
  Additional data collected at the European Southern Observatory 
  and Telescopio Nazionale Galileo under programme ID
  GAPS,
  082.C-0718(B),
  072.C-0488(E),
  183.C-0437(A),
  180.C-0886(A),
  198.C-0838(A),
  0101.C-0516(A),
  1102.C-0339(A),
  183.C-0972(A),
  096.C-0460(A),
  082.C-0718(A),
  191.C-0505(A),
  096.C-0499(A),
  191.C-0873,
  191.C-0873(A),
  088.C-0662(B),
  089.C-0497(A),
  090.C-0395(A),
  087.C-0831(A),
  093.C-0409(A),
  60.A-9036(A),
  192.C-0224(G),
  192.C-0224(H),
  192.C-0224(C),
  CAT14B\_162,
  CAT16A\_99,
  198.C-0873(A),
  CAT17A\_95,
  CAT14A\_43,
  OPT16A\_11,
  CAT16A\_134,
  092.C-0721(A),
  0100.C-0097(A),
  0101.C-0379(A),
  0102.C-0558(A),
  CAT13B\_170,
  CAT14A\_128,
  185.D-0056(A),
  185.D-0056(B),
  185.D-0056(K),
  076.C-0155(A),
  086.C-0284(A),
  089.C-0732(A),
  090.C-0421(A),
  0101.D-0494(A),
  192.C-0224(B),
  099.C-0205(A),
  085.C-0019(A),
  495.L-0963(A),
  097.C-0390(B),
  099.C-0798(A),
  0100.C-0487(A),
  0101.D-0494(B),
  095.C-0718(A),
  077.C-0364(E),
  097.C-0864(B),
  099.C-0880(A),
  075.C-0202(A),
  192.C-0224,
  098.C-0739(A),
  60.A-9709(G),
  060.A-9709(G),
  074.C-0364(A),
  078.C-0044(A),
  097.C-0561(A),
  075.D-0614(A),
  have been used. 
  We sincerely appreciate the careful reading of the manuscript and the constructive comments of the anonymous referee.

\end{acknowledgements}

%
\bibliographystyle{aa}
\bibliography{abundances_mdwarfs.bib}


\begin{appendix}
\section{Additional tables}


\longtab[1]{
\begin{landscape}
\begin{longtable}{lcccccccc}
\caption{
 { Basic properties of the stars considered in this work.}
  }\label{parameters_table_full}\\
\hline
 Star & T$_{\rm eff}$ & SpType & [Fe/H]  & M$_{\star}$   & R$_{\star}$   & $\log(g)$ & $\log(L_{\star}/L_{\odot})$ & Age   \\
      &   (K)         &        &  (dex)  & (M$_{\odot}$) & (R$_{\odot}$) &  (cm s$^{\rm -2}$)    &                             & (Gyr) \\
\hline
\endfirsthead
\caption{Continued.} \\
\hline
 Star & T$_{\rm eff}$ & SpType & [Fe/H]  & M$_{\star}$   & R$_{\star}$    & $\log(g)$ &  $\log(L_{\star}/L_{\odot})$  & Age   \\
       &   (K)         &        &  (dex)  & (M$_{\odot}$) & (R$_{\odot}$) & (cm s$^{\rm -2}$)    &                            & (Gyr) \\
\hline
\endhead
\hline
\endfoot
\hline
\endlastfoot
\hline\noalign{\smallskip}
GJ 2	&	3709	$\pm$	68	&	M1.0	&	-0.14	$\pm$	0.09	&	0.50	$\pm$	0.05	&	0.49	$\pm$	0.05	&	4.76	$\pm$	0.04	&	-1.385	$\pm$	0.087	&	8.28	$\pm$	4.11	\\
Gl 1	&	3478	$\pm$	68	&	M2.5	&	-0.26	$\pm$	0.09	&	0.33	$\pm$	0.07	&	0.33	$\pm$	0.06	&	4.91	$\pm$	0.06	&	-1.833	$\pm$	0.169	&	8.27	$\pm$	4.11	\\
GJ  1002	&	3378	$\pm$	69	&	M4.0	&	-0.25	$\pm$	0.10	&	0.22	$\pm$	0.10	&	0.24	$\pm$	0.09	&	5.01	$\pm$	0.10	&	-2.176	$\pm$	0.344	&				\\
GJ 3014	&	3692	$\pm$	69	&	M1.5	&	-0.19	$\pm$	0.09	&	0.48	$\pm$	0.05	&	0.47	$\pm$	0.05	&	4.79	$\pm$	0.04	&	-1.437	$\pm$	0.090	&	8.22	$\pm$	4.10	\\
Gl 12	&	3387	$\pm$	68	&	M3.0	&	-0.29	$\pm$	0.09	&	0.22	$\pm$	0.10	&	0.24	$\pm$	0.09	&	5.01	$\pm$	0.09	&	-2.180	$\pm$	0.327	&	7.20	$\pm$	4.23	\\
GJ 16	&	3679	$\pm$	68	&	M1.5	&	-0.16	$\pm$	0.09	&	0.48	$\pm$	0.05	&	0.47	$\pm$	0.05	&	4.78	$\pm$	0.04	&	-1.434	$\pm$	0.089	&	8.23	$\pm$	4.10	\\
GJ 15A	&	3610	$\pm$	68	&	M1.0	&	-0.35	$\pm$	0.09	&	0.38	$\pm$	0.05	&	0.38	$\pm$	0.05	&	4.87	$\pm$	0.04	&	-1.659	$\pm$	0.112	&				\\
GJ 15B	&	3386	$\pm$	68	&	M3.0	&	-0.43	$\pm$	0.09	&	0.16	$\pm$	0.10	&	0.19	$\pm$	0.09	&	5.06	$\pm$	0.09	&	-2.387	$\pm$	0.416	&				\\
\end{longtable}
\tablefoot{$^{\star}$ 2MASS J22353504+3712131; $^{\dag}$ evolutionary parameters computed from photometry.
}%
\end{landscape}
}


\longtab[2]{
\begin{landscape}
\begin{longtable}{lcccccccc}
\caption{
 { Kinematic data for the observed stars.}
  }\label{parameters_table_kinematics}\\
\hline
 Star & $\pi$    & $\mu_{\alpha}$      & $\mu_{\delta}$      & V$_{\rm rad}$     &    $U$            & $V$               & $W$               & Notes \\
      &  (mas)   & (mas yr$^{\rm -1}$) & (mas yr$^{\rm -1}$) & (km s$^{\rm -1}$) & (km s$^{\rm -1}$) & (km s$^{\rm -1}$) & (km s$^{\rm -1}$) &       \\ 
\hline
\endfirsthead
\caption{Continued.} \\
\hline
 Star & $\pi$    & $\mu_{\alpha}$      & $\mu_{\delta}$      & V$_{\rm rad}$     &    $U$            & $V$               & $W$               & Notes \\
       &  (mas)   & (mas yr$^{\rm -1}$) & (mas yr$^{\rm -1}$) & (km s$^{\rm -1}$) & (km s$^{\rm -1}$) & (km s$^{\rm -1}$) & (km s$^{\rm -1}$) &       \\
\hline
\endhead
\hline
\endfoot
\hline
\endlastfoot
\hline\noalign{\smallskip}
GJ 2	&	86.9567	$\pm$	0.0407	&	870.753	$\pm$	0.057	&	-151.267	$\pm$	0.036	&	-0.23	$\pm$	0.0013	&	-38.96	$\pm$	0.02	&	-23.28	$\pm$	0.01	&	-16.18	$\pm$	0.01	&	D	\\
Gl 1	&	230.1331	$\pm$	0.0600	&	5633.374	$\pm$	0.170	&	-2334.794	$\pm$	0.075	&	25.53	$\pm$	0.0030	&	-74.51	$\pm$	0.02	&	-97.38	$\pm$	0.03	&	-37.35	$\pm$	0.00	&	TD	\\
GJ  1002	&	206.2134	$\pm$	0.1281	&	-811.480	$\pm$	0.206	&	-1892.905	$\pm$	0.125	&	-33.70	$\pm$	3.8000	&	38.14	$\pm$	0.07	&	-39.17	$\pm$	1.44	&	19.72	$\pm$	3.52	&	D	\\
GJ 3014	&	51.9505	$\pm$	0.0270	&	250.623	$\pm$	0.051	&	184.890	$\pm$	0.048	&	-15.61	$\pm$	0.0028	&	-11.69	$\pm$	0.01	&	-29.24	$\pm$	0.01	&	7.72	$\pm$	0.01	&	D	\\
Gl 12	&	81.8711	$\pm$	0.0866	&	618.003	$\pm$	0.163	&	329.319	$\pm$	0.139	&	51.21	$\pm$	0.0036	&	-51.03	$\pm$	0.04	&	26.66	$\pm$	0.01	&	-30.85	$\pm$	0.01	&	TR	\\
GJ 16	&	59.5952	$\pm$	0.0950	&	2.385	$\pm$	0.185	&	-29.318	$\pm$	0.084	&	-15.10	$\pm$	0.4000	&	4.04	$\pm$	0.08	&	-10.41	$\pm$	0.23	&	10.43	$\pm$	0.31	&	D	\\
GJ 15A	&	280.6902	$\pm$	0.0429	&	2891.525	$\pm$	0.061	&	411.903	$\pm$	0.034	&	11.82	$\pm$	0.0007	&	-49.15	$\pm$	0.01	&	-12.04	$\pm$	0.00	&	-3.45	$\pm$	0.00	&	D	\\
GJ 15B	&	280.7866	$\pm$	0.0519	&	2863.284	$\pm$	0.069	&	336.529	$\pm$	0.039	&	11.00	$\pm$	0.0007	&	-48.04	$\pm$	0.01	&	-12.78	$\pm$	0.00	&	-4.31	$\pm$	0.00	&	D	\\
\end{longtable}
\tablefoot{$^{\star}$ 2MASS J22353504+3712131; D: Thin disk, TD: Thick disck, TR: Transition, H: Halo
}%
\end{landscape}
}


\longtab[3]{
\begin{longtable}{llrrrr}
\caption{
 {  Planet host stars in the sample, along with the planetary properties.}
  }\label{table_planets}\\
\hline
Star    &              Planet  &       P (days) &              a (au)  &       eccentricity    & m$\sin i$$^{\dag}$ (M$_{\oplus}$)             \\
\hline
\endfirsthead
\caption{Continued.} \\
\hline
Star    &              Planet  &       P (days) &              a (au)  &       eccentricity    & m$\sin i$$^{\dag}$ (M$_{\oplus}$)             \\
\hline
\endhead
\hline
\endfoot
\hline
\endlastfoot
GJ 15A	&	b	&	11.4407	&	0.072	&	0.094	&	3.03	  \\  
GJ 15A	&	c	&	7600	&	5.4	&	0.27	&	36	  \\  
GJ 27.1	&	b	&	15.819	&	0.101	&	0.08	&	13	  \\  
GJ 49	&	b	&	13.8508	&	0.0905	&	0.363	&	5.63	  \\  
Gl 54.1	&	b	&	1.96876	&	0.01557	&	0	&	0.75	  \\  
Gl 54.1	&	c	&	3.06008	&	0.0209	&	0.04	&	0.98	  \\  
Gl 54.1	&	d	&	4.65627	&	0.02764	&	0.129	&	1.14	  \\  
LP 771-96&	b	&	5.35882	&	0.03807	&	0.19	&	2.2$^{\dag}$ \\
\end{longtable}
\tablefoot{$^{\dag}$ True mass
}%
}


\longtab[4]{
\begin{landscape}
\begin{longtable}{lrrrrrrrrrrrrrrrr}
\caption{
 { Derived abundances, [X/H] (dex), for the M stars analyzed in this work.}
  }\label{abundances_table_full}\\
\hline
Star & 	[Fe/H]	&	[C/H]	&		[Na/H]	&	[Mg/H]	&	[Al/H]	&	[Si/H]	&	[Ca/H]	&	[Sc/H]	&	[Ti/H]	&	[V/H]	&	[Cr/H]	&	[Mn/H]	&	[Co/H]	&	[Ni/H]	&	[Zn/H]	\\
\hline
\endfirsthead
\caption{Continued.} \\
\hline
Star &	[Fe/H]	&	[C/H]	&		[Na/H]	&	[Mg/H]	&	[Al/H]	&	[Si/H]	&	[Ca/H]	&	[Sc/H]	&	[Ti/H]	&	[V/H]	&	[Cr/H]	&	[Mn/H]	&	[Co/H]	&	[Ni/H]	&	[Zn/H]	\\
\hline
\endhead
\hline
\endfoot
\hline
\endlastfoot
\hline\noalign{\smallskip}
GJ 2	&	-0.12	&	-0.02	&	-0.05	&	-0.43	&	0.03	&	-0.01	&	-0.21	&	-0.31	&	-0.03	&	0.09	&	-0.05	&	-0.09	&	0.10	&	-0.08	&	-0.27	\\
Gl 1	&	-0.33	&	-0.38	&	-0.12	&	-0.44	&	-0.07	&	-0.36	&	-0.27	&	-0.63	&	0.01	&	0.20	&	-0.26	&	-0.59	&	-0.29	&	-0.51	&	-0.74	\\
GJ  1002	&	-0.01	&	-0.07	&	-0.33	&	0.01	&	-0.28	&	0.23	&	-0.04	&	0.32	&	-0.19	&	0.05	&	-0.02	&	0.24	&	0.03	&	0.27	&	0.64	\\
GJ 3014	&	-0.36	&	0.01	&	-0.29	&	-0.51	&	-0.02	&	-0.01	&	-0.26	&	-0.28	&	-0.21	&	-0.10	&	-0.24	&	-0.22	&	0.08	&	-0.07	&	-0.50	\\
Gl 12	&	0.07	&	0.00	&	-0.04	&	-0.22	&	-0.11	&	0.10	&	-0.14	&	-0.03	&	-0.04	&	0.10	&	-0.05	&	-0.06	&	0.10	&	0.03	&	-0.14	\\
GJ 16	&	-0.01	&	0.01	&	0.03	&	-0.33	&	0.15	&	0.04	&	-0.25	&	-0.25	&	0.03	&	0.10	&	-0.01	&	0.04	&	0.26	&	-0.04	&	-0.19	\\
GJ 15A	&	-0.34	&	-0.48	&	-0.06	&	-0.69	&	-0.18	&	-0.29	&	-0.20	&	-0.73	&	-0.03	&	0.17	&	-0.17	&	-0.53	&	-0.22	&	-0.45	&	-0.58	\\
GJ 15B	&	-0.30	&	-0.33	&	-0.27	&	-0.47	&	-0.31	&	-0.08	&	-0.11	&	-0.32	&	-0.17	&	0.04	&	-0.26	&	-0.29	&	-0.10	&	-0.17	&	-0.28	\\
\end{longtable}
\tablefoot{$^{\star}$ 2MASS J22353504+3712131.
}%
\end{landscape}
}

\end{appendix}

\end{document}